\PassOptionsToPackage{unicode}{hyperref}
\PassOptionsToPackage{hyphens}{url}
\PassOptionsToPackage{dvipsnames,svgnames,x11names}{xcolor}
\documentclass[
  12pt]{article}

\usepackage{amsmath,amssymb}

\RequirePackage{placeins}
\RequirePackage[OT1]{fontenc}
\RequirePackage{amsthm,amsmath,amsfonts,amssymb}
\RequirePackage[colorlinks]{hyperref}
\RequirePackage{float}
\RequirePackage{epsfig, graphicx, color}
\RequirePackage{times}
\RequirePackage{latexsym}
\RequirePackage{psfrag}
\RequirePackage{color}
\usepackage{natbib}
\usepackage{xcolor}
\usepackage{cleveref}
\usepackage{multirow}
\usepackage{multicol}
\usepackage{makecell}
\usepackage{epsf,epsfig}
\usepackage{graphics,graphicx}
\usepackage{enumerate}
\usepackage{algorithmicx,algorithm}
\usepackage{bbm}
\usepackage{bm}
\usepackage{mathtools}

\usepackage{iftex}
\ifPDFTeX
  \usepackage[T1]{fontenc}
  \usepackage[utf8]{inputenc}
  \usepackage{textcomp} 
\else 
  \usepackage{unicode-math}
  \defaultfontfeatures{Scale=MatchLowercase}
  \defaultfontfeatures[\rmfamily]{Ligatures=TeX,Scale=1}
\fi
\usepackage{lmodern}
\ifPDFTeX\else  
\fi
\IfFileExists{upquote.sty}{\usepackage{upquote}}{}
\IfFileExists{microtype.sty}{
  \usepackage[]{microtype}
  \UseMicrotypeSet[protrusion]{basicmath} 
}{}
\makeatletter
\@ifundefined{KOMAClassName}{
  \IfFileExists{parskip.sty}{%
    \usepackage{parskip}
  }{
    \setlength{\parindent}{0pt}
    \setlength{\parskip}{6pt plus 2pt minus 1pt}}
}{
  \KOMAoptions{parskip=half}}
\makeatother
\usepackage{xcolor}
\makeatletter
\ifx\paragraph\undefined\else
  \let\oldparagraph\paragraph
  \renewcommand{\paragraph}{
    \@ifstar
      \xxxParagraphStar
      \xxxParagraphNoStar
  }
  \newcommand{\xxxParagraphStar}[1]{\oldparagraph*{#1}\mbox{}}
  \newcommand{\xxxParagraphNoStar}[1]{\oldparagraph{#1}\mbox{}}
\fi
\ifx\subparagraph\undefined\else
  \let\oldsubparagraph\subparagraph
  \renewcommand{\subparagraph}{
    \@ifstar
      \xxxSubParagraphStar
      \xxxSubParagraphNoStar
  }
  \newcommand{\xxxSubParagraphStar}[1]{\oldsubparagraph*{#1}\mbox{}}
  \newcommand{\xxxSubParagraphNoStar}[1]{\oldsubparagraph{#1}\mbox{}}
\fi
\makeatother

\usepackage{longtable,booktabs,array}
\usepackage{calc} 
\usepackage{etoolbox}
\makeatletter
\patchcmd\longtable{\par}{\if@noskipsec\mbox{}\fi\par}{}{}
\makeatother
\IfFileExists{footnotehyper.sty}{\usepackage{footnotehyper}}{\usepackage{footnote}}
\makesavenoteenv{longtable}
\usepackage{graphicx}
\makeatletter
\def\maxwidth{\ifdim\Gin@nat@width>\linewidth\linewidth\else\Gin@nat@width\fi}
\def\maxheight{\ifdim\Gin@nat@height>\textheight\textheight\else\Gin@nat@height\fi}
\makeatother
\setkeys{Gin}{width=\maxwidth,height=\maxheight,keepaspectratio}
\makeatletter
\def\fps@figure{htbp}
\makeatother

\makeatletter
\@ifpackageloaded{caption}{}{\usepackage{caption}}
\AtBeginDocument{%
\ifdefined\contentsname
  \renewcommand*\contentsname{Table of contents}
\else
  \newcommand\contentsname{Table of contents}
\fi
\ifdefined\listfigurename
  \renewcommand*\listfigurename{List of Figures}
\else
  \newcommand\listfigurename{List of Figures}
\fi
\ifdefined\listtablename
  \renewcommand*\listtablename{List of Tables}
\else
  \newcommand\listtablename{List of Tables}
\fi
\ifdefined\figurename
  \renewcommand*\figurename{Figure}
\else
  \newcommand\figurename{Figure}
\fi
\ifdefined\tablename
  \renewcommand*\tablename{Table}
\else
  \newcommand\tablename{Table}
\fi
}
\@ifpackageloaded{float}{}{\usepackage{float}}
\floatstyle{ruled}
\@ifundefined{c@chapter}{\newfloat{codelisting}{h}{lop}}{\newfloat{codelisting}{h}{lop}[chapter]}
\floatname{codelisting}{Listing}

\makeatother
\makeatletter
\@ifpackageloaded{caption}{}{\usepackage{caption}}
\@ifpackageloaded{subcaption}{}{\usepackage{subcaption}}
\makeatother

\ifLuaTeX
  \usepackage{selnolig}  
\fi
\usepackage[]{natbib}
\usepackage{bookmark}

\IfFileExists{xurl.sty}{\usepackage{xurl}}{} 
\hypersetup{
  pdftitle={Title},
  pdfauthor={Author 1; Author 2},
  pdfkeywords={3 to 6 keywords, that do not appear in the title},
  colorlinks=true,
  linkcolor={blue},
  filecolor={Maroon},
  citecolor={Blue},
  urlcolor={Blue},
  pdfcreator={LaTeX via pandoc}}

\newcommand{\anon}{1}

\newcommand{\cN}{\mathcal{N}}
\newcommand{\R}{\mathbb{R}}

\newcommand{\E}{\mathbb{E}}

\newcommand{\one}{\mathbbm{1}}

\newcommand{\cov}{\text{cov}}
\newcommand{\var}{\text{var}}

\def\bI{\bm{I}}
\def\bepsilon{\bm{\epsilon}}

\def\cC{\mathcal{C}}

\def\pr{\mathbb{P}}

\def\var{\textrm{var}}
\def\cov{\textrm{cov}}
\def\bx{\mathbf{x}}
\def\bX{\boldsymbol{X}}
\def\bY{\boldsymbol{Y}}
\def\bV{\boldsymbol{V}}
\def\bS{\boldsymbol{S}}
\def\bs{\boldsymbol}
\def\bq{\boldsymbol{q}}

\newcommand{\vertiii}[1]{{\left\vert\kern-0.25ex\left\vert\kern-0.25ex\left\vert #1 
    \right\vert\kern-0.25ex\right\vert\kern-0.25ex\right\vert}}
    
\newtheorem{theorem}{Theorem}

\newtheorem{lemma}{Lemma}

\newtheorem{corollary}{Corollary}

\theoremstyle{definition}

\newtheorem{assumption}{Assumption}

\newtheorem{proposition}{Proposition}

\theoremstyle{remark}
\newtheorem{remark}{Remark}

\begin{document}

\def\spacingset#1{\renewcommand{\baselinestretch}%
{#1}\small\normalsize} \spacingset{1}


\if1\anon
{
  \title{\bf Rerandomization for quantile treatment effects}
  \author{Tingxuan Han\hspace{.2cm}\\
    Department of Statistics and Data Science, Tsinghua University\\
    and \\
    Yuhao Wang\thanks{Corresponding author: yuhaow@tsinghua.edu.cn} \\
    Institute for Interdisciplinary Information Sciences, Tsinghua University\\
    Shanghai Qi Zhi Institute}
  \maketitle
} \fi

\if0\anon
{
  \bigskip
  \bigskip
  \bigskip
  \begin{center}
    {\LARGE\bf Rerandomization for quantile treatment effects}
\end{center}
  \medskip
} \fi

\bigskip
\begin{abstract}
Although complete randomization is widely regarded as the gold standard for causal inference, covariate imbalance can still arise by chance in finite samples.
Rerandomization has emerged as an effective tool to improve covariate balance across treatment groups and enhance the precision of causal effect estimation. 
While existing work focuses on average treatment effects, quantile treatment effects (QTEs) provide a richer characterization of treatment heterogeneity by capturing distributional shifts in outcomes, which is crucial for policy evaluation and equity-oriented research. 
In this article, we establish the asymptotic properties of the QTE estimator under rerandomization within a finite-population framework, without imposing any distributional or modeling assumptions on the covariates or outcomes.
The estimator exhibits a non-Gaussian asymptotic distribution, represented as a linear combination of Gaussian and truncated Gaussian random variables. 
To facilitate inference, we propose a conservative variance estimator and construct corresponding confidence interval. 
Our theoretical analysis demonstrates that rerandomization improves efficiency over complete randomization under mild regularity conditions. 
Simulation studies further support the theoretical findings and illustrate the practical advantages of rerandomization for QTE estimation.
\end{abstract}

\noindent%
\vfill

\newpage
\spacingset{1.8} 

\section{Introduction}

Randomized experiments are widely regarded as the gold standard for causal inference, as they create comparable treatment groups and enable credible causal conclusions \citep{fisher1925,fisher1935,cox1982randomization,Cox2009,Imbens2015}. 
Yet, despite balanced in expectation, individual randomizations often suffer from chance covariate imbalances \citep{fisher1926arrangement,student1938comparison,greevy2004optimal,hansen2008covariate,bruhn2009pursuit,krieger2019nearly}. 
To address this concern, as suggested by \cite{cox1982randomization,Cox2009}, \citet{morgan2012rerandomization} formally introduced \textit{rerandomization} (ReM), a design-based procedure that repeatedly samples treatment assignments until a prespecified covariate balance criterion, typically based on the Mahalanobis distance, is satisfied.

Subsequent works have established that ReM reduces the sampling variance of the difference-in-means estimator of the average treatment effect (ATE) under both superpopulation and finite-population frameworks \citep{morgan2012rerandomization,li2018asymptotic,wang2022rerandomization}. 
Building on these foundational results, later research has adapted ReM to more complex experimental designs \citep{Branson2016,Zhou2018,li2020rerandomization2,yang2021rejective,Branson2021,wang2023rerandomization,lu2023design} and has proposed alternative rerandomization criteria beyond the original ReM framework \citep{Kallus2018,Li_2021kde,zhu2022pair,zhang2021pca,liu2025bayesian}. 
Nevertheless, existing research has overly focused on ATE estimation, leaving the effects of ReM on other causal estimands relatively unexplored.

In many applications, particularly those concerned with fairness, inequality, or tail behavior, the ATE, which captures only average effects, may obscure important distributional heterogeneity \citep{chernozhukov2018sorted,cui2025policy}.
\textit{Quantile treatment effects} (QTEs) address this limitation by comparing outcome quantiles under treatment and control \citep{doksum1974empirical,lehmann1975statistical}. 
By capturing treatment effects across different points of the outcome distribution, QTEs offer a more nuanced understanding of how interventions impact individuals, highlighting whether benefits or harms are concentrated in specific subgroups or the tails of the distribution \citep{djebbari2008heterogeneous}.
Consequently, QTEs play a central role in policy evaluation \citep{bitler2006mean,finkelstein2008did}, as well as in a wide range of empirical applications in economics \citep{djebbari2008heterogeneous,banerjee2015miracle}, education \citep{muralidharan2011teacher,shea2021examining,thompson2024examining}, and related fields \citep{willke2012concepts}.

Although a rich literature has developed robust and efficient QTE estimators, especially for observational data \citep[e.g.,][]{abadie2002instrumental,chernozhukov2005iv,firpo2007efficient,horowitz2007nonparametric,frolich2013unconditional,donald2014estimation,callaway2018quantile,callaway2019quantile,chiang2019robust,tang2021new,zongwu2021estimation,zhou2022role,ai2022estimation,hsu2022estimation,han2023quantile,zhan2024estimation,kallus2024localized,cai2025nonparametric}, research on QTEs in randomized experiments remains limited. 
Recent advances have considered heavy-tailed outcomes \citep{athey2023semi}, covariate-adaptive designs \citep{zhang2020quantile,jiang2023regression,jiang2024bootstrap}, and optimal designs for QTEs \citep{li2024efficient}. 
Yet, to the best of our knowledge, no theoretical results exist for QTE estimation under rerandomization. 
Moreover, the existing QTE studies adopt a superpopulation framework, while the finite-population perspective, where potential outcomes and covariates are fixed and randomness arises solely from treatment assignment, remains unexplored.

This paper addresses these gaps. 
We establish the first asymptotic theory for quantile treatment effects (QTEs) estimation under rerandomization (ReM), filling an important gap in the causal inference literature where ReM has been primarily studied for average treatment effects. 
The theory is developed within the finite-population framework, without imposing distributional assumptions on potential outcomes or covariates. 
Our analysis reveals a non-Gaussian asymptotic distribution of the estimator, expressed as a linear combination of Gaussian and truncated Gaussian random variables. 
Building on this theory, we further develop a conservative variance estimator and corresponding confidence interval, thereby enabling rigorous inference on distributional treatment effects. 
The results demonstrate that ReM preserves design-based asymptotic unbiasedness and improves efficiency for QTE estimation by enhancing covariate balance. 
Finally, numerical simulations and real-data-based studies confirm the theoretical findings and highlight the practical advantages of applying ReM to QTE analysis.

The remainder of the paper is organized as follows. In Section~\ref{sec:notation}, we introduce the notation and review the rerandomization criterion. Section~\ref{sec:cumulant} presents key properties of empirical cumulants, which serve as building blocks for subsequent derivations. In Section~\ref{sec:QTE}, we derive the asymptotic distribution of the QTE estimator under ReM. Section~\ref{sec:CI} develops conservative variance estimators and corresponding confidence intervals. Simulation results are reported in Section~\ref{sec:simulation}, and Section~\ref{sec:discussion} concludes with a summary and discussion of potential extensions. Technical proofs are provided in the Supplementary Material.

\section{Problem set up}\label{sec:notation}

\subsection{Finite population and QTE estimation}
Consider a finite population of $n$ units, where each unit $i \in \{1,\ldots\}$ has two potential outcomes: $Y_i(1)$ if assigned to the treatment group, and $Y_i(0)$ if assigned to the control group.
Each unit is also associated with a covariate vector $\bX_i \in \mathbb{R}^{K_n}$, where the dimensionality $K_n$ may grow with the sample size $n$.
Let $\bY(1)=(Y_1(1),\ldots,Y_n(1))^\top$ and $\bY(0)=(Y_1(0),\ldots,Y_n(0))^\top$ denote the vectors of potential outcomes under treatment and control, respectively.
For $z=0,1$, define the empirical distribution of $\bY(z)$ as 
$$F_z(q) = \frac{1}{n} \sum_{i = 1}^n \one \{Y_i(z) \leq q\}.$$
For $\alpha \in (0,1)$,
the parameter of interest is the $\alpha$-th \textit{quantile treatment effect} (QTE):
\begin{equation}\label{eq:estimand}
\tau_{\alpha}:= q_{1, \alpha} - q_{0, \alpha},
\end{equation}
where $q_{z, \alpha} :=\inf\{q: F_z(q) \ge \alpha\}$ is the $\alpha$-th quantile of $\bY(z)$ for 
$z \in \{0,1\}$. 
The QTE summarizes distributional treatment effects by comparing the $\alpha$-th quantiles of the potential outcome distributions under treatment and control, thereby revealing heterogeneity that may be obscured by average effects.
This quantity is unobservable and must be estimated using data from a randomized experiment.

For a randomized experiment, denote the treatment assignment as $\bs Z = (Z_1,\ldots,Z_n)^\top \in \{0,1\}^n$, where $Z_i=1$ if unit $i$ is assigned to treatment and $Z_i=0$ otherwise.
Under this assignment, the observed outcome for unit $i$ follows the standard potential outcomes representation,
$Y_i = Z_iY_i(1) + (1-Z_i)Y_i(0)$.
Let $r_1$ and $r_0$ denote the treatment and control proportions, respectively, with $r_1+r_0=1$, and corresponding group sizes $n_1 = nr_1$ and $n_0 = nr_0$.
Conditional on the assignment vector $\bs Z$, the empirical distribution of observed outcomes in group $z \in \{0,1\}$ is given by
\begin{equation}\label{eq:empirical}
\hat{F}_z(q) = \frac{1}{n_z} \sum_{Z_i=z} \one \{Y_i \le q\}.
\end{equation}
The empirical $\alpha$-th quantile in group $z$ is then defined as $\hat{q}_{z, \alpha} : = \inf\{q: \hat{F}_z(q) \ge \alpha\}$, leading to the natural plug-in estimator of the quantile treatment effect (QTE),
\begin{equation}\label{eq:estimator}
\hat{\tau}_{\alpha} := \hat{q}_{1, \alpha}-\hat{q}_{0, \alpha}.
\end{equation}
Throughout this paper, we adopt a finite-population perspective and treat the potential outcomes $\{Y_i(0),Y_i(1)\}_{i=1}^n$ and covariates $\{\bX_i\}_{i=1}^n$ as fixed. 
Consequently, all randomness arises solely from the treatment assignment $\bs Z$, which is decided by the experimental design.

\subsection{Completely randomized experiments and rerandomization}
Under a \textit{completely randomized experiment} (CRE), $n_1$ units are randomly assigned to the treatment group, and the remaining $n_0$ units to the control group, so that each assignment $\bs z = (z_1,\ldots,z_n)^\top \in \{0,1\}^n$ satisfying $\sum_{i=1}^n z_i = n_1$ occurs with probability $\binom{n}{n_1}^{-1}$.
The average treatment effect (ATE) is commonly estimated by the difference in sample means of the observed outcomes, $\hat{\tau} = n_1^{-1}\sum_{i=1}^n Z_i Y_i - n_0^{-1}\sum_{i=1}^n (1-Z_i) Y_i$.
Although CRE guarantees covariate balance in expectation, individual randomizations may exhibit chance imbalances, particularly when covariates are high dimensional, which can substantially reduce the precision of the difference-in-means estimator \citep{morgan2012rerandomization}.

To improve covariate balance, \citet{morgan2012rerandomization} formalized \textit{rerandomization} based on the Mahalanobis distance (ReM), which repeatedly draws treatment assignments until a prespecified balance criterion is satisfied.
Under ReM, covariate imbalance is quantified by the Mahalanobis distance of the covariate difference-in-means vector between the treatment and control groups.
Specifically, the difference in covariate means is defined as
\begin{equation}\label{eq:Mdist}
\hat{\bs{\tau}}_{\bx} = \frac{1}{n_1}\sum_{i=1}^n Z_i \bX_i - \frac{1}{n_0}\sum_{i=1}^n (1-Z_i) \bX_i,
\end{equation}
and the corresponding Mahalanobis distance is given by
\[\begin{split}
M = \hat{\bs\tau}_{\bx}^\top
(\cov(\hat{\bs\tau}_{\bx}))
^{-1} \hat{\bs\tau}_{\bx}
=nr_1r_0\cdot\hat{\bs\tau}_{\bx}^\top \bs{S}_{\bx\bx}^{-1} \hat{\bs\tau}_{\bx},
\end{split}\]
where $\bs{S}_{\bx\bx} = (n-1)^{-1}\sum_{i=1}^n (\bX_i-\bar{\bX})(\bX_i-\bar{\bX})^\top$ is the finite-population covariance matrix of the covariates and $\bar{\bX} = n^{-1}\sum_{i=1}^n \bX_i$.
A treatment assignment is accepted under rerandomization if and only if
$M \le a_n$, where $a_n$ is a prespecified threshold.
The corresponding theoretical acceptance probability is $p_n = \pr(\chi^2_{K_n} \le a_n)$, and the realized acceptance probability under rerandomization is $\tilde p_n = \pr(M \le a_n)$, where $K_n$ denotes the dimension of the covariates. 
Throughout, we allow the covariate dimension $K_n$, the acceptance probability $p_n$, and the rerandomization threshold $a_n$ to vary with the sample size $n$.
Asymptotic theory \citep{li2018asymptotic,wang2022rerandomization} shows that ReM improves covariate balance and enhances the efficiency of the difference-in-means estimator of ATE, $\hat{\tau}$.
Specifically, let $S_1^2$ and $S_0^2$ denote the sample variances of the potential outcomes $Y_i(1)$ and $Y_i(0)$, respectively, 
and let $S^2_\tau$ denotes the sample variance of the individual treatment effects $\tau_i = Y_i(1)-Y_i(0)$. 
Under CRE, the asymptotic distribution of $\hat{\tau}$ is Gaussian with variance $V_{\tau\tau} = n_1^{-1}S_1^2 + n_0^{-1} S_0^2 - n^{-1}S^2_\tau$. In contrast, under ReM the asymptotic distribution of $\hat{\tau}$ is non-Gaussian and can be expressed as a weighted sum of an independent Gaussian and a non-Gaussian component,
\[\begin{split}
V_{\tau\tau}^{1/2}\left(\sqrt{1-R^2_n}\cdot \varepsilon_0 + \sqrt{R^2_n}\cdot L_{K_n,a_n}\right),
\end{split}\]
where $R^2_n$ is the squared multiple correlation between $\hat{\tau}$ and the covariate mean-difference vector $\hat{\bs \tau}_\bx$, and $\varepsilon_0 \sim \cN(0,1)$ is independent of 
$L_{K_n, a_n} := D_1 \mid \bs{D}_{K_n}^\top \bs{D}_{K_n} \le a_n$, with $\bs{D}_{K_n}$ being a $K_n$ dimensional standard normal random vector.

Intuitively, improved covariate balance under rerandomization should benefit not only the estimation of average treatment effects but also other causal estimands, including the quantile treatment effect (QTE). 
In this paper, we develop a comprehensive asymptotic theory for the QTE estimator under ReM and rigorously show that rerandomization continues to improve estimation efficiency in this setting.

\section{Non-asymptotic convergence of empirical cumulant estimators}\label{sec:cumulant}

By definition (Eq.~\eqref{eq:estimand}), the quantile treatment effect (QTE) at level $\alpha \in (0,1)$ is the difference between the $\alpha$-th quantiles of the potential outcomes under treatment and control, denoted by $q_{1,\alpha}$ and $q_{0,\alpha}$. 
Its estimator, $\hat{\tau}_{\alpha}$, is obtained from the corresponding empirical quantiles of the observed outcome distributions, $\hat{F}_1(\cdot)$ and $\hat{F}_0(\cdot)$, as shown in Eq.~\eqref{eq:estimator}.
As a first step toward establishing the asymptotic behavior of $\hat{\tau}_{\alpha}$, we characterize the joint convergence of the empirical distribution functions evaluated at the population quantiles, $(\hat{F}_1(q_{1,\alpha}), \hat{F}_0(q_{0,\alpha}))$, to their population counterparts $(F_1(q_{1,\alpha}),F_0(q_{0,\alpha}))$ for a fixed $\alpha$ as the sample size $n$ grows.

For each unit $i = 1,\ldots,n$, we define $\bs{u}_{i} := (r_0 \one(Y_i(1) \le q_{1,\alpha}), - r_1 \one(Y_i(0) \le q_{0,\alpha}), \bs{X}_i^\top)^\top \in \mathbb{R}^{K_n+2}$, which incorporates both the potential outcomes and covariates.
Using this construction, the empirical quantities of interest can be compactly expressed as
\[
\left(
\begin{matrix}
	\hat{F}_1(q_{1,\alpha}) \\ \hat{F}_0(q_{0,\alpha}) \\ \hat{\bs{\tau}}_{\bx} 
\end{matrix}
\right) = \frac{n}{n_1 n_0} \sum_{i=1}^n Z_i \bs{u}_{i} - \frac{n}{n_0} \left(
\begin{matrix}
	0 \\ -F_0(q_{0,\alpha}) \\ \bar{\bX} 
\end{matrix}
\right).
\]
In other words, the observed empirical vector is a linear transformation of the sum of a simple random sample of size $n_1$ drawn from the finite population $\{\bm u_i\}_{i=1}^n$.
To study its sampling properties, let $\bq = (q_{1,\alpha},q_{0,\alpha})^\top$ and consider the finite-population covariance matrix of $\bs{u}_i$,
$\bs{S}_{\bs u\bs u} := \frac{1}{n - 1} \sum_{i = 1}^n (\bs{u}_{i} - \bar{\bs{u}}) (\bs{u}_{i} - \bar{\bs{u}})^\top,\ \bar{\bs{u}} = n^{-1}\sum_{i=1}^n \bs{u}_{i}.$
This matrix can be naturally partitioned into blocks corresponding to the potential outcomes and covariates:
\[
\bs{S}_{\bs u\bs u} = \left(
\begin{matrix}
	\bs{S}_{\bq\bq} & \bs{S}_{\bq\bx}\\
	\bs{S}_{\bx\bq} & \bs{S}_{\bx\bx}
\end{matrix}
\right).
\]
The block for the potential outcomes is given by
\begin{equation}\label{eq:Sqq}
\begin{split}
\bs{S}_{\bq\bq} &= \cov\left(\begin{pmatrix}
r_0 \one(Y_i(1) \le q_{1,\alpha})\\- r_1 \one(Y_i(0) \le q_{0,\alpha})
\end{pmatrix}\right)=\begin{pmatrix}
S_{q_1 q_1} & S_{q_1 q_0}\\
S_{q_1 q_0} & S_{q_0 q_0}
\end{pmatrix},
\end{split}
\end{equation}
where 
$S_{q_1 q_1} = \frac{n r_0^2}{n-1} F_1(q_{1,\alpha})(1-F_1(q_{1,\alpha}))$, 
$S_{q_1 q_0} = S_{q_0 q_1} = \frac{n r_0r_1}{n-1}\left(F_1(q_{1,\alpha})F_0(q_{0,\alpha})-F(q_{1,\alpha},q_{0,\alpha})\right)$, 
$S_{q_0 q_0} = \frac{n r_1^2}{n-1}F_0(q_{0,\alpha})(1-F_0(q_{0,\alpha}))$, with $F(q_{1,\alpha},q_{0,\alpha}) = n^{-1}\sum_{i=1}^n \one(Y_i(1)\leq q_{1,\alpha},
Y_i(0)\leq q_{0,\alpha})$.
The covariance between the potential outcomes and covariates is
\[\begin{split}
\bS_{\bq\bx} & = \bS_{\bx\bq}^\top = \cov\left(\begin{pmatrix}
r_0 \one(Y_i(1) \le q_{1,\alpha})\\- r_1 \one(Y_i(0) \le q_{0,\alpha})
\end{pmatrix},\bX_i\right)=\begin{pmatrix}
\bS_{q_1\bx}\\
\bS_{q_0\bx}
\end{pmatrix}.
\end{split}\]
For convenience, we scale the covariance matrix as $\bs{V} := (n r_1 r_0)^{-1} \bs{S}_{\bs u\bs u}$ and partition it similarly:
\[
\bs{V} = \left(
\begin{matrix}
	\bs{V}_{\bq\bq} & \bs{V}_{\bq\bx}\\
	\bs{V}_{\bx\bq} & \bs{V}_{\bx\bx}
\end{matrix}
\right),\  
\bV_{qq} = \begin{pmatrix}
V_{q_1q_1} & V_{q_1q_0}\\
V_{q_0q_1} & V_{q_0q_0}
\end{pmatrix},\ 
\bV_{\bq\bx} = \bV_{\bx\bq}^\top = \begin{pmatrix}
\bV_{q_1 \bx}\\
\bV_{q_0 \bx}
\end{pmatrix}.
\]
Finally, the multiple squared correlation between the potential-outcome indicators $(r_0 \one(Y_i(1) \le q_{1,\alpha}), - r_1 \one(Y_i(0) \le q_{0,\alpha}))^\top$ and the covariates $\bX_i$ is defined as 
\begin{equation}\label{eq:R2}
\bs{R}_{\bq}^2 := \bs{S}_{\bq\bq}^{-1/2} \bs{S}_{\bq\bx} \bs{S}_{\bx\bx}^{-1} \bs{S}_{\bx\bq} \bs{S}_{\bq\bq}^{-1/2} = \bs{V}_{\bq\bq}^{-1/2}\bs{V}_{\bq\bx}\bs{V}_{\bx\bx}^{-1}\bs{V}_{\bx\bq}\bs{V}_{\bq\bq}^{-1/2},
\end{equation}
where $\bs{M}^{-1/2}$ denotes the inverse of the square root of a symmetric matrix $\bs{M}$.

Define
\[
\gamma_n := \frac{(K_n + 2)^{1/4}}{\sqrt{n r_1 r_0}} \frac{1}{n} \sum_{i = 1}^n \|\bs{S}_{\bs{u}\bs u}^{-1/2} (\bs{u}_{i} - \bar{\bs{u}})\|_2^3,
\]
which can be interpreted as a normalized third moment of the standardized vectors $\bs{u}_i$, capturing the magnitude of potential skewness or “heaviness” in the tails.
For completeness, we set $\gamma_n := \infty$ in degenerate cases where either $r_1 = 0$, $r_0 = 0$, or $\bs{S}_{\bs{u}\bs{u}}$ is singular.
Next, let $\cC_{K_n+2}$ denote the collection of all measurable convex sets in $\mathbb{R}^{K_n+2}$. 
To quantify the distance between the finite-sample distribution of the standardized difference-in-means vector and the multivariate standard Gaussian distribution, define the Kolmogorov distance
\begin{equation}\label{eq:delta_n}
\Delta_n \equiv \sup_{C \in \cC_{K_n + 2}}\left|\pr\left(\bs{V}^{-1/2} \left(
	\begin{matrix}
		\hat{F}_1(q_{1,\alpha}) - F_1(q_{1,\alpha}) \\ \hat{F}_0(q_{0,\alpha}) - F_0(q_{0,\alpha}) \\ \hat{\bs{\tau}}_{\bx} 
	\end{matrix}
	\right) \in C\right) - \pr\left(\bs\varepsilon_{K_n+2} \in C\right)\right|,
\end{equation}
where $\bs{\varepsilon}_{K_n+2} \sim \cN(\bs0,\bs I_{K_n+2})$ is a standard Gaussian vector.
Here, $\Delta_n$ is set to $1$ whenever $\gamma_n = \infty$.

The following lemma, which is a direct consequence of Theorem 1 in \citet{wang2022rerandomization}, provides the asymptotic justification for treating the  empirical vector $(\hat{F}_1(q_{1,\alpha}) - F_1(q_{1,\alpha}), \hat{F}_0(q_{0,\alpha}) - F_0(q_{0,\alpha}), \hat{\bs{\tau}}_{\bx}^\top)^\top$ as approximately Gaussian under CRE.

\begin{lemma}\label{lem:wang&li}
Under a completely randomized experiment, if $\bs{V}$ is nonsingular, then the Kolmogorov distance $\Delta_n$ defined in \eqref{eq:delta_n} converges to zero if $\gamma_n \to 0$ as $n \to \infty$.
\end{lemma}

This lemma provides the foundation for establishing the asymptotic convergence of $(\hat{F}_1(q_{1,\alpha}) - F_1(q_{1,\alpha}), \hat{F}_0(q_{0,\alpha}) - F_0(q_{0,\alpha}))^\top$ under ReM. 
Building on this result, we impose the following assumptions, which were also introduced in \citet{wang2022rerandomization} to develop the asymptotic theory of ReM for the estimation of the average treatment effect (ATE).

\begin{assumption}\label{assumption:gamma_n}
As $n\rightarrow\infty$, the sequence of finite populations satisfies that $\gamma_n \rightarrow 0$.
\end{assumption}

\begin{assumption}\label{assumption:delta_p}
As $n\rightarrow\infty$, 
$\Delta_n/p_n \rightarrow 0$, where $p_n = \pr(\chi^2_{K_n} \leq a_n)$ is the theoretical acceptance probability of ReM.
\end{assumption}

Intuitively, Assumption~\ref{assumption:gamma_n} requires that, for sufficiently large $n$, both treatment and control groups contain positive proportions of units, and that the covariance matrix $\bs{V}$ is nonsingular. 
Consequently, the covariates are not perfectly collinear, and the matrix $\bm R^2_{\bq}$ defined in \eqref{eq:R2} is well-defined, and is strictly smaller than the identity matrix $\bs I$ in the sense of symmetric matrices. This ensures that the vector $(r_0 \one(Y_i(1) \le q_{1,\alpha}), - r_1 \one(Y_i(0) \le q_{0,\alpha}))$ cannot be fully explained by the covariates, a condition that is likely to hold in most practical applications.
According to \cite{wang2022rerandomization}, this assumption also restricts the growth rate of the number of covariates $K_n$ relative to the sample size $n$. 
For instance, $\gamma_n = o(1)$ implies that $K_n = o(n^{2/7})$, while if the standardized deviations are uniformly bounded, i.e., $\|\bs{S}_{\bs{u}\bs u}^{-1/2} (\bs{u}_{i} - \bar{\bs{u}})\|_\infty \leq C$ for some finite constant $C$, then $K_n = o(n^{4/7})$ suffices to guarantee $\gamma_n = o(1)$.
Assumption~\ref{assumption:delta_p} further ensures that the theoretical acceptance probability $p_n$ does not decay too rapidly as $n\rightarrow\infty$, allowing the asymptotic results for ReM to hold.

Based on the assumptions above, we can characterize the large-sample distribution of $(\hat{F}_1(q_{1,\alpha}) - F_1(q_{1,\alpha}), \hat{F}_0(q_{0,\alpha}) - F_0(q_{0,\alpha}))^\top$ under ReM as follows.
\begin{theorem}\label{thm:AsympDis}
Let $\mathcal{C}_2$ denote the collection of all measurable convex sets in $\R^{2}$.
Under ReM, and assuming Assumptions \ref{assumption:gamma_n} and \ref{assumption:delta_p} hold,
as $n\rightarrow \infty$,
\[\begin{split}
&\sup_{C \in \cC_2}
\left|
\pr\left(
\bs{V}_{\bq\bq}^{-1 / 2}
\left(
\begin{matrix}
    \hat{F}_1(q_{1,\alpha}) - F_1(q_{1,\alpha}) \\ \hat{F}_0(q_{0,\alpha}) - F_0(q_{0,\alpha})
\end{matrix}
\right) \in C \mid M \le a_n
\right) - 
\pr\left(
(\bs{I} - \bs{R}_{\bq}^2)^{1/2} \bs{\varepsilon}_2 + \bs{R}_{\bq} \bs{L}_{K_n, a_n} \in C
\right)\right|\\ &= \mathcal{O}\left(\Delta_n/p_n\right) = o(1),
\end{split}
\]
where $\bs{\varepsilon}_2$ is a 2-dimensional standard Gaussian random vector, independent of 
$\bs{L}_{K_n, a_n} := (D_1, D_2)^\top \mid \bs{D}_{K_n}^\top \bs{D}_{K_n} \le a_n$, with $\bs{D}_{K_n} \sim \cN(\bm 0,\bI_{K_n})$, and 
$\bm R_{\bq}^2$ is defined in \eqref{eq:R2}.
\end{theorem}

The proof of Theorem \ref{thm:AsympDis} follows a similar strategy as Theorem 3 in \cite{wang2022rerandomization}.
It shows that under ReM, the asymptotic distribution of  $(\hat{F}_1(q_{1,\alpha}) - F_1(q_{1,\alpha}), \hat{F}_0(q_{0,\alpha}) - F_0(q_{0,\alpha}))^\top$ can be represented as the independent sum of a Gaussian component and a truncated Gaussian component, with the relative contribution of each part determined by the correlation matrix $\bs{R}_{\bq}^2$.
In the special case where $a_n = \infty$, the truncation disappears and $\bm L_{K_n, a_n}$ reduces to a standard Gaussian vector. 
This immediately yields a corollary for completely randomized experiments (CRE):
\[\begin{split}
&\sup_{C \in \cC_2}
\left|
\pr\left(
\bs{V}_{\bq\bq}^{-1 / 2}
\begin{pmatrix}
    \hat{F}_1(q_{1,\alpha}) - F_1(q_{1,\alpha}) \\ \hat{F}_0(q_{0,\alpha}) - F_0(q_{0,\alpha})
\end{pmatrix} \in C
\right) - 
\pr\left(
\bs{\varepsilon} \in C
\right)\right|= \mathcal{O}\left(\Delta_n/p_n\right) = o(1),
\end{split}
\]
where $\bs\varepsilon \sim \cN(\bm 0,\bI_2)$.
This result shows that under CRE, the joint distribution of the empirical quantiles converges to a bivariate normal distribution with covariance $\bs{V}_{\bq\bq}$.

In a similar manner, we can derive the asymptotic distribution of $\hat{F}_z(q_{z,\alpha}) - F_z(q_{z,\alpha})$ under ReM for each $z \in \{0,1\}$.
For $z\in \{0,1\}$, define the multiple squared correlation between $(-1)^{1-z} r_{1-z}\one(Y_i(z) \le q_{z})$ and $\bX_i$ as 
\begin{equation}\label{eq:R22}
R_{q_{z}}^2 :=  \bs{S}_{q_{z}\bx} \bs{S}_{\bx\bx}^{-1} \bs{S}_{\bx q_{z}}/S_{q_{z}q_z} = \bs{V}_{q_{z}\bx}\bs{V}_{\bx\bx}^{-1}\bs{V}_{\bx q_{z}}/V_{q_{z}q_z}.
\end{equation}
The following theorem gives the asymptotic distribution of
$\hat{F}_z(q_{z,\alpha}) - F_z(q_{z,\alpha})$ under ReM, and
the proof proceeds analogously to that of Theorem~3 in \cite{wang2022rerandomization}.

\begin{theorem}\label{thm:AsympDis2}
Under ReM, and Assumptions \ref{assumption:gamma_n} and \ref{assumption:delta_p},
for $z \in \{0,1\}$, as $n\rightarrow\infty$,
\[\begin{split}
&\sup_{x \in \mathbb{R}}
\left|
\pr\left(
V_{q_{z}q_z}^{-1 / 2}
\left(
\hat{F}_z(q_{z,\alpha}) - F_z(q_{z,\alpha})
\right) \leq x \mid M \le a_n
\right) - 
\pr\left(
(1 - R_{q_{z}}^2)^{1/2} \varepsilon + R_{q_{z}} L_{K_n, a_n} \leq x
\right)\right|\\
&= \mathcal{O}\left(\Delta_n/p_n\right) = o(1),
\end{split}\]
where $\varepsilon$ is a 1-dimensional standard normal random variable, independent of 
$L_{K_n, a_n} := D_1 \mid \bs{D}_{K_n}^\top \bs{D}_{K_n} \le a_n$, with $\bs{D}_{K_n}$ being a $K_n$ dimensional standard normal random vector.
\end{theorem}


\section{Asymptotic convergence of QTE estimators}\label{sec:QTE}

The central objective of this study is to estimate and conduct inference on the quantile treatment effect (QTE), ${\tau}_{\alpha}:={q}_{1, \alpha}-{q}_{0, \alpha}$, which measures the difference between the $\alpha$th quantiles of the potential outcome distributions under treatment and control. 
The corresponding estimator is
$\hat{\tau}_{\alpha}:=\hat{q}_{1, \alpha}-\hat{q}_{0, \alpha}$.
To study the large-sample behavior of $\hat{\tau}_{\alpha}$, it is therefore essential to first characterize the asymptotic properties of the component quantile estimators $\hat{q}_{z,\alpha}$ for $z \in \{0,1\}$.
We begin by introducing a set of assumptions under which a rigorous asymptotic theory can be developed.

\begin{assumption}\label{assumption:derivativeF}
Fix any $\alpha \in (0,1)$. For each $z \in \{0,1\}$, there exist a constant $s < 0$ and two {bounded non-negative} functions $f_{z, n}(\cdot)$ and $r_{z, n}(\cdot)$, depending on $z$ and $n$, such that for any $t \in [-2 n^\gamma, 2 n^\gamma]$ with some $0<\gamma < 1/2$,
\[
|\sqrt{n}(F_z(q + t / \sqrt{n}) - F_z(q)) - t f_{z, n}(q)| \le  r_{z, n}(q),
\]
where $r_{z, n}(q_{z,\alpha})= \mathcal{O}(n^s)$, and $\max\{f_{z, n}(q_{z,\alpha}),1/f_{z, n}(q_{z,\alpha})\} = \mathcal{O}(1)$.
\end{assumption}
Assumption~\ref{assumption:derivativeF} imposes a local smoothness condition on the distribution of potential outcomes.
The remainder term $r_{z,n}(q_{z,\alpha})$ ensures that $f_{z,n}(\cdot)$ serves as an effective local density in a shrinking neighborhood of the quantile $q_{z,\alpha}$, acting as an analogue of the derivative of $F_z$ even when $F_z$ is not globally smooth.
Further discussion of this assumption is provided in Remark~\ref{rem:assumption}.

\begin{assumption}\label{assumption:replicate}
For each $z \in \{0,1\}$, 
there exists a constant $\delta > 0$ s.t. $F_z(q_{z,\alpha}) \leq \alpha + \delta/n$.
\end{assumption}

Assumption~\ref{assumption:replicate} ensures that the distribution of $Y_i(z)$ places sufficient probability mass in a neighborhood of $q_{z,\alpha}$.
This condition rules out large flat regions or abrupt jumps of $F_z$ near the target quantile, which would otherwise destabilize the empirical quantile estimator.
Since $F_z(q_{z,\alpha}) \ge \alpha$ by definition, this assumption implies
$F_z(q_{z,\alpha}) \to \alpha$ as $n\to\infty$.
Consequently,
$$S_{q_zq_z} = \frac{n r_{1-z}^2}{n-1} F_z(q_{z,\alpha})(1-F_z(q_{z,\alpha})) \rightarrow r_{1-z}^2\alpha(1-\alpha) > 0,$$
for any $\alpha \in (0,1)$.
Define the rescaled variance
\[\begin{split}
\Tilde{V}_{q_{z}q_z} :=V_{q_{z}q_z}/ f_{z, n}^2(q_{z, \alpha})= (n r_1 r_0)^{-1} S_{q_{z}q_z} / f_{z, n}^2(q_{z, \alpha}).
\end{split}\]
Under Assumptions~\ref{assumption:derivativeF} and~\ref{assumption:replicate}, we have $\liminf_n n\cdot{V}_{q_{z}q_z} > 0$ and $\liminf_n n\cdot\Tilde{V}_{q_{z}q_z} > 0$,
which guarantees that the relevant variances remain well behaved asymptotically.

\begin{assumption}\label{assumption:pn}
As $n\rightarrow \infty$,
$-\log p_n = \mathcal{O}(n^m)$ for some $0 < m < 1/2-\gamma$.
\end{assumption}

Assumption~\ref{assumption:pn} restricts the rate at which the acceptance probability $p_n$ under rerandomization may decay.
It prevents rerandomization from becoming excessively restrictive as the sample size increases, thereby preserving the validity of the asymptotic approximation.
This assumption is automatically satisfied when $p_n$ is fixed and does not depend on $n$.

\begin{assumption}\label{assumption:R2_1}
For each $z\in\{0,1\}$, as $n\rightarrow\infty$,
$\limsup_{n} R_{q_{z}}^2 < 1$.
\end{assumption}

By the definition in Eq.~\eqref{eq:R22}, Assumption~\ref{assumption:R2_1} ensures that the indicator
$\one(Y_i(z) \le q_{z,\alpha})$ cannot be perfectly explained by the covariates.
This condition rules out degeneracy and is expected to hold in most practical applications.


Under the above assumptions, the following theorem characterizes the asymptotic distribution of $\hat{q}_{z,\alpha}$ for $z \in \{0,1\}$ under rerandomization.

\begin{theorem}\label{thm:qz}
Under ReM and Assumptions \ref{assumption:gamma_n}-\ref{assumption:R2_1}, for $z \in \{0,1\}$ and $t \in [-2 n^\gamma, 2 n^\gamma]$, 
\[\begin{split}
&\sup_{t \in [-2 n^\gamma, 2 n^\gamma]} \left|\pr\left(\sqrt{n}(\hat{q}_{z, \alpha} - q_{z, \alpha}) \le t \mid M \le a_n\right) - \pr\left(\sqrt{n} \Tilde{V}_{q_{z}q_z}^{1/2} \left(\sqrt{1 - R^2_{q_{z}}}\varepsilon + R_{q_{z}} L_{K_n, a_n}\right) \le t\right)\right|\\ & = \mathcal{O}(\Delta_n/p_n) + \mathcal{O}(n^{k}) = o(1),
\end{split}
\]
where $\varepsilon\sim \cN(0,1)$ is independent of  $L_{K_n, a_n}$, and
$k = \max\{s,m/2+\gamma/2-1/4\} < 0$.
\end{theorem}


For a vector $\bs t = (t_1,t_0)^\top$ and a random vector $\bs r = (r_1,r_0)^\top$,
we use the shorthand notation $\pr(\bs r \le \bs t)$ to denote $\pr(r_1 \le t_1, r_0 \le t_0)$.
Moreover, define  $\bs{\Lambda}_{\bq} = \text{diag}(f_{1, n}^{-2}(q_{1, \alpha}),f_{0, n}^{-2}(q_{0, \alpha}))$ and $\tilde{\bV}_{\bq\bq}^{1/2} = \bm\Lambda_{\bq}^{1/2}\bV_{\bq\bq}^{1/2}$.
Building on the marginal asymptotic results of $\hat q_{z,\alpha}$ in Theorem~\ref{thm:qz}, 
we now extend the analysis to the joint behavior of the quantile estimators $(\hat q_{1,\alpha}, \hat q_{0,\alpha})^\top$ under rerandomization, as stated in the following theorem.

\begin{theorem}\label{thm:joint}
Under ReM and Assumptions \ref{assumption:gamma_n}-\ref{assumption:R2_1},
for any $\bs{t} = (t_1,t_0)^\top \in [-2 n^\gamma, 2 n^\gamma]^2$,
\[\begin{split}
&\sup_{\bm t \in [-2 n^\gamma, 2 n^\gamma]^2}\left|\pr\left(
\left(
\begin{matrix}
\sqrt{n}(\hat{q}_{1, \alpha} - q_{1, \alpha}) \\
\sqrt{n}(\hat{q}_{0, \alpha} - q_{0, \alpha})
\end{matrix}
\right) \le \bs{t} \mid M \le a_n\right) - \pr\left(\sqrt{n} \tilde{\bV}_{\bq\bq}^{1/2} ((\bs{I} - \bs{R}_{\bq}^2)^{1/2} \bs{\varepsilon}_2 + \bs{R}_{\bq} \bs{L}_{K_n, a_n}) \le \bs{t}\right)\right|\\
&= \mathcal{O}(\Delta_n/p_n) + \mathcal{O}(n^{k}) = o(1),
\end{split}\]
where $k = \max\{s,m/2+\gamma/2-1/4\} < 0$,  $\bs{\varepsilon}_2 \sim \mathcal{N}(\bs 0,\bs I_2)$ is independent of 
$\bs{L}_{K_n, a_n} := (D_1, D_2)^\top \mid \bs{D}_{K_n}^\top \bs{D}_{K_n} \le a_n$, $\bs{D}_{K_n}\sim \mathcal{N}(\bs 0,\bs I_{K_n})$.
\end{theorem}

Equipped with the above results, we are now ready to establish the asymptotic distribution of the empirical QTE estimator
$\hat{\tau}_{\alpha}:=\hat{q}_{1, \alpha}-\hat{q}_{0, \alpha}$ under rerandomization.
To this end, we introduce two additional assumptions.

\begin{assumption}\label{assumption:gamma}
There exists a constant $k' < 0$ such that
$\Delta_n/p_n = \mathcal{O}(n^{k'})$.
\end{assumption}


Assumption~\ref{assumption:gamma} strengthens Assumption~\ref{assumption:delta_p} by requiring that the Gaussian approximation error $\Delta_n$ vanishes sufficiently fast relative to the acceptance probability $p_n$.
This condition guarantees that conditioning on the rerandomization event does not distort the asymptotic convergence of the QTE estimator.
Under Assumption \ref{assumption:gamma_n},
both Assumptions~\ref{assumption:delta_p} and~\ref{assumption:gamma} are automatically satisfied when $p_n$ is fixed and does not depend on $n$.

Next, let $\bs{c} = (1,-1)^\top$ and define the scaling matrix
$\bs{\Lambda}_{\bq} = \text{diag}(f_{1, n}^{-2}(q_{1, \alpha}),f_{0, n}^{-2}(q_{0, \alpha}))$.
Using this notation, we introduce the rescaled covariance matrix, variance, and multiple squared correlation associated with the QTE estimator:
\[\begin{split}
\tilde{\bV}_{\bq\bq} := \bs{\Lambda}_{\bq}^{1/2} \bV_{\bq\bq}\bs{\Lambda}_{\bq}^{1/2},\ 
\tilde{V}_{\bq\bq} := \bs{c}^\top\tilde{\bV}_{\bq\bq}\bs{c},\
\tilde{R}_{\bq}^2 := \frac{\bs{c}^\top \bs{\Lambda}_{\bq}^{1/2} \bs{V}_{\bq\bx} \bs{V}_{\bx\bx}^{-1}\bs{V}_{\bx\bq}\bs{\Lambda}_{\bq}^{1/2}\bs{c}}{\bs{c}^\top \bs{\Lambda}_{\bq}^{1/2} \bs{V}_{\bq\bq}\bs{\Lambda}_{\bq}^{1/2} \bs{c}}.
\end{split}\]
We impose the following regularity condition to ensure that the asymptotic distribution of $\hat{\tau}_{\alpha}$ is nondegenerate.
\begin{assumption}\label{assumption:R2_2}
As $n\rightarrow\infty$, $\liminf_{n} n\cdot \tilde{V}_{\bq\bq} > 0$, and  $\limsup_{n}\tilde{R}_{\bq}^2 < 1$.
\end{assumption}

A sufficient condition for this assumption to hold is that
$\liminf_{n} \lambda_{\min} (n\cdot \bm V_{\bq\bq}) > 0$ and 
$\limsup_{n} \lambda_{\max} (\bm R^2_{\bq}) < 1$. 
Under these conditions, the vector $(r_0 \one(Y_i(1) \le q_{1,\alpha}), - r_1 \one(Y_i(0) \le q_{0,\alpha}))$ cannot be perfectly predicted by the covariates, and its covariance matrix is nonsingular.
Both conditions are mild and are expected to hold in most empirical applications.

Combining the preceding results with the assumptions above, we establish the asymptotic distribution of the QTE estimator $\hat{\tau}_\alpha$, summarized in the following theorem.

\begin{theorem}\label{thm:QTE}
Under Assumptions \ref{assumption:gamma_n}-\ref{assumption:R2_2},
for $t \in [-n^\gamma,n^{\gamma}]$,
the empirical quantile treatment effect estimator
$\hat{\tau}_{\alpha} = \hat{q}_{1, \alpha} - \hat{q}_{0, \alpha}$ satisfies
\[\begin{split}
\sup_{t \in [-n^\gamma,n^{\gamma}]}\left|\pr\left(\sqrt{n}\left(\hat{\tau}_{\alpha} - {\tau}_{\alpha}\right) \le t \mid M \leq a_n\right) -  \pr\left(\sqrt{n} \tilde{V}_{\bq\bq}^{1/2} \left(\sqrt{1 - \tilde{R}_{\bq}^2}\varepsilon + \tilde{R}_{\bq} L_{K_n, a_n}\right) \le t \right)\right| = o(1),
\end{split}\]
where $\varepsilon\sim \cN(0,1)$ is independent of  $L_{K_n, a_n}$.
\end{theorem}

Theorem~\ref{thm:QTE} shows that the QTE estimator $\hat{\tau}_\alpha$ is asymptotically distributed as the sum of an independent Gaussian component and a truncated Gaussian component. 
The relative contribution of these two components is governed by the rescaled variance $\tilde{V}_{\bq\bq}$ and the multiple squared correlation $\tilde{R}_{\bq}^2$.
As a direct consequence, the estimator $\hat{\tau}_\alpha$ is asymptotically unbiased for the quantile treatment effect $\tau_\alpha$.
Since complete randomization (CRE) is equivalent to rerandomization with $a_n = \infty$, Theorem~\ref{thm:QTE} immediately implies that
\begin{equation}\label{eq:CRE}
\sup_{t \in [-n^\gamma,n^{\gamma}]}\left|\pr\left(\sqrt{n}\left(\hat{\tau}_{\alpha} - {\tau}_{\alpha}\right) \le t\right) -  \pr\left(\sqrt{n} \tilde{V}_{\bq\bq}^{1/2} \varepsilon \le t \right)\right| = o(1),
\end{equation}
that is, under CRE, $\hat{\tau}_{\alpha}$ is asymptotically Gaussian with mean zero and variance $\tilde{V}_{\bq\bq}$.

Based on Theorem \ref{thm:QTE}, we can quantify the efficiency gain of rerandomization (ReM) in estimating the quantile treatment effect.
Let $v_{K_n,a_n} = \var(L_{K_n,a_n})$ denote the variance of the truncated normal random variable $L_{K_n,a_n}$ defined in Theorem \ref{thm:AsympDis2}.
As shown in \cite{morgan2012rerandomization}, this variance admits the closed-form expression
$v_{K_n,a_n} = \pr(\chi^2_{K_n+2} \leq a_n)/\pr(\chi^2_{K_n} \leq a_n)$, where $\chi^2_{K_n}$ denotes a chi-square random variable with $K_n$ degrees of freedom.
The following corollary characterizes the extent to which ReM reduces the asymptotic sampling variance of the QTE estimator relative to complete randomization (CRE), and shows that this reduction is governed by the generalized squared multiple correlation $\tilde{R}_{\bq}^2$.

\begin{corollary}\label{cor:PRIASV}
Under Assumptions \ref{assumption:gamma_n}-\ref{assumption:R2_2},
the percent reduction in the asymptotic sampling variance (PRIASV) of $\hat{\tau}_{\alpha}$, relative to its asymptotic variance under CRE given in Eq.~\eqref{eq:CRE}, is 
\[\begin{split}
\text{PRIASV} = 1-\frac{\var_a(\hat{\tau}_{\alpha} \mid M \leq a_n)}{\var_a(\hat{\tau}_{\alpha})}
= (1-v_{K_n,a_n})\tilde{R}_{\bq}^2,
\end{split}\]
where $\var_a$ denotes asymptotic variance.
\end{corollary}

Corollary \ref{cor:PRIASV} shows that the precision gain from rerandomization is non-decreasing in $\tilde{R}_{\bq}^2$, since $v_{K_n,a_n} < 1$ whenever $a_n < \infty$. 
Moreover, the percentage reduction in asymptotic variance is nonincreasing in the acceptance probability $p_n$, and attains its maximum value $\tilde{R}_{\bq}^2$ in the limit as $p_n \to 0$.

\begin{remark}\label{rem:assumption}
Assumption \ref{assumption:derivativeF} plays a crucial role in establishing the asymptotic properties of $\hat{\tau}_{\alpha}$. 
To illustrate its validity, consider the case where, for each $z \in \{0,1\}$, the potential outcomes $Y_i(z)$ are independent and identically distributed (i.i.d.) with cumulative distribution function (c.d.f.) $F^z$ and probability density function (p.d.f.) $f^z$. 
Suppose that $f^z$ is continuously differentiable and that both $f^z$ and its derivative $(f^z)'$ are bounded. 
Furthermore, assume that $f^z(t)$ is strictly positive for all $t \in \mathbb{R}$.
For a constant $q$, define $U_i^z(t,q) = \one\{q < Y_i(z) \leq q + t / \sqrt{n}\}$. 
Then, the difference in the empirical c.d.f. can be expressed as $F_z(q + t / \sqrt{n}) - F_z(q) = \Bar{U}^z(t,q)$, where $\bar{U}^z(t,q)$ denotes the sample mean of $U_i^z(t,q)$ across units $i=1,\dots,n$.
By a Taylor expansion of the c.d.f. $F^z$ around $q$, for $0 < \gamma < 1/2$, we have
\[\begin{split}
\E [U_i^z(t,q)] &= F^z(q + t/\sqrt{n}) - F^z(q) = \frac{t}{\sqrt{n}}f^z(q) + \mathcal{O}\left(\frac{t^2}{n}\right),\\
\var (U_i^z(t,q)) &= \E U_i^z(t,q) - (\E U_i^z(t,q))^2 = \mathcal{O}\left(\frac{t}{\sqrt{n}}\right).
\end{split}\]
Applying the Central Limit Theorem (CLT) gives
$\sqrt{n}(\Bar{U}^z(t,q) - \E U_i^z(t,q)) = \mathcal{O}_p\big(\sqrt{\var (U_i^z(t,q))}\big)$.
Define $f_{z, n}(q) = f^z(q)$, and $r_{z, n}(t;q) := |\sqrt{n}\Bar{U}^z(t,q) - t\cdot f^z(q)|$, then we have $r_{z, n}(t;q) = \mathcal{O}_p(|t|^{1/2}n^{-1/4}) + \mathcal{O}(t^2 n^{-1/2}).$
Let $r_{z, n}(q) = \sup_{t \in [-2n^{\gamma},2n^{\gamma}]}|r_{z, n}(t;q)|$,
then for any $t \in [-2 n^\gamma, 2 n^\gamma]$, we have: $|\sqrt{n}(F_z(q + t / \sqrt{n}) - F_z(q)) - t f_{z, n}(q)| \le  r_{z, n}(q)$, where 
\[\begin{split}
r_{z, n}(q) = 
\mathcal{O}_p(n^{\gamma/2-1/4}) + \mathcal{O}(n^{2\gamma-1/2})
= \begin{cases}
\mathcal{O}_p(n^{2\gamma-1/2}), \text{ if } 1/6 < \gamma < 1/4,\\
\mathcal{O}_p(n^{\gamma/2-1/4}), \text{ if } 0 < \gamma \leq 1/6.
\end{cases}
\end{split}
\]
Consequently, in this i.i.d. smooth density setting, Assumption~\ref{assumption:derivativeF} holds in probability with $s = 2\gamma-1/2$ for $1/6 < \gamma < 1/4$, and $s = \gamma/2-1/4$ for $0 < \gamma \leq 1/6$.
\end{remark}

\section{Large-sample inference under ReM}\label{sec:CI}

Theorem~\ref{thm:QTE} provides an asymptotic approximation to the distribution of the estimator $\hat{\tau}_{\alpha}$ under rerandomization, which in turn enables the construction of large-sample confidence intervals for the quantile treatment effect (QTE).
Specifically, Theorem~\ref{thm:QTE} implies that the asymptotic distribution of $\sqrt{n}(\hat{\tau}_{\alpha}-\tau_\alpha)$ under ReM can be written as
\[\begin{split}
\sqrt{n} \tilde{V}_{\bq\bq}^{1/2} \left(\sqrt{1 - \tilde{R}_{\bq}^2}\varepsilon + \tilde{R}_{\bq} L_{K_n, a_n}\right) = A_n^{1/2}\varepsilon + B_n^{1/2}L_{K_n, a_n},
\end{split}\]
where $\varepsilon \sim \mathcal{N}(0,1)$ is independent of $L_{K_n,a_n}$,
$A_n = n\cdot \tilde{V}_{\bq\bq}(1-\tilde{R}_{\bq}^2)$, $B_n = n\cdot \tilde{V}_{\bq\bq}\tilde{R}_{\bq}^2$, and $C_n := A_n + B_n = n\cdot \tilde{V}_{\bq\bq}$ denotes the total asymptotic variance.
For each $z \in \{0,1\}$,
define 
$S_{q_{z}|\bx}^2 = \bs S_{q_{z}\bx}\bs S_{\bx\bx}^{-1} \bs S_{\bx q_{z}}$. 
Then $B_n$ admits the representation
\[\begin{split}
B_n &= \frac{1}{r_1r_0}\left[\frac{S_{q_{1}|\bx}^2}{f_{1,n}^2(q_{1,\alpha})}+\frac{S_{q_{0}|\bx}^2}{f_{0,n}^2(q_{0,\alpha})}-\frac{2\cdot \bs S_{q_{1}\bx}\bs S_{\bx\bx}^{-1} \bs S_{\bx q_{0}}}{f_{1,n}(q_{1,\alpha})f_{0,n}(q_{0,\alpha})}\right],
\end{split}\]
and $C_n$ can be expressed as
\[\begin{split}
C_n 
&= \frac{1}{r_1r_0}\left[\frac{S_{q_{1}q_1}}{f_{1,n}^2(q_{1,\alpha})}+\frac{S_{q_{0}q_0}}{f_{0,n}^2(q_{0,\alpha})}-\frac{2\cdot S_{q_{1}q_{0}}}{f_{1,n}(q_{1,\alpha})f_{0,n}(q_{0,\alpha})}\right],
\end{split}\]
where $S_{q_{1}q_1}$, $S_{q_{1}q_0}$ and $S_{q_{0}q_0}$ are defined in Eq.~\eqref{eq:Sqq}.
Unlike $B_n$, the term $C_n$ involves the cross-covariance component $S_{q_{1}q_0}$, which depends on the joint distribution
$F(q_{1,\alpha},q_{0,\alpha}) = \frac{1}{n}\sum_{i=1}^n \one\{Y_i(1)\leq q_{1,\alpha},Y_i(0)\leq q_{0,\alpha}\}$.
Because this quantity involves counterfactual outcomes, it is not identifiable from the observed data. 
As a result, $C_n$, and hence $A_n = C_n - B_n$, cannot be consistently estimated.
To address this issue, we construct estimable upper bounds for $C_n$ and $A_n$, denoted by $\tilde{C}_n$ and $\tilde{A}_n$, respectively, as formalized in the following proposition.

\begin{proposition}\label{pro:upper}
Define 
\[\begin{split}
\Tilde{C}_n =  \frac{1}{r_1r_0}\left[\frac{S_{q_{1}q_1}}{f_{1,n}^2(q_{1,\alpha})}+\frac{S_{q_{0}q_0}}{f_{0,n}^2(q_{0,\alpha})}+\frac{2\min \left\{\frac{r_1}{r_0}S_{q_{1}q_1},\frac{r_0}{r_1}S_{q_{0}q_0}\right\}}{f_{1,n}(q_{1,\alpha})f_{0,n}(q_{0,\alpha})}\right],
\end{split}\]
and 
$$\Tilde{A}_n = \Tilde{C}_n-B_n
= \frac{1}{r_1r_0}\left[\frac{S_{q_{1}\backslash \bx}^2}{f_{1,n}^2(q_{1,\alpha})}+\frac{S_{q_{0}\backslash \bx}^2}{f_{0,n}^2(q_{0,\alpha})}+\frac{2\left(\min \left\{\frac{r_1}{r_0}S_{q_{1}q_1},\frac{r_0}{r_1}S_{q_{0}q_0}\right\} + \bs S_{q_{1}\bx}\bs S_{\bx\bx}^{-1} \bs S_{\bx q_{0}}\right)}{f_{1,n}(q_{1,\alpha})f_{0,n}(q_{0,\alpha})}\right],$$ 
where $S_{q_{1}\backslash \bx}^2 = S_{q_1q_1}-S_{q_{1}\mid \bx}^2$, and $S_{q_{0}\backslash \bx}^2 = S_{q_0q_0}-S_{q_{0}\mid \bx}^2$.
Then, $C_n \leq \tilde{C}_n$ and $A_n \leq \tilde{A}_n$.
\end{proposition}

Proposition~\ref{pro:upper} provides estimable upper bounds for the unobservable quantities $C_n$ and $A_n$, which in turn allow us to construct an asymptotically conservative estimator of the covariance and a corresponding confidence interval.

Let the sample analogs of the population variances and covariances be
\[\begin{split}
\hat{s}_{q_{1}q_1} &= \frac{n_1 r_0^2}{n_1-1}\big[\hat{F}_1(\hat{q}_{1,\alpha})-\hat{F}_1^2(\hat{q}_{1,\alpha})\big],\ 
\hat{s}_{q_{0}q_0} = \frac{n_0 r_1^2}{n_0-1}\big[\hat{F}_0(\hat{q}_{0,\alpha})-\hat{F}_0^2(\hat{q}_{0,\alpha})\big],\\
\hat{\bs s}_{q_{1}\bx} &= \frac{r_0}{n_1-1}\sum_{i=1}^n Z_i (\one \{Y_i(1) \leq \hat{q}_{1,\alpha}\}-\hat{F}_1(\hat{q}_{1,\alpha}))(\bs X_i-\hat{\Bar{\bs X}}_1)^\top,\\
\hat{\bs s}_{q_{0}\bx} &= \frac{-r_1}{n_0-1}\sum_{i=1}^n (1-Z_i) (\one \{Y_i(0) \leq \hat{q}_{0,\alpha}\}-\hat{F}_0(\hat{q}_{0,\alpha}))(\bs X_i-\hat{\Bar{\bs X}}_0)^\top,\\
\hat{\bs s}_{q_{1}\mid \bx}^2 &=\hat{\bs s}_{q_{1}\bx}\bS_{\bx\bx}^{-1}\hat{\bs s}_{\bx q_{1}},\ 
\hat{\bs s}_{q_{0}\mid \bx}^2 =\hat{\bs s}_{q_{0}\bx}\bS_{\bx\bx}^{-1}\hat{\bs s}_{\bx q_{0}},
\end{split}\]
where $\hat{\Bar{\bs X}}_1 = n_1^{-1}\sum_{i=1}^n Z_i \bs X_i$ and $\hat{\Bar{\bs X}}_0 = n_0^{-1}\sum_{i=1}^n (1-Z_i) \bs X_i$ are the sample means of covariates in the treatment and control groups, respectively.
Using these quantities, we estimate the upper bound $\tilde{C}_n$ as
\[\begin{split}
\hat{C}_n = \frac{1}{r_1r_0}\left[\frac{\hat{s}_{q_{1}q_1}}{\hat{f}_{1,n}^2(\hat{q}_{1,\alpha})}+\frac{\hat{s}_{q_{0}q_0}}{\hat{f}_{0,n}^2(\hat{q}_{0,\alpha})}+\frac{2\min \left\{\frac{r_1}{r_0}\hat{s}_{q_{1}q_1},\frac{r_0}{r_1}\hat{s}_{q_{0}q_0}\right\}}{\hat{f}_{1,n}(\hat{q}_{1,\alpha})\hat{f}_{0,n}(\hat{q}_{1,\alpha})}\right],
\end{split}\]
and estimate $B_n$ by
\[\begin{split}
\hat{B}_n &= \frac{1}{r_1r_0}\left[\frac{\hat{\bs s}_{q_{1}\mid \bx}^2}{\hat{f}_{1,n}^2(\hat{q}_{1,\alpha})}+\frac{\hat{\bs s}_{q_{0}\mid \bx}^2}{\hat{f}_{0,n}^2(\hat{q}_{0,\alpha})}-\frac{2\cdot \hat{\bs s}_{q_{1}\bx}\bS_{\bx\bx}^{-1}\hat{\bs s}_{\bx q_{0}}}{\hat{f}_{1,n}(\hat{q}_{1,\alpha})\hat{f}_{0,n}(\hat{q}_{1,\alpha})}\right],
\end{split}\]
so that $\hat{A}_n = \hat{C}_n - \hat{B}_n$.
Here, the densities at the empirical quantiles are estimated via kernel density estimation:
\[\begin{split}
\hat{f}_{z,n}(\hat{q}_{z,\alpha}) =\frac{1}{n_z}\sum_{i:Z_i=z} \frac{1}{h_n}K\left(\frac{Y_i(z) - \hat{q}_{z,\alpha}}{h_n}\right),\ z\in \{0,1\},
\end{split}\] 
where $K(\cdot)$ is the Gaussian kernel and $h_n$ is the bandwidth. 
The bandwidth is chosen such that $\sqrt{n} h_n \to \infty$ and $n^{\gamma-1/2}/h_n = \Theta(n^v)$ for some $v>0$; for instance, when $\gamma > 1/6$, a valid choice is $h_n = n^{-1/3}$.

Finally, an asymptotically conservative $1-\alpha$ confidence interval for the QTE $\tau_\alpha$ is constructed as
\begin{equation}\label{eq:CI}
\hat{C}_{\alpha} = [\hat{\tau}_{\alpha}-n^{-1/2}\cdot\nu_{1-\alpha/2,K_n,a_n}(\hat{A}_n,\hat{B}_n),\hat{\tau}_{\alpha}+n^{-1/2}\cdot \nu_{1-\alpha/2,K_n,a_n}(\hat{A}_n,\hat{B}_n)],
\end{equation}
where $\nu_{1-\alpha/2, K_n, a_n}(\hat{A}_n, \hat{B}_n)$ denotes the $(1-\alpha/2)$th quantile of the distribution of $\hat{A}_n^{1/2} \varepsilon + \hat{B}_n^{1/2} L_{K_n, a_n}$.
To guarantee the asymptotic validity of the confidence interval in \eqref{eq:CI}, we impose the following additional regularity condition.

\begin{assumption}\label{assumption:step1}
As $n \rightarrow \infty$,
\[\begin{split}
\frac{K_n}{\Tilde{A}_n} \cdot\sqrt{\frac{\max\{\sqrt{n},\log K_n, -\log p_n\}}{n}} \rightarrow 0.
\end{split}\]
\end{assumption}

This assumption ensures that the covariate dimension $K_n$ does not grow too rapidly and that the rerandomization acceptance probability $p_n$ does not decay too quickly relative to the variance bound $\tilde{A}_n$. 
When both $K_n$ and $p_n$ are constant, it reduces to the condition $n^{1/4}\tilde{A}_n \to \infty$. 
Moreover, under Assumption~\ref{assumption:R2_2}, we have $\liminf_n A_n > 0$, so the condition is automatically satisfied in this case.

The following theorem establishes the asymptotic conservativeness of the confidence interval defined in \eqref{eq:CI}.

\begin{theorem}\label{thm:CI}
Under ReM and Assumptions \ref{assumption:gamma_n}-\ref{assumption:step1}, as $n \rightarrow \infty$, 
\begin{enumerate}
\item[(i)] The estimators $\hat{A}_n$ and $\hat{B}_n$ are asymptotically conservative in the sense that 
\begin{equation}\label{eq:targetCI}
\max\{|\hat{A}_n-\tilde{A}_n|,|\hat{B}_n-B_n|\} = o_p(\tilde{A}_n);
\end{equation}
\item[(ii)] For any $\alpha\in (0, 1)$, the $1-\alpha$ confidence interval $\hat{C}_\alpha$ in \eqref{eq:CI} is asymptotically conservative, in the sense that 
$\liminf_{n\rightarrow\infty} \pr(\tau_{\alpha} \in \hat{C}_{\alpha}) \geq 1-\alpha$.
\end{enumerate}
\end{theorem}

\section{Simulation analysis}\label{sec:simulation}
This section presents a series of numerical and real-data-based simulations to evaluate the performance of rerandomization (ReM) relative to complete randomization (CRE) in estimating quantile treatment effects (QTE) across various quantile levels. 
While the theoretical results established earlier concern the asymptotic distribution of the QTE estimator as the sample size tends to infinity, these simulations allow us to examine the finite-sample performance of ReM for QTE estimation. 
A range of scenarios, including both linear and nonlinear outcome models, are considered to assess how the methods perform under different underlying data-generating mechanisms.

\subsection{Numerical simulations}
We begin by evaluating the performance of ReM and CRE using data generated from the following linear potential outcome models:
 \[\begin{split}
Y_i(1) &= \mu_1 + \bm \beta_1^\top \bX_i + \epsilon_{1,i},\ \ i=1,\ldots,n,\\
Y_i(0) &= \mu_0 + \bm \beta_0^\top \bX_i + \epsilon_{0,i},\ \ i=1,\ldots,n,
\end{split}\]
where $\mu_0=0$, $\mu_1=5$, and the regression coefficients are set as $\bm \beta_1 = \bm \beta_0 =0.1 \cdot \bm 1_p$. 
The errors $\epsilon_1,\epsilon_0$ are independently drawn from $\mathcal{N}(0, \sigma^2_{\bepsilon})$.
Covariates $\bX$ follow a multivariate normal distribution, $\mathcal{N}(\bm 0_p,(1-\rho)\bI_p + \rho \bm 1_p \bm 1_p^T)$ with $\rho = 0.5$.
We consider $p = 10$ covariates, a sample size of $n = 1000$, and evaluate quantile levels $\alpha \in \{0.25, 0.5, 0.75\}$.
The error variance $\sigma^2_{\bepsilon}$ is chosen so that the corresponding $\Tilde{R}^2_{\bq}$ values are approximately 0.2 or 0.5.
Both balanced ($r_1 = 0.5$) and imbalanced ($r_1 = 0.2$) treatment allocations are examined.
For each scenario, a single dataset is generated and held fixed while 10,000 treatment assignments are simulated under both CRE and ReM.
The theoretical acceptance probability for rerandomization is set to $p_n = 0.001$.
The $\alpha$-th quantile treatment effect is estimated for each simulated assignment, and the following metrics are computed: bias, percent reduction in variance (PRIV), percent reduction in mean squared error (PRIMSE), average length of the 95\% confidence interval, and empirical coverage probability.
Confidence intervals are constructed according to \eqref{eq:CI} with kernel bandwidth $h_n = n^{-1/3}$.
The results are presented in Table \ref{tab:simulation}.

\begin{longtable}[]{@{}ccccccccc@{}}
\caption{Results of numerical simulations under linear models.}\label{tab:simulation}\tabularnewline
\toprule\noalign{}
& & & & Bias & PRIV & PRIMSE & CI Length & Coverage \\
\midrule\noalign{}
\endfirsthead
\toprule\noalign{}
& & & & Bias & PRIV & PRIMSE & CI Length & Coverage \\
\midrule\noalign{}
\endhead
\bottomrule\noalign{}
\endlastfoot
\multirow{12}{*}{$r_1 = 0.5$} & 
\multirow{4}{*}{$\alpha = 0.25$} & \multirow{2}{*}{$\Tilde{R}^2_{\alpha} = 0.2$} & CRE & 0.000 & 0.000 & 0.000 & 0.541 & 0.996 \\ 
& & & ReM & -0.001 & 16.817 & 16.811 & 0.507 & 0.997 \\ 
& & \multirow{2}{*}{$\Tilde{R}^2_{\alpha} = 0.5$} & CRE & -0.024 & 0.000 & 0.000 & 0.336 & 0.973 \\ 
& & & ReM & -0.023 & 44.242 & 40.570 & 0.280 & 0.984 \\ 
\cline{2-9} 
& \multirow{4}{*}{$\alpha = 0.5$} & \multirow{2}{*}{$\Tilde{R}^2_{\alpha} = 0.2$} & CRE & 0.012 & 0.000 & 0.000 & 0.583 & 0.993 \\ 
& & & ReM & 0.008 & 16.276 & 16.670 & 0.548 & 0.995 \\ 
& & \multirow{2}{*}{$\Tilde{R}^2_{\alpha} = 0.5$} & CRE & 0.003 & 0.000 & 0.000 & 0.367 & 0.981 \\ 
& & & ReM & 0.003 & 42.463 & 42.426 & 0.309 & 0.992 \\ 
\cline{2-9}
& \multirow{4}{*}{$\alpha = 0.75$} & \multirow{2}{*}{$\Tilde{R}^2_{\alpha} = 0.2$} & CRE & -0.013 & 0.000 & 0.000 & 0.618 & 0.990 \\ 
& & & ReM & -0.016 & 12.412 & 11.770 & 0.581 & 0.989 \\ 
& & \multirow{2}{*}{$\Tilde{R}^2_{\alpha} = 0.5$} & CRE & -0.000 & 0.000 & 0.000 & 0.299 & 0.962 \\ 
& & & ReM & -0.002 & 43.734 & 43.632 & 0.248 & 0.976 \\ 
\hline 
\multirow{12}{*}{$r_1 = 0.2$} & 
\multirow{4}{*}{$\alpha = 0.25$} & \multirow{2}{*}{$\Tilde{R}^2_{\alpha} = 0.2$} & CRE & -0.027 & 0.000 & 0.000 & 0.652 & 0.961 \\ 
& & & ReM & -0.026 & 17.943 & 17.663 & 0.595 & 0.965 \\ 
& & \multirow{2}{*}{$\Tilde{R}^2_{\alpha} = 0.5$} & CRE & -0.007 & 0.000 & 0.000 & 0.368 & 0.956 \\ 
& & & ReM & -0.008 & 43.481 & 43.100 & 0.292 & 0.966 \\ 
\cline{2-9} 
& \multirow{4}{*}{$\alpha = 0.5$} & \multirow{2}{*}{$\Tilde{R}^2_{\alpha} = 0.2$} & CRE & 0.002 & 0.000 & 0.000 & 0.725 & 0.968 \\ 
& & & ReM & 0.001 & 17.318 & 17.320 & 0.664 & 0.972 \\ 
& & \multirow{2}{*}{$\Tilde{R}^2_{\alpha} = 0.5$} & CRE & 0.006 & 0.000 & 0.000 & 0.436 & 0.956 \\ 
& & & ReM & 0.004 & 41.669 & 41.697 & 0.352 & 0.972 \\ 
\cline{2-9}
& \multirow{4}{*}{$\alpha = 0.75$} & \multirow{2}{*}{$\Tilde{R}^2_{\alpha} = 0.2$} & CRE & -0.024 & 0.000 & 0.000 & 0.720 & 0.958 \\ 
& & & ReM & -0.021 & 14.940 & 15.102 & 0.661 & 0.954 \\ 
& & \multirow{2}{*}{$\Tilde{R}^2_{\alpha} = 0.5$} & CRE & 0.003 & 0.000 & 0.000 & 0.347 & 0.950 \\ 
& & & ReM & 0.005 & 45.168 & 44.938 & 0.274 & 0.958 \\ 
\end{longtable}

The results provide several key insights.
First, they confirm the theoretical findings that the estimator $\hat{\tau}_{\alpha}$ is asymptotically unbiased and that the percent reduction in asymptotic variance (PRIASV) is proportional to $\tilde{R}^2_{\bq}$, as predicted by Corollary \ref{cor:PRIASV}.
Second, rerandomization consistently produces shorter (yet conservatively estimated) confidence intervals while maintaining empirical coverage rates above 95\%.
Finally, imbalanced treatment allocations (e.g., $r_1 = 0.2$) are more challenging than balanced allocations ($r_1 = 0.5$), as evidenced by the longer confidence intervals observed under both CRE and ReM.

To further evaluate the performance of ReM, we consider datasets generated from the following nonlinear potential outcome model:
 \[\begin{split}
Y_i(1) &= \mu_1 + \bm \beta_1^\top \exp(\bX_i) + \epsilon_{1,i},\ \ i=1,\ldots,n,\\
Y_i(0) &= \mu_0 + \bm \beta_0^\top \exp(\bX_i) + \epsilon_{0,i},\ \ i=1,\ldots,n,
\end{split}\]
while keeping all other parameters the same as in the linear setting.
For each quantile level $\alpha \in \{0.25,0.5,0.75\}$,
the error variance $\sigma^2_{\bepsilon}$ is adjusted to ensure that $\Tilde{R}^2_{\bq} \in \{0.2,0.5\}$.
The results are summarized in the first two blocks of Table~\ref{tab:simulation_nonlinear}.
Overall, the findings are consistent with those observed under the linear model. 
In particular, ReM continues to produce shorter, conservatively estimated confidence intervals and maintains good performance in terms of bias and variance reduction.
However, for the imbalanced design with $r_1 = 0.2$, the empirical coverage probability occasionally falls slightly below 95\%.
This can be attributed to the combined effects of the imbalanced treatment allocation and the nonlinear outcome model, both of which require a larger sample size to achieve the asymptotic conservativeness of the confidence intervals.
To investigate this, we increase the sample size to $n = 2000$ for the imbalanced design. 
As reported in the last block of Table~\ref{tab:simulation_nonlinear}, the empirical coverage probabilities improve and reach the nominal level, confirming that a sufficiently large sample size restores the desired conservativeness.

\begin{longtable}[]{@{}ccccccccc@{}}
\caption{Results of numerical simulations under nonlinear models.}\label{tab:simulation_nonlinear}\tabularnewline
\toprule\noalign{}
& & & & Bias & PRIV & PRIMSE & CI Length & Coverage \\
\midrule\noalign{}
\endfirsthead
\toprule\noalign{}
& & & & Bias & PRIV & PRIMSE & CI Length & Coverage \\
\midrule\noalign{}
\endhead
\bottomrule\noalign{}
\endlastfoot
\multirow{12}{*}{\makecell{$r_1 = 0.5$\\($n = 1000$)}} & 
\multirow{4}{*}{$\alpha = 0.25$} & \multirow{2}{*}{$\Tilde{R}^2_{\alpha} = 0.2$} & CRE & -0.014 & 0.000 & 0.000 & 0.683 & 0.981 \\ 
& & & ReM & -0.014 & 17.125 & 16.832 & 0.642 & 0.985 \\ 
& & \multirow{2}{*}{$\Tilde{R}^2_{\alpha} = 0.5$} & CRE & -0.000 & 0.000 & 0.000 & 0.277 & 0.969 \\ 
& & & ReM & -0.000 & 45.291 & 45.291 & 0.231 & 0.983 \\ 
\cline{2-9} 
& \multirow{4}{*}{$\alpha = 0.5$} & \multirow{2}{*}{$\Tilde{R}^2_{\alpha} = 0.2$} & CRE & 0.003 & 0.000 & 0.000 & 1.041 & 0.980 \\ 
& & & ReM & -0.002 & 13.997 & 14.004 & 0.979 & 0.983 \\ 
& & \multirow{2}{*}{$\Tilde{R}^2_{\alpha} = 0.5$} & CRE & 0.015 & 0.000 & 0.000 & 0.480 & 0.980 \\ 
& & & ReM & 0.014 & 44.186 & 43.632 & 0.402 & 0.988 \\ 
\cline{2-9}
& \multirow{4}{*}{$\alpha = 0.75$} & \multirow{2}{*}{$\Tilde{R}^2_{\alpha} = 0.2$} & CRE & -0.032 & 0.000 & 0.000 & 1.066 & 0.979 \\ 
& & & ReM & -0.037 & 12.420 & 11.488 & 1.004 & 0.978 \\ 
& & \multirow{2}{*}{$\Tilde{R}^2_{\alpha} = 0.5$} & CRE & -0.011 & 0.000 & 0.000 & 0.593 & 0.961 \\ 
& & & ReM & -0.011 & 42.442 & 42.099 & 0.490 & 0.976 \\ 
\hline 
\multirow{12}{*}{\makecell{$r_1 = 0.2$\\($n = 1000$)}} & 
\multirow{4}{*}{$\alpha = 0.25$} & \multirow{2}{*}{$\Tilde{R}^2_{\alpha} = 0.2$} & CRE & -0.015 & 0.000 & 0.000 & 0.689 & 0.962 \\ 
& & & ReM & -0.017 & 20.302 & 19.862 & 0.630 & 0.963 \\ 
& & \multirow{2}{*}{$\Tilde{R}^2_{\alpha} = 0.5$} & CRE & 0.007 & 0.000 & 0.000 & 0.315 & 0.956 \\ 
& & & ReM & 0.005 & 46.910 & 46.871 & 0.251 & 0.970 \\ 
\cline{2-9} 
& \multirow{4}{*}{$\alpha = 0.5$} & \multirow{2}{*}{$\Tilde{R}^2_{\alpha} = 0.2$} & CRE & 0.016 & 0.000 & 0.000 & 1.113 & 0.935 \\ 
& & & ReM & 0.015 & 16.526 & 16.519 & 1.027 & 0.938 \\ 
& & \multirow{2}{*}{$\Tilde{R}^2_{\alpha} = 0.5$} & CRE & -0.001 & 0.000 & 0.000 & 0.584 & 0.941 \\ 
& & & ReM & -0.006 & 44.245 & 44.071 & 0.469 & 0.948 \\ 
\cline{2-9}
& \multirow{4}{*}{$\alpha = 0.75$} & \multirow{2}{*}{$\Tilde{R}^2_{\alpha} = 0.2$} & CRE & -0.028 & 0.000 & 0.000 & 1.158 & 0.961 \\ 
& & & ReM & -0.026 & 19.337 & 19.246 & 1.043 & 0.960 \\ 
& & \multirow{2}{*}{$\Tilde{R}^2_{\alpha} = 0.5$} & CRE & 0.011 & 0.000 & 0.000 & 0.718 & 0.909 \\ 
& & & ReM & 0.009 & 45.962 & 45.939 & 0.567 & 0.931 \\ 
\hline 
\multirow{12}{*}{\makecell{$r_1 = 0.2$\\($n = 2000$)}} & 
\multirow{4}{*}{$\alpha = 0.25$} & \multirow{2}{*}{$\Tilde{R}^2_{\alpha} = 0.2$} & CRE & -0.009 & 0.000 & 0.000 & 0.463 & 0.965 \\ 
& & & ReM & -0.009 & 12.676 & 12.628 & 0.429 & 0.964 \\ 
& & \multirow{2}{*}{$\Tilde{R}^2_{\alpha} = 0.5$} & CRE & -0.004 & 0.000 & 0.000 & 0.201 & 0.974 \\ 
& & & ReM & -0.004 & 41.293 & 40.904 & 0.160 & 0.979 \\ 
\cline{2-9} 
& \multirow{4}{*}{$\alpha = 0.5$} & \multirow{2}{*}{$\Tilde{R}^2_{\alpha} = 0.2$} & CRE & 0.001 & 0.000 & 0.000 & 0.565 & 0.966 \\ 
& & & ReM & 0.002 & 14.356 & 14.332 & 0.524 & 0.966 \\ 
& & \multirow{2}{*}{$\Tilde{R}^2_{\alpha} = 0.5$} & CRE & 0.000 & 0.000 & 0.000 & 0.304 & 0.973 \\ 
& & & ReM & 0.001 & 43.043 & 43.035 & 0.245 & 0.984 \\  
\cline{2-9}
& \multirow{4}{*}{$\alpha = 0.75$} & \multirow{2}{*}{$\Tilde{R}^2_{\alpha} = 0.2$} & CRE & -0.030 & 0.000 & 0.000 & 0.829 & 0.956 \\ 
& & & ReM & -0.026 & 15.929 & 16.172 & 0.763 & 0.953 \\ 
& & \multirow{2}{*}{$\Tilde{R}^2_{\alpha} = 0.5$} & CRE & 0.001 & 0.000 & 0.000 & 0.442 & 0.964 \\ 
& & & ReM & -0.002 & 43.364 & 43.321 & 0.348 & 0.972 \\ 
\end{longtable}

\subsection{Real-data-based simulations}
To further assess the performance of ReM in comparison with CRE beyond the idealized settings considered above, we apply both methods to a real-world dataset.
While the preceding simulation studies were conducted under controlled designs with covariates generated from multivariate normal distributions, this analysis leverages covariates in an actual empirical study, thereby providing a more realistic evaluation of the methods.


Specifically, the dataset is constructed by adapting covariates from the Infant Health and Development Program (IHDP) and simulating potential outcomes following the procedure described in \cite{Hill2011}. 
The IHDP was designed to evaluate the effects of intensive child care and home visits on low-birth-weight and premature infants. 
The dataset contains 747 units and 25 pretreatment covariates, including 6 standardized continuous variables and 19 binary indicators.
Potential outcomes are generated according to the linear models
$$\bm Y(0) \sim N(\bm X\bm \beta,3)\text{ and } \bm Y(1) \sim N(\bm X\bm \beta+4,3),$$
where $\bm X$ denotes an expanded covariate matrix of dimension $747 \times 26$, with the first column consisting of ones. 
The coefficient vector $\bm\beta$ is a 26-dimensional random vector, whose elements are independently drawn from the discrete support $\{0,1,2,3,4\}$ with corresponding probabilities $(0.5, 0.2, 0.15, 0.1, 0.05)$.

We consider an experimental design in which the last unit is discarded and the remaining 746 units are evenly assigned to treatment and control groups, yielding $n_0 = n_1 = 373$ and a balanced treatment proportion $r_1 = 0.5$.
For each quantile level $\alpha \in {0.25, 0.5, 0.75}$, we generate three pseudo-datasets of potential outcomes by varying the residual variance in the fitted model so that $\tilde{R}^2_{\bq}$ takes values in $\{0.2, 0.3, 0.4\}$.
This results in a total of nine pseudo-datasets constructed from the same set of real covariates.
On each pseudo-dataset, we implement both CRE and ReM, with the acceptance probability under rerandomization fixed at $p_n = 0.001$.
The results, reported in 
Table~\ref{tab:real2}, confirm the advantage of rerandomization over complete randomization in terms of efficiency gains.
Consistent with the numerical simulations in the previous subsection, confidence interval estimation under this balanced design exhibits robustness, reinforcing the empirical stability of ReM when applied to realistic covariate distributions.

\begin{longtable}[]{@{}cccccccc@{}}
\caption{Results of real data analysis for IHDP data.}\label{tab:real2}\tabularnewline
\toprule\noalign{}
& & & Bias & PRIV & PRIMSE & CI Length & Coverage \\
\midrule\noalign{}
\endfirsthead
\toprule\noalign{}
& & & Bias & PRIV & PRIMSE & CI Length & Coverage \\
\midrule\noalign{}
\endhead
\bottomrule\noalign{}
\endlastfoot
\multirow{6}{*}{$\alpha = 0.25$} & \multirow{2}{*}{$\Tilde{R}^2_{\alpha} = 0.2$} & CRE & 0.034 & 0.000 & 0.000 & 3.310 & 0.942 \\ 
& & ReM & 0.046 & 12.449 & 12.221 & 3.013 & 0.933 \\ 
& \multirow{2}{*}{$\Tilde{R}^2_{\alpha} = 0.3$} & CRE & 0.050 & 0.000 & 0.000 & 2.077 & 0.968 \\ 
& & ReM & 0.058 & 17.426 & 16.732 & 1.807 & 0.961 \\ 
& \multirow{2}{*}{$\Tilde{R}^2_{\alpha} = 0.4$} & CRE & -0.020 & 0.000 & 0.000 & 1.871 & 0.965 \\ 
& & ReM & -0.016 & 22.952 & 22.982 & 1.589 & 0.963 \\ 
\hline 
\multirow{6}{*}{$\alpha = 0.5$} & \multirow{2}{*}{$\Tilde{R}^2_{\alpha} = 0.2$} & CRE & 0.069 & 0.000 & 0.000 & 3.273 & 0.960 \\ 
& & ReM & 0.073 & 11.622 & 11.368 & 2.966 & 0.955 \\ 
& \multirow{2}{*}{$\Tilde{R}^2_{\alpha} = 0.3$} & CRE & -0.025 & 0.000 & 0.000 & 2.211 & 0.980 \\ 
& & ReM & -0.021 & 21.836 & 21.863 & 1.935 & 0.976 \\ 
& \multirow{2}{*}{$\Tilde{R}^2_{\alpha} = 0.4$} & CRE & -0.066 & 0.000 & 0.000 & 2.039 & 0.972 \\ 
& & ReM & -0.061 & 28.937 & 28.617 & 1.702 & 0.971 \\ 
\hline 
\multirow{6}{*}{$\alpha = 0.75$} & \multirow{2}{*}{$\Tilde{R}^2_{\alpha} = 0.2$} & CRE & -0.102 & 0.000 & 0.000 & 2.965 & 0.961 \\ 
& & ReM & -0.079 & 12.961 & 13.653 & 2.689 & 0.959 \\ 
& \multirow{2}{*}{$\Tilde{R}^2_{\alpha} = 0.3$} & CRE & 0.018 & 0.000 & 0.000 & 2.205 & 0.967 \\ 
& & ReM & 0.031 & 19.160 & 18.823 & 1.934 & 0.963 \\ 
& \multirow{2}{*}{$\Tilde{R}^2_{\alpha} = 0.4$} & CRE & -0.027 & 0.000 & 0.000 & 1.501 & 0.982 \\ 
& & ReM & -0.025 & 26.917 & 26.834 & 1.263 & 0.980 \\ 
\end{longtable}

\section{Summary and Discussion}\label{sec:discussion}

In this paper, we investigate the asymptotic properties of the empirical quantile treatment effect (QTE) estimator under rerandomization (ReM) within a finite-population framework, where potential outcomes and covariates are treated as fixed and no superpopulation distributional assumptions are imposed.
Under mild regularity conditions, including smoothness of the empirical distributions of potential outcomes, appropriate rates for the acceptance probability, and stable limiting behavior of the multiple squared correlations, we show that the asymptotic distribution of the QTE estimator under ReM can be characterized as the sum of an independent Gaussian component and a truncated Gaussian component.
Based on this characterization, we derive the percent reduction in asymptotic sampling variance (PRIASV) under rerandomization and propose an asymptotically conservative confidence interval for the QTE.
Numerical simulations and real-data-based experiments further corroborate our theoretical results, demonstrating the finite-sample advantages of rerandomization over complete randomization in QTE estimation.

In addition to rerandomization, regression adjustment is another widely used strategy for enhancing the precision of treatment effect estimation.
For the average treatment effect, \citet{li2020rerandomization} show that combining regression adjustment with rerandomization leads to improved coverage properties and more reliable statistical inference.
This naturally motivates extending regression adjustment to the estimation of quantile treatment effects (QTE).
Recent work has investigated regression-adjusted QTE estimators under the super-population framework, particularly in settings involving covariate-adaptive designs and randomized experiments \citep{jiang2023regression, byambadalai2024estimating}.
However, regression adjustment for QTE, especially its interaction with rerandomization, have not yet been studied within the finite-population framework.
Developing regression-adjusted QTE estimators under rerandomization in the finite-population setting, and establishing their asymptotic distributions and inferential guarantees, therefore constitutes an important and promising direction for future research.

\section{Disclosure statement}\label{disclosure-statement}

No potential conflict of interest was reported by the authors.

\section{Data Availability Statement}\label{data-availability-statement}

Deidentified data used in the illustration of rerandomization in Section~\ref{sec:simulation} is available at \url{https://www.scidb.cn/en/anonymous/NnpFN2Zx}.

\phantomsection\label{supplementary-material}
\bigskip

\begin{center}

{\large\bf SUPPLEMENTARY MATERIAL}

\end{center}

\begin{description}
\item[Supplementary Material for ``Rerandomization for quantile treatment effects'':]
This file includes all the technical proofs. (.pdf file)
\end{description}



\bibliography{bibliography.bib}

\newpage
\appendix

\setcounter{equation}{0}
\renewcommand{\theequation}{A.\arabic{equation}}
\setcounter{theorem}{0}
\renewcommand{\thetheorem}{A\arabic{theorem}}
\setcounter{lemma}{0}
\renewcommand{\thelemma}{A\arabic{lemma}}

\section{Proof}
\subsection{Proof of Theorem \ref{thm:qz}}
\begin{proof}
Without loss of generality, we consider strata $z = 1$. Then for any $t \in [-2 n^\gamma, 2 n^\gamma]$,
\begin{align*}
&\pr(\sqrt{n}(\hat{q}_{1, \alpha} - q_{1, \alpha}) > t \mid M \le a_n) = \pr(\hat{q}_{1, \alpha} > q_{1, \alpha} + t/\sqrt{n}  \mid M \le a_n) \\ = &\pr\left(\hat{F}_1 (t / \sqrt{n} + q_{1, \alpha}) < \alpha \mid M \le a_n\right) \\
= &\pr\left(\hat{F}_1 (t / \sqrt{n} + q_{1, \alpha}) - F_1 (t / \sqrt{n} + q_{1, \alpha}) < \alpha - F_1 (t / \sqrt{n} + q_{1, \alpha})  \mid M \le a_n\right).
\end{align*}
Assumption \ref{assumption:derivativeF}
and the fact $\alpha \leq F_1(q_{1,\alpha})$ implies that 
\begin{align*}
&  \pr(\sqrt{n}(\hat{q}_{1, \alpha} - q_{1, \alpha}) > t \mid M \le a_n)  \\ 
= &\pr\left(\hat{F}_1 (t / \sqrt{n} + q_{1, \alpha}) - F_1 (t / \sqrt{n} + q_{1, \alpha}) < \alpha - F_1 (t / \sqrt{n} + q_{1, \alpha})  \mid M \le a_n\right)\\
\le &\pr\left(\hat{F}_1 (t / \sqrt{n} + q_{1, \alpha}) - F_1 (t / \sqrt{n} + q_{1, \alpha}) < F_1(q_{1,\alpha}) - F_1 (t / \sqrt{n} + q_{1, \alpha})\mid M \le a_n\right)\\
\le &\pr\left(\sqrt{n}\left(\hat{F}_1 (t / \sqrt{n} + q_{1, \alpha}) - F_1 (t / \sqrt{n} + q_{1, \alpha})\right) < - t \cdot f_{1, n}(q_{1, \alpha})+ r_{1, n}(q_{1, \alpha}) \mid M \le a_n\right) \\
= &\pr\left(\sqrt{n}\left(\hat{F}_1 (q_{1, \alpha}) - F_1 (q_{1, \alpha})\right) < - t \cdot f_{1, n}(q_{1, \alpha})+ \tilde{r}_{1,n} \mid M \le a_n\right),
\end{align*}
where 
\[\begin{split}
\tilde{r}_{1,n} := & \sqrt{n}\left(F_1 (q_{1, \alpha} + t/\sqrt{n}) - F_1 (q_{1, \alpha})\right) - 
\sqrt{n}\left(\hat{F}_1 (q_{1, \alpha} + t/\sqrt{n}) - \hat{F}_1 (q_{1, \alpha})\right) + r_{1, n}(q_{1, \alpha}).
\end{split}\]
Let $U_i^1(t,q_{1,\alpha}) = \one(Y_i(1) \le t / \sqrt{n} + q_{1, \alpha})-\one(Y_i(1) \le q_{1, \alpha})$, then 
\[\begin{split}
\tilde{r}_{1,n} = -\sqrt{n} \left(\frac{1}{n_1} \sum_{i=1}^n Z_iU_i^1(t,q_{1,\alpha})-\bar{U}^1(t,q_{1,\alpha})\right) + r_{1,n}(q_{1,\alpha}).
\end{split}\]
For any constants $k < 0$ and $c > 0$,
\[\begin{split}
&\pr\left(\sqrt{n}\left(\hat{F}_1 (q_{1, \alpha}) - F_1 (q_{1, \alpha})\right) < - t \cdot f_{1, n}(q_{1, \alpha})+ \tilde{r}_{1,n} \mid M \le a_n\right)\\ 
=&\pr\left(\sqrt{n}\left(\hat{F}_1 (q_{1, \alpha}) - F_1 (q_{1, \alpha})\right) < - t \cdot f_{1, n}(q_{1, \alpha})+ \tilde{r}_{1,n},\tilde{r}_{1,n} < c \cdot n^k \mid M \le a_n\right)\\
&+\pr\left(\sqrt{n}\left(\hat{F}_1 (q_{1, \alpha}) - F_1 (q_{1, \alpha})\right) < - t \cdot f_{1, n}(q_{1, \alpha})+ \tilde{r}_{1,n},\tilde{r}_{1,n} \geq c \cdot n^k \mid M \le a_n\right)
\\
\leq &\pr\left(\sqrt{n}\left(\hat{F}_1 (q_{1, \alpha}) - F_1 (q_{1, \alpha})\right) < - t \cdot f_{1, n}(q_{1, \alpha})+ c \cdot n^{k} \mid M \le a_n\right) + \pr(\tilde{r}_{1,n} \geq c \cdot n^{k} \mid M \le a_n) \\
=: & \mathrm{I} + \mathrm{II}.
\end{split}\]

We now control the two terms $\mathrm{I}$ and $\mathrm{II}$ separately.
\hspace{\fill}\\
\textbf{Part 1. control of $\mathrm{I}$}:
According to Theorem \ref{thm:AsympDis2}, for any $t \in [-2 n^\gamma, 2 n^\gamma]$, we have
\[\begin{split}
\sup_{t \in [-2 n^\gamma, 2 n^\gamma]} \big[&\pr\left(\sqrt{n}\left(\hat{F}_1 (q_{1, \alpha}) - F_1 (q_{1, \alpha})\right) < - t \cdot f_{1, n}(q_{1, \alpha}) + c \cdot n^{k} \mid M \le a_n\right)\\
& -\pr\left(G < - t \cdot f_{1, n}(q_{1, \alpha})+ c\cdot n^{k}
\right)\big] \leq O(\Delta_n/p_n),
\end{split}\]
where $G \sim \sqrt{n}
V_{q_{1}q_1}^{1/2}\left((1 - R_{q_{1}}^2)^{1/2} \varepsilon + R_{q_{1}} L_{K_n, a_n}\right)$ is the sum of independent Gaussian and truncated Gaussian distributions. Denote $G = X + Y$, where $X$ and $Y$ are the Gaussian and  truncated Gaussian components respectively, then 
\[\begin{split}
\pr\left(G < - t \cdot f_{1, n}(q_{1, \alpha})+  c\cdot n^{k}
\right) = \int_{-\infty}^{\infty} \pr(X < - t \cdot f_{1, n}(q_{1, \alpha})+  c\cdot n^{k} - y) f_Y(y) dy.
\end{split}\]
Assumption \ref{assumption:derivativeF} and Assumption \ref{assumption:replicate} imply that $\liminf_n n\cdot{V}_{q_{1}q_1} > 0$,
and Assumption \ref{assumption:R2_1} implies that 
$\limsup_{n} R_{q_{1}}^2 < 1$.
Hence, $\liminf_n \var(X) > 0$.
Then by Taylor expansion, there exists $c' > 0$ s.t. for any $y\in\mathbb{R}$, 
\[\begin{split}
|\pr(X < - t \cdot f_{1, n}(q_{1, \alpha})+ c\cdot n^{k} - y) - \pr(X < - t \cdot f_{1, n}(q_{1, \alpha}) - y)| \leq  c'\cdot n^{k}.
\end{split}\]
Hence, 
\[\begin{split}
&\left|\pr\left(G < - t \cdot f_{1, n}(q_{1, \alpha})+ c\cdot n^{k}
\right) - \pr\left(G < - t \cdot f_{1, n}(q_{1, \alpha})
\right)\right|\\
= & \left|\int_{-\infty}^{\infty}\left[\pr(X < - t \cdot f_{1, n}(q_{1, \alpha})+ c\cdot n^{k} - y) - \pr(X < - t \cdot f_{1, n}(q_{1, \alpha}) - y)\right]f_Y(y) dy
\right|\\
\leq & \int_{-\infty}^{\infty}\left|\pr(X < - t \cdot f_{1, n}(q_{1, \alpha})+ c\cdot n^{k} - y) - \pr(X < - t \cdot f_{1, n}(q_{1, \alpha}) - y)\right| f_Y(y) dy \leq c'\cdot n^{k},
\end{split}\]
i.e.,
\[\begin{split}
\pr\left(G < - t \cdot f_{1, n}(q_{1, \alpha})+ c\cdot n^{k}
\right) \leq  \pr\left(G < - t \cdot f_{1, n}(q_{1, \alpha})
\right) + c'\cdot n^{k}.
\end{split}\]
Then we have 
\[
\begin{split}
\sup_{t \in [-2 n^\gamma, 2 n^\gamma]}\big[&\pr\left(\sqrt{n}\left(\hat{F}_1 (q_{1, \alpha}) - F_1 (q_{1, \alpha})\right) < - t \cdot f_{1, n}(q_{1, \alpha})+ c\cdot n^{k} \mid M \le a_n\right) \\ - &\pr\left(\sqrt{n}
V_{q_{1}q_1}^{1/2}\left((1 - R_{q_{1}}^2)^{1/2} \varepsilon + R_{q_{1}} L_{K_n, a_n}\right) < - t \cdot f_{1, n}(q_{1, \alpha})
\right)\big] \leq O\left(\frac{\Delta_n}{p_n}\right) + c'\cdot n^{k}.
\end{split}\]
Let $\Tilde{V}_{q_{1}q_1} = V_{q_{1}q_1}/f^2_{1,n}(q_{1,\alpha})$, then we have 
\begin{equation}\label{eq:result1}
\begin{split}
\sup_{t \in [-2 n^\gamma, 2 n^\gamma]}\big[&\pr\left(\sqrt{n}\left(\hat{F}_1 (q_{1, \alpha}) - F_1 (q_{1, \alpha})\right) < - t \cdot f_{1, n}(q_{1, \alpha})+ c\cdot n^{k} \mid M \le a_n\right) \\ - &\pr\left(\sqrt{n}
\tilde{V}_{q_{1}q_1}^{1/2}\left((1 - R_{q_{1}}^2)^{1/2} \varepsilon + R_{q_{1}} L_{K_n, a_n}\right) > t
\right)\big] \leq O\left(\Delta_n/p_n\right) + O(n^{k}).
\end{split}
\end{equation}


\hspace{\fill}\\
\textbf{Part 2. control of $\mathrm{II}$:}
We first recall Lemma 1 in \cite{bloniarz2016lasso}.
\begin{lemma}[\cite{bloniarz2016lasso}]\label{lem:bloniarz2016lasso}
Let $\{z_i, i = 1,\ldots, n\}$ be a finite population of real numbers. Let $A \subset \{i,\ldots,n\}$ be a subset of deterministic size $|A| = n_A$ that is selected randomly without replacement. Define $p_A = n_A/n$, $\sigma^2 = (n-1)^{-1} \sum_{i=1}^n (z_i-\bar{z})^2$. Then, for any $t > 0$,
\begin{equation}\label{eq:bloniarz2016lasso}
\pr(\bar{z}_A - \bar{z} \geq t) \leq \exp\left\{-\frac{p_An_A t^2}{(1+\tau)^2\sigma^2}\right\},
\end{equation}
where $\tau = \min \{1/70,(3p_A)^2/70, (3 - 3p_A)^2/70\}$.
\end{lemma}

By definition, 
\[\begin{split}
\tilde{r}_{1,n} = -\sqrt{n} \left(\frac{1}{n_1} \sum_{i=1}^n Z_iU_i^1(t,q_{1,\alpha})-\bar{U}^1(t,q_{1,\alpha})\right) + r_{1,n}(q_{1,\alpha}),
\end{split}\]
and 
\[\begin{split}
\pr(\tilde{r}_{1,n} \geq c\cdot n^k) &= \pr\left(\left(\frac{1}{n_1}\sum_{i=1}^n Z_i U_i^1(t,q_{1,\alpha}) - \bar{U}^1(t,q_{1,\alpha})\right) \leq \frac{r_{1,n}(q_{1,\alpha})-c\cdot n^k}{\sqrt{n}}\right)\\
&= \pr\left(\left(\frac{1}{n_0}\sum_{i=1}^n (1-Z_i) U_i^1(t,q_{1,\alpha}) - \bar{U}^1(t,q_{1,\alpha})\right) \geq \frac{n_1}{n_0}\cdot\frac{c\cdot n^k-r_{1,n}(q_{1,\alpha})}{\sqrt{n}}\right).
\end{split}\]
For $k \geq s$, Assumption \ref{assumption:derivativeF} implies that for sufficiently large $n$ and $c > 0$, we have
$$c\cdot n^k-r_{1,n}(q_{1,\alpha}) \geq (c/2) \cdot n^k > 0.$$
This, together with Lemma \ref{lem:bloniarz2016lasso}, imply that 
\[\begin{split}
&\pr\left(\left(\frac{1}{n_0}\sum_{i=1}^n (1-Z_i) U_i^1(t,q_{1,\alpha}) - \bar{U}^1(t,q_{1,\alpha})\right) \geq \frac{n_1}{n_0}\cdot\frac{c\cdot n^k-r_{1,n}(q_{1,\alpha})}{\sqrt{n}}\right)\\
\leq & 
\exp\left\{-\frac{n_1^2n^{-2}(c\cdot n^k - r_{1,n}(q_{1,\alpha}))^2}{(1+\tau)^2 \cdot S_{U^1,n}^2(t,q_{1,\alpha})}\right\},
\end{split}\]
where
\[\begin{split}
S_{U^1,n}^2(t,q_{1,\alpha}) &= (n-1)^{-1} \sum_{i=1}^n (U_i^1(t,q_{1,\alpha}) - \Bar{U}^1(t,q_{1,\alpha}))^2\\ & = n(n-1)^{-1} \Bar{U}^1(t,q_{1,\alpha})(1-\Bar{U}^1(t,q_{1,\alpha})).
\end{split}\]
By Assumption \ref{assumption:derivativeF}, for $t \in [-2 n^\gamma, 2 n^\gamma]$,
\[\begin{split}
\Bar{U}^1(t,q_{1,\alpha}) =F_1(q_{1,\alpha} + t/\sqrt{n}) - F_1(q_{1,\alpha}) = O(n^{\gamma-1/2}),
\end{split}\]
and thus, $S_{U^1,n}^2(t,q_{1,\alpha}) = O(n^{\gamma-1/2})$.
Therefore, there a exist constant $a_c > 0$ depending on $c > 0$ such that for sufficiently large $n$, 
\[\begin{split}
\frac{n_1^2n^{-2}(c\cdot n^k - r_{1,n}(q_{1,\alpha}))^2}{(1+\tau)^2 \cdot S_{U^1,n}^2(t,q_{1,\alpha})} \geq a_c \cdot n^{2k-\gamma+1/2}.
\end{split}\]
Moreover, the constant $a_c$ satisfies that $a_c \rightarrow \infty$ as $c \rightarrow \infty$.
Hence,
\[\begin{split}
\pr(\tilde{r}_{1,n} \geq c\cdot n^k) \leq \exp\left\{- a_c \cdot n^{2k-\gamma+1/2}\right\}.
\end{split}\]
Let $\Tilde{p}_n = \pr(M \leq a_n)$,
then 
\[\begin{split}
\pr(\tilde{r}_{1,n} \geq c\cdot n^{k} \mid M \le a_n) &= \frac{\pr(\tilde{r}_n > c\cdot n^{k}, M \le a_n)}{\pr(M \le a_n)}\leq \frac{\pr(\tilde{r}_n > c\cdot n^{k})}{\pr(M \le a_n)}\\
&\leq \exp\big\{-a_c\cdot n^{2k-\gamma+1/2} - \log \Tilde{p}_n\big\}.
\end{split}\]
By the definition of $\Delta_n$, we have $|\Tilde{p}_n - p_n|\leq \Delta_n$. 
Under Assumption \ref{assumption:delta_p}, we have
\[\begin{split}
-\log \Tilde{p}_n \leq -\log (p_n - \Delta_n) = -\log p_n - \log(1-\Delta_n/p_n) =  -\log p_n + o(1).
\end{split}\]
Hence, 
\[\begin{split}
\pr(\tilde{r}_{1,n}\geq  c\cdot n^{k} \mid M \le a_n) \leq \exp\big\{-a_c\cdot n^{2k-\gamma+1/2}-\log p_n + o(1)\big\}.
\end{split}\]
By Assumption \ref{assumption:pn}, we have $-\log {p}_n = O(n^m)$ for some $0 < m < 1/2-\gamma$. 
Choose $s \leq k < 0$ s.t. $k \geq m/2 + \gamma/2 - 1/4$, then $2k-\gamma+1/2 > 0$ and 
\[\begin{split}
-\log {p}_n = O\left(n^{2k-\gamma+1/2}\right). 
\end{split}\]
Hence, for large $c$,
\begin{equation}\label{eq:result2}
\pr(\tilde{r}_{1,n} \geq c\cdot n^k\mid M \le a_n) = O(n^k).
\end{equation}
In light of our control of terms $\mathrm{I}$ and $\mathrm{II}$ in \eqref{eq:result1} and \eqref{eq:result2}, respectively, for $t \in [-2n^\gamma,2n^\gamma]$, 
\[\begin{split}
\sup_{t \in [-2n^\gamma,2n^\gamma]} \left[\pr(\sqrt{n}(\hat{q}_{1, \alpha} - q_{1, \alpha}) > t  \mid M \le a_n)\right. &\left.-\pr\left(\sqrt{n}\Tilde{V}_{q_{1}q_1}^{1 / 2}\left(
(1 - R_{q_{1}}^2)^{1/2} \varepsilon + R_{q_{1}} L_{K_n, a_n}\right) > t
\right)\right]\\ &\leq O(\Delta_n/p_n) + O(n^{k}),
\end{split}\]
i.e., 
\[\begin{split}
\inf_{t \in [-2n^\gamma,2n^\gamma]} \left[\pr(\sqrt{n}(\hat{q}_{1, \alpha} - q_{1, \alpha}) \leq t  \mid M \le a_n)\right. &\left.-\pr\left(\sqrt{n}\Tilde{V}_{q_{1}q_1}^{1 / 2}\left(
(1 - R_{q_{1}}^2)^{1/2} \varepsilon + R_{q_{1}} L_{K_n, a_n}\right) \leq t
\right)\right]\\ &\geq -O(\Delta_n/p_n) - O(n^{k}),
\end{split}\]
The above analysis provides a lower bound of  $\pr(\sqrt{n}(\hat{q}_{1, \alpha} - q_{1, \alpha}) \le t  \mid M \le a_n)$; we now consider its upper bound. Assumption \ref{assumption:replicate} implies that 
\[\begin{split}
& \pr(\sqrt{n}(\hat{q}_{1, \alpha} - q_{1, \alpha}) \leq t \mid M \le a_n)  \\ 
= &\pr\left(\hat{F}_1 (t / \sqrt{n} + q_{1, \alpha}) - F_1 (t / \sqrt{n} + q_{1, \alpha}) \geq \alpha - F_1 (t / \sqrt{n} + q_{1, \alpha})  \mid M \le a_n\right)\\
\leq & \pr\left(\hat{F}_1 (t / \sqrt{n} + q_{1, \alpha}) - F_1 (t / \sqrt{n} + q_{1, \alpha}) \geq F_1(q_{1, \alpha}) - F_1 (t / \sqrt{n} + q_{1, \alpha})-\delta/n  \mid M \le a_n\right)\\
\leq &\pr\left(\sqrt{n}\left(\hat{F}_1 (t / \sqrt{n} + q_{1, \alpha}) - F_1 (t / \sqrt{n} + q_{1, \alpha})\right) \geq - t \cdot f_{1, n}(q_{1, \alpha}) - r_{1, n}(q_{1, \alpha})-\delta/\sqrt{n} \mid M \le a_n\right) \\
= &\pr\left(\sqrt{n}\left(\hat{F}_1 (q_{1, \alpha}) - F_1 (q_{1, \alpha})\right) \geq - t \cdot f_{1, n}(q_{1, \alpha})- \tilde{r}_{1,n}' \mid M \le a_n\right),
\end{split}\]
where 
\[\begin{split}
\tilde{r}_{1,n}' := & -\sqrt{n}\left(F_1 (q_{1, \alpha} + t/\sqrt{n}) - F_1 (q_{1, \alpha})\right) +
\sqrt{n}\left(\hat{F}_1 (q_{1, \alpha} + t/\sqrt{n}) - \hat{F}_1 (q_{1, \alpha})\right) + r_{1, n}(q_{1, \alpha}) + \delta/\sqrt{n} \\
= & \sqrt{n} \left(\frac{1}{n_1} \sum_{i=1}^n Z_iU_i^1(t,q_{1,\alpha})-\bar{U}^1(t,q_{1,\alpha})\right)+r_{1, n}(q_{1, \alpha})+\delta/\sqrt{n}.
\end{split}\]
Similarly, we can prove that for any $k < 0$ and $c > 0$, 
\[\begin{split}
&\pr\left(\sqrt{n}\left(\hat{F}_1 (q_{1, \alpha}) - F_1 (q_{1, \alpha})\right) \geq - t \cdot f_{1, n}(q_{1, \alpha})- \tilde{r}_{1,n}' \mid M \le a_n\right)\\ 
= &\pr\left(\sqrt{n}\left(\hat{F}_1 (q_{1, \alpha}) - F_1 (q_{1, \alpha})\right) \geq - t \cdot f_{1, n}(q_{1, \alpha}) - \tilde{r}_{1,n}',\tilde{r}_{1,n}' < c \cdot n^k \mid M \le a_n\right)\\
&+ \pr\left(\sqrt{n}\left(\hat{F}_1 (q_{1, \alpha}) - F_1 (q_{1, \alpha})\right) \geq - t \cdot f_{1, n}(q_{1, \alpha}) - \tilde{r}_{1,n}',\tilde{r}_{1,n}' \geq c \cdot n^k \mid M \le a_n\right)\\
\leq &\pr\left(\sqrt{n}\left(\hat{F}_1 (q_{1, \alpha}) - F_1 (q_{1, \alpha})\right) \geq - t \cdot f_{1, n}(q_{1, \alpha}) - c \cdot n^{k} \mid M \le a_n\right) + \pr\left(\tilde{r}_{1,n}' \geq c \cdot n^k \mid M \le a_n\right).
\end{split}\]
Moreover, since
\[\begin{split}
\sup_{t \in [-2 n^\gamma, 2 n^\gamma]}\big[&\pr\left(\sqrt{n}\left(\hat{F}_1 (q_{1, \alpha}) - F_1 (q_{1, \alpha})\right) \geq - t \cdot f_{1, n}(q_{1, \alpha})- c\cdot n^{k} \mid M \le a_n\right) \\ - &\pr\left(\sqrt{n}
\tilde{V}_{q_{1}q_1}^{1/2}\left((1 - R_{q_{1}}^2)^{1/2} \varepsilon + R_{q_{1}} L_{K_n, a_n}\right) \leq t
\right)\big] \leq O\left(\Delta_n/p_n\right) + O(n^{k}),
\end{split}\]
and 
\[\pr(\tilde{r}_{1,n}' \geq c\cdot n^k\mid M \le a_n) = O(n^k),\]
we have 
\[\begin{split}
\sup_{t \in [-2n^\gamma,2n^\gamma]} \left[\pr(\sqrt{n}(\hat{q}_{1, \alpha} - q_{1, \alpha}) \leq t  \mid M \le a_n)\right. &\left.-\pr\left(\sqrt{n}\Tilde{V}_{q_{1}q_1}^{1 / 2}\left(
(1 - R_{q_{1}}^2)^{1/2} \varepsilon + R_{q_{1}} L_{K_n, a_n}\right) \leq t
\right)\right]\\ &\leq O(\Delta_n/p_n) + O(n^{k}).
\end{split}\]

In light of our control of both the lower and upper bounds, we have
\[\begin{split}
&\sup_{t \in [-2n^\gamma,2n^\gamma]} \left| \pr(\sqrt{n}(\hat{q}_{1, \alpha} - q_{1, \alpha}) \le t  \mid M \le a_n) - \pr\left(\sqrt{n}\Tilde{V}_{q_{1}q_1}^{1 / 2}\left(
(1 - R_{q_{1}}^2)^{1/2} \varepsilon + R_{q_{1}} L_{K_n, a_n}\right) \le t
\right)\right| \\ &\le O(\Delta_n/p_n) + O(n^{k}),
\end{split}\]
where $k = \max\{s,m/2+\gamma/2-1/4\}$.
\end{proof}

\subsection{Proof of Theorem \ref{thm:joint}}
\begin{proof}
For $k = \max\{s,m/2 + \gamma/2-1/4\} < 0$,
similar to the proof of Theorem \ref{thm:qz},
for $\bm t = (t_1,t_0)^\top \in [-2n^\gamma,2n^\gamma]^2$,
\[\begin{split}
&\pr\left(
\left(
\begin{matrix}
\sqrt{n}(\hat{q}_{1, \alpha} - q_{1, \alpha}) \\
\sqrt{n}(\hat{q}_{0, \alpha} - q_{0, \alpha})
\end{matrix}
\right) \le \bs{t} \mid M \le a_n\right) \\ \le &\pr\left(\sqrt{n}(\hat{F}_1(q_{1,\alpha})-F_1(q_{1,\alpha})) \ge -t_1 f_{1,n}(q_{1,\alpha}) - \tilde{r}_{1,n}',\right.\\ 
&\indent\indent\left. \sqrt{n}(\hat{F}_0(q_{0,\alpha})-F_0(q_{0,\alpha})) \ge -t_0 f_{0,n}(q_{0,\alpha}) - \tilde{r}_{0,n}' \mid M \leq a_n\right)\\
\leq& \pr\left(\sqrt{n}(\hat{F}_1(q_{1,\alpha})-F_1(q_{1,\alpha})) \ge -t_1 f_{1,n}(q_{1,\alpha}) - \tilde{r}_{1,n}',\tilde{r}_{1,n}' \leq c\cdot n^k,\tilde{r}_{0,n}' \leq c\cdot n^k,\right.\\ 
&\indent\indent\left. \sqrt{n}(\hat{F}_0(q_{0,\alpha})-F_0(q_{0,\alpha})) \ge -t_0 f_{0,n}(q_{0,\alpha}) - \tilde{r}_{0,n}' \mid M \leq a_n\right)\\
& +  \pr\left(\tilde{r}_{1,n}' > c\cdot n^k \mid M \leq a_n\right) + \pr\left(\tilde{r}_{0,n}' > c\cdot n^k \mid M \leq a_n\right)\\
\leq & \pr\left(\sqrt{n}(\hat{F}_1(q_{1,\alpha})-F_1(q_{1,\alpha})) \ge -t_1 f_{1,n}(q_{1,\alpha}) - c\cdot n^k,\right.\\ 
&\indent\indent\left. \sqrt{n}(\hat{F}_0(q_{0,\alpha})-F_0(q_{0,\alpha})) \ge -t_0 f_{0,n}(q_{0,\alpha}) - c\cdot n^k \mid M \leq a_n\right) + O(n^k),
\end{split}\]
where 
\[\begin{split}
\tilde{r}_{1,n}' := & -\sqrt{n}\left(F_1 (q_{1, \alpha} + t/\sqrt{n}) - F_1 (q_{1, \alpha})\right) +
\sqrt{n}\left(\hat{F}_1 (q_{1, \alpha} + t/\sqrt{n}) - \hat{F}_1 (q_{1, \alpha})\right) + r_{1, n}(q_{1, \alpha}) + \delta/\sqrt{n},\\
\tilde{r}_{0,n}' := & -\sqrt{n}\left(F_0 (q_{0, \alpha} + t/\sqrt{n}) - F_0 (q_{0, \alpha})\right) +
\sqrt{n}\left(\hat{F}_0 (q_{0, \alpha} + t/\sqrt{n}) - \hat{F}_0 (q_{0, \alpha})\right) + r_{0, n}(q_{0, \alpha}) + \delta/\sqrt{n}.
\end{split}\]
Similar to the proof of Theorem \ref{thm:qz}, we can prove that for $z\in\{0,1\}$, and any $t_z \in [-2n^\gamma,2n^\gamma]$,
\[\begin{split}
\sup_{t_z \in [-2 n^\gamma, 2 n^\gamma]}\big[&\pr\left(\sqrt{n}\left(\hat{F}_z (q_{z, \alpha}) - F_z (q_{z, \alpha})\right) \geq - t_z \cdot f_{z, n}(q_{z, \alpha})- c\cdot n^{k} \mid M \le a_n\right) \\ - &\pr\left(\sqrt{n}
V_{q_{z}q_z}^{1/2}\left((1 - R_{q_{z}}^2)^{1/2} \varepsilon + R_{q_{z}} L_{K_n, a_n}\right) \geq -t_z f_{z,n}(q_{z,\alpha})
\right)\big] \leq O\left(\Delta_n/p_n\right) + O(n^{k}).
\end{split}\]
By Theorem \ref{thm:AsympDis2}, we have 
\[\begin{split}
\sup_{t_z \in [-2 n^\gamma, 2 n^\gamma]}\big[&\pr\left(\sqrt{n}\left(\hat{F}_z (q_{z, \alpha}) - F_z (q_{z, \alpha})\right) \geq - t_z \cdot f_{z, n}(q_{z, \alpha})- c\cdot n^{k} \mid M \le a_n\right) \\ - &\pr\left(\sqrt{n}(\hat{F}_z(q_{z,\alpha})-F_z(q_{z,\alpha})) \ge -t_z f_{z,n}(q_{z,\alpha})\mid M \leq a_n\right)\big] \leq O\left(\Delta_n/p_n\right) + O(n^{k}).
\end{split}\]
Since 
\[\begin{split}
\left\{\sqrt{n}(\hat{F}_z(q_{z,\alpha})-F_z(q_{z,\alpha})) \ge -t_z f_{z,n}(q_{z,\alpha})\right\}
\subseteq \left\{\sqrt{n}(\hat{F}_z(q_{z,\alpha})-F_z(q_{z,\alpha})) \ge -t_z f_{z,n}(q_{z,\alpha}) - c\cdot n^k\right\},
\end{split}\]
we have 
\[\begin{split}
\sup_{t_0,t_1 \in [-2 n^\gamma, 2 n^\gamma]}\big[&\pr\left(\sqrt{n}\left(\hat{F}_z (q_{z, \alpha}) - F_z (q_{z, \alpha})\right) \geq - t_z \cdot f_{z, n}(q_{z, \alpha})- c\cdot n^{k},\forall z \in \{0,1\} \mid M \le a_n\right) \\ - &\pr\left(\sqrt{n}(\hat{F}_z(q_{z,\alpha})-F_z(q_{z,\alpha})) \ge -t_z f_{z,n}(q_{z,\alpha}),\forall z \in \{0,1\}\mid M \leq a_n\right)\big] \leq O\left(\Delta_n/p_n\right) + O(n^{k}).
\end{split}\]
By definition,
\[\begin{split}
&\pr\left(\sqrt{n}(\hat{F}_1(q_{1,\alpha})-F_1(q_{1,\alpha})) \ge -t_1 f_{1,n}(q_{1,\alpha}),\sqrt{n}(\hat{F}_0(q_{0,\alpha})-F_0(q_{0,\alpha})) \ge -t_0 f_{0,n}(q_{0,\alpha})\mid M \leq a_n\right)\\
=& \pr\left(\sqrt{n}\left(\begin{matrix}
\hat{F}_1(q_{1,\alpha})-F_1(q_{1,\alpha})\\ \hat{F}_0(q_{0,\alpha})-F_0(q_{0,\alpha})  
\end{matrix}\right)\ge -\bs{\Lambda}_{\bq}^{-1/2}\bs{t} \mid M \leq a_n \right),
\end{split}\]
where 
$\bs{\Lambda}_{\bq} = \text{diag}(f_{1,n}^{-2}(q_{1,\alpha}),f_{0,n}^{-2}(q_{0,\alpha}))$.
By Theorem \ref{thm:AsympDis}, we have 
\[\begin{split}
\sup_{\bs t\in [-2n^\gamma,2n^\gamma]^2}
\Bigg|&\pr\left(\sqrt{n}\left(\begin{matrix}
\hat{F}_1(q_{1,\alpha})-F_1(q_{1,\alpha})\\ \hat{F}_0(q_{0,\alpha})-F_0(q_{0,\alpha})  
\end{matrix}\right)\ge -\bs{\Lambda}_{\bq}^{-1/2}\bs{t} \mid M \leq a_n \right)\\ &-\pr\left(\sqrt{n}\bs{V}_{\bq\bq}^{1/2}((\bs{I} - \bs{R}_{\bq}^2)^{1/2} \bs{\varepsilon} + \bs{R}_{\bq} \bs{L}_{K_n, a_n}) \le \bs{\Lambda}_{\bq}^{-1/2}\bs{t}\right)\Bigg|
\leq O(\Delta_n/p_n).
\end{split}\]
Therefore, 
\[\begin{split}
\sup_{\bs t\in [-2n^\gamma,2n^\gamma]^2}
\Bigg[&\pr\left(
\left(
\begin{matrix}
\sqrt{n}(\hat{q}_{1, \alpha} - q_{1, \alpha}) \\
\sqrt{n}(\hat{q}_{0, \alpha} - q_{0, \alpha})
\end{matrix}
\right) \le \bs{t} \mid M \le a_n\right)\\ &-\pr\left(\sqrt{n}\bs{V}_{\bq\bq}^{1/2}((\bs{I} - \bs{R}_{\bq}^2)^{1/2} \bs{\varepsilon} + \bs{R}_{\bq} \bs{L}_{K_n, a_n}) \le \bs{\Lambda}_{\bq}^{-1/2}\bs{t}\right)\Bigg]
\leq O(\Delta_n/p_n) + O(n^k).
\end{split}\]
Since $\tilde{\bV}_{\bq\bq}^{1/2} := \bs{\Lambda}_{\bq}^{1/2} \bV_{\bq\bq}^{1/2}$, we have 
\[\begin{split}
\sup_{\bs t\in [-2n^\gamma,2n^\gamma]^2}
\Bigg[&\pr\left(
\left(
\begin{matrix}
\sqrt{n}(\hat{q}_{1, \alpha} - q_{1, \alpha}) \\
\sqrt{n}(\hat{q}_{0, \alpha} - q_{0, \alpha})
\end{matrix}
\right) \le \bs{t} \mid M \le a_n\right)\\ &-\pr\left(\sqrt{n}\tilde{\bs{V}}_{\bq\bq}^{1/2}((\bs{I} - \bs{R}_{\bq}^2)^{1/2} \bs{\varepsilon} + \bs{R}_{\bq} \bs{L}_{K_n, a_n}) \le \bs{t}\right)\Bigg]
\leq O(\Delta_n/p_n) + O(n^k).
\end{split}\]
Using an analogous argument we can also derive the lower bound. Putting together we obtain the desired result.
\end{proof}

\subsection{Proof of Theorem \ref{thm:QTE}}
\begin{proof}
By definition,
\begin{align*}
\pr(\sqrt{n}((\hat{q}_{1, \alpha} - \hat{q}_{0, \alpha}) - (q_{1, \alpha} - q_{0, \alpha})) \le t) = \pr(\sqrt{n}(\hat{q}_{1, \alpha} - q_{1, \alpha}) - \sqrt{n} (\hat{q}_{0, \alpha} - q_{0, \alpha}) \le t).
\end{align*}
Let $0 < \gamma' \leq \gamma$ s.t. $\gamma' < \min\{-k/2,-k'/2\}$ for $k = \max\{s,m/2+\gamma/2-1/4\}$, and $k'$ defined in Assumption \ref{assumption:gamma}, then
Assumption \ref{assumption:gamma} implies that $n^{-2\gamma'} \gg \Delta_n/p_n$.
Then we have
\begin{align*}
& \pr(\sqrt{n}((\hat{q}_{1, \alpha} - \hat{q}_{0, \alpha}) - (q_{1, \alpha} - q_{0, \alpha})) \le t \mid M \leq a_n)  \\
& \le \sum_{\zeta = - [n^{2\gamma'}] + 1}^{[n^{2\gamma'}]}\pr(\sqrt{n}(\hat{q}_{1, \alpha} - q_{1, \alpha}) - \sqrt{n} (\hat{q}_{0, \alpha} - q_{0, \alpha}) \le t, (\zeta - 1)  n^{-\gamma'} < \sqrt{n} (\hat{q}_{0, \alpha} - q_{0, \alpha}) \le \zeta  n^{-\gamma'} \mid M \leq a_n) \\
& \qquad + \pr(|\sqrt{n} (\hat{q}_{0, \alpha} - q_{0, \alpha})| > [n^{2\gamma'}] \cdot n^{-\gamma'} \mid M \leq a_n) \\
& \le \sum_{\zeta = - [n^{2\gamma'}] + 1}^{[n^{2\gamma'}]}\pr(\sqrt{n}(\hat{q}_{1, \alpha} - q_{1, \alpha}) \le t + \zeta \cdot n^{-\gamma'}, (\zeta - 1) \cdot n^{-\gamma'} < \sqrt{n} (\hat{q}_{0, \alpha} - q_{0, \alpha}) \le \zeta \cdot n^{-\gamma'} \mid M \leq a_n) \\
& \qquad + \pr(|\sqrt{n} (\hat{q}_{0, \alpha} - q_{0, \alpha})| > n^{\gamma'}-n^{-\gamma'} \mid M \leq a_n),
\end{align*}
where $[x]$ denotes the floor of $x$.
Since $1-[n^{2\gamma'}]\leq\zeta\leq [n^{2\gamma'}]$, then $\zeta\cdot n^{-\gamma'} \in [n^{-\gamma'}-n^{\gamma'},n^{\gamma'}]$, $(\zeta-1)\cdot n^{-\gamma'} \in [-n^{\gamma'},n^{\gamma'}-n^{-\gamma'}]$.
For $t \in [-n^{\gamma}, n^{\gamma}]$, when $n$ is large enough, we have 
\[\begin{split}
t+\zeta\cdot n^{-\gamma'} \in [-2\cdot n^\gamma,2\cdot n^\gamma].
\end{split}\]
According to the convergence results in Theorem \ref{thm:qz} and Theorem \ref{thm:joint},
\begin{align*}
& \pr(\sqrt{n}((\hat{q}_{1, \alpha} - \hat{q}_{0, \alpha}) - (q_{1, \alpha} - q_{0, \alpha})) \le t \mid M \leq a_n)  \\
& \le \sum_{\zeta = - [n^{2\gamma'}] + 1}^{[n^{2\gamma'}]}\pr\left(
\left(
\begin{matrix}
    - \infty \\
    (\zeta - 1) \cdot n^{-\gamma'}
\end{matrix}
\right) < \sqrt{n} \tilde{\bV}_{\bq\bq}^{1/2} ((\bs{I} - \bs{R}_{\bq}^2)^{1/2} \bs{\varepsilon} + \bs{R}_{\bq} \bs{L}_{K_n, a_n}) \le
\left(
\begin{matrix}
    t + \zeta \cdot n^{-\gamma'} \\
    \zeta \cdot n^{-\gamma'}
\end{matrix}
\right)
\right) \\
& \qquad + \pr(|\sqrt{n} (\hat{q}_{0, \alpha} - q_{0, \alpha})| > n^{\gamma'}-n^{-\gamma'} \mid M \leq a_n)+ 2 [n^{2\gamma'}] (O(\Delta_n/p_n) + O(n^{k})) \\
& \le \pr\left(
\bs{c}^\top \sqrt{n} \tilde{\bV}_{\bq\bq}^{1/2} ((\bs{I} - \bs{R}_{\bq}^2)^{1/2} \bs{\varepsilon} + \bs{R}_{\bq} \bs{L}_{K_n, a_n}) \le t + n^{-\gamma'}
\right) + o(1) \\
& = \pr\left(\sqrt{n} \tilde{V}_{\bq\bq}^{1/2} (\sqrt{1 - \tilde{R}_{\bq}^2}\varepsilon + \tilde{R}_{\bq} L_{K_n, a_n}) \le t + n^{-\gamma'} \right) + o(1),
\end{align*}
where $\bm c = (1,-1)^\top$, and $o(1)$ holds uniformly for $t \in [-n^\gamma,n^\gamma]$.
Since 
Assumption \ref{assumption:R2_2} implies that $\limsup_{n\rightarrow\infty}\tilde{R}_{\bq}^2 < 1$, and $\liminf_n n\cdot\Tilde{V}_{\bq\bq} > 0$, then for $t \in [-n^\gamma,n^\gamma]$,
\[\begin{split}
&\sup_{t \in [-n^\gamma,n^\gamma]} \Big[ \pr(\sqrt{n}((\hat{q}_{1, \alpha} - \hat{q}_{0, \alpha}) - (q_{1, \alpha} - q_{0, \alpha})) \le t \mid M \leq a_n)  - \pr\left(\sqrt{n} \tilde{V}_{\bq\bq}^{1/2} (\sqrt{1 - \tilde{R}_{\bq}^2}\varepsilon + \tilde{R}_{\bq} L_{K_n, a_n}) \le t \right)\Big]\\ &\leq o(1).
\end{split}\]
On the other hand, 
\begin{align*}
& \pr(\sqrt{n}((\hat{q}_{1, \alpha} - \hat{q}_{0, \alpha}) - (q_{1, \alpha} - q_{0, \alpha})) \le t \mid M \leq a_n)  \\
& \ge \sum_{\zeta = - [n^{2\gamma'}] + 1}^{[n^{2\gamma'}]}\pr(\sqrt{n}(\hat{q}_{1, \alpha} - q_{1, \alpha}) - \sqrt{n} (\hat{q}_{0, \alpha} - q_{0, \alpha}) \le t, (\zeta - 1)  n^{-\gamma'} < \sqrt{n} (\hat{q}_{0, \alpha} - q_{0, \alpha}) \le \zeta \cdot n^{-\gamma'} \mid M \leq a_n) \\
& \ge \sum_{\zeta = - [n^{2\gamma'}] + 1}^{[n^{2\gamma'}]}\pr(\sqrt{n}(\hat{q}_{1, \alpha} - q_{1, \alpha}) \le t + (\zeta-1) \cdot n^{-\gamma'}, (\zeta - 1) \cdot n^{-\gamma'} < \sqrt{n} (\hat{q}_{0, \alpha} - q_{0, \alpha}) \le \zeta \cdot n^{-\gamma'} \mid M \leq a_n).
\end{align*}
Since 
\[\begin{split}
t+(\zeta-1)\cdot n^{-\gamma'} \in  [-2\cdot n^\gamma,2\cdot n^\gamma],
\end{split}\]
we have 
\begin{align*}
& \pr(\sqrt{n}((\hat{q}_{1, \alpha} - \hat{q}_{0, \alpha}) - (q_{1, \alpha} - q_{0, \alpha})) \le t \mid M \leq a_n)  \\
& \ge \sum_{\zeta = - [n^{2\gamma'}] + 1}^{[n^{2\gamma'}]}\pr\left(
\left(
\begin{matrix}
    - \infty \\
    (\zeta - 1) \cdot n^{-\gamma'}
\end{matrix}
\right) < \sqrt{n} \tilde{\bV}_{\bq\bq}^{1/2} ((\bs{I} - \bs{R}_{\bq}^2)^{1/2} \bs{\varepsilon} + \bs{R}_{\bq} \bs{L}_{K_n, a_n}) \le
\left(
\begin{matrix}
    t + (\zeta - 1) \cdot n^{-\gamma'} \\
    \zeta \cdot n^{-\gamma'}
\end{matrix}
\right)
\right) \\
& \qquad - 2 [n^{2\gamma'}] (O(\Delta_n/p_n) + O(n^{k})) \\
& \ge \pr\left(
\bs{c}^\top \sqrt{n} \tilde{\bV}_{\bq\bq}^{1/2} ((\bs{I} - \bs{R}_{\bq}^2)^{1/2} \bs{\varepsilon} + \bs{R}_{\bq} \bs{L}_{K_n, a_n}) \le t - n^{-\gamma'},\right. \\
\indent &\left.
\left(\begin{matrix}
-\infty \\
-[n^{2\gamma'}]\cdot n^{-\gamma'}
\end{matrix}\right)< 
\sqrt{n} \tilde{\bV}_{\bq\bq}^{1/2} ((\bs{I} - \bs{R}_{\bq}^2)^{1/2} \bs{\varepsilon} + \bs{R}_{\bq} \bs{L}_{K_n, a_n}) \leq \left(\begin{matrix}
+\infty \\
[n^{2\gamma'}]\cdot n^{-\gamma'}
\end{matrix}\right)
\right) - o(1) \\
&\geq \pr\left(
\bs{c}^\top \sqrt{n} \tilde{\bV}_{\bq\bq}^{1/2} ((\bs{I} - \bs{R}_{\bq}^2)^{1/2} \bs{\varepsilon} + \bs{R}_{\bq} \bs{L}_{K_n, a_n}) \le t - n^{-\gamma'}\right)\\
&\ \ \ \ \ \ \ \  - 
\pr\left(\left|\sqrt{n} \Tilde{V}_{q_{0}q_0}^{1/2} \left(\sqrt{1 - R^2_{q_{0}}}\varepsilon + R_{q_{0}} L_{K_n, a_n}\right)\right|
> [n^{2\gamma'}]\cdot n^{-\gamma'}\right)
- o(1)
\\
& = \pr\left(\sqrt{n} \tilde{V}_{\bq\bq}^{1/2} (\sqrt{1 - \tilde{R}_{\bq}^2}\varepsilon + \tilde{R}_{\bq} L_{K_n, a_n}) \le t-n^{-\gamma'} \right) - o(1),
\end{align*}
where $o(1)$ holds uniformly for $t \in [-n^\gamma,n^\gamma]$, and the last equation holds 
since Assumption \ref{assumption:derivativeF} and Assumption \ref{assumption:replicate} imply that $\limsup_n n\cdot\Tilde{V}_{q_{z}q_z} < \infty$, Assumption \ref{assumption:R2_1} implies that $\limsup_n R^2_{q_z} < 1$,
and thus,
\[\begin{split}
\pr\left(\left|\sqrt{n} \Tilde{V}_{q_{z}q_z}^{1/2} \left(\sqrt{1 - R^2_{q_{z}}}\varepsilon + R_{q_{z}} L_{K_n, a_n}\right)\right|
> [n^{2\gamma'}]\cdot n^{-\gamma'}\right) = o(1).
\end{split}\]
Since 
Assumption \ref{assumption:R2_2} implies that $\limsup_{n\rightarrow\infty}\tilde{R}_{\bq}^2 < 1$, and $\liminf_n n\cdot\Tilde{V}_{\bq\bq} > 0$, then 
\[\begin{split}
&\inf_{t \in [-n^\gamma,n^\gamma]} \Big[ \pr(\sqrt{n}((\hat{q}_{1, \alpha} - \hat{q}_{0, \alpha}) - (q_{1, \alpha} - q_{0, \alpha})) \le t \mid M \leq a_n)  - \pr\left(\sqrt{n} \tilde{V}_{\bq\bq}^{1/2} (\sqrt{1 - \tilde{R}_{\bq}^2}\varepsilon + \tilde{R}_{\bq} L_{K_n, a_n}) \le t \right)\Big]\\ &\geq -o(1).
\end{split}\]
Putting together, we obtain that
\[\begin{split}
\sup_{t \in [-n^\gamma,n^{\gamma}]}\Big|&\pr\left(\sqrt{n}\left((\hat{q}_{1, \alpha} - \hat{q}_{0, \alpha}) - (q_{1, \alpha} - q_{0, \alpha})\right) \le t \mid M \leq a_n\right) \\ & \indent \indent  -  \pr\left(\sqrt{n} \tilde{V}_{\bq\bq}^{1/2} \left(\sqrt{1 - \tilde{R}_{\bq}^2}\varepsilon + \tilde{R}_{\bq} L_{K_n, a_n}\right) \le t \right)\Big| = o(1).
\end{split}\]
\end{proof}

\subsection{Proof of Proposition \ref{pro:upper}}
\begin{proof}
Since  
\[\begin{split}
F(q_{1,\alpha},q_{0,\alpha}) &= \frac{1}{n}\sum_{i=1}^n \one\{Y_i(1)\leq q_{1,\alpha},Y_i(0)\leq q_{0,\alpha}\} \leq \min\{F_1(q_{1,\alpha}),F_0(q_{0,\alpha})\},
\end{split}\]
we have 
\[\begin{split}
-S_{q_{1}q_{0}}
&= \frac{nr_0r_1}{n-1}\big[F(q_{1,\alpha},q_{0,\alpha})-F_1(q_{1,\alpha})F_0(q_{0,\alpha})\big]\\
&\leq \frac{nr_0r_1}{n-1}\big[\min\{F_1(q_{1,\alpha}),F_0(q_{0,\alpha})\}-F_1(q_{1,\alpha})F_0(q_{0,\alpha})\big]\\
&\leq \frac{nr_0r_1}{n-1} \cdot\min\big\{F_1(q_{1,\alpha})(1-F_1(q_{1,\alpha})),F_0(q_{0,\alpha})(1-F_0(q_{0,\alpha}))\big\}\\
&= \min \left\{\frac{r_1}{r_0}S_1^2,\frac{r_0}{r_1}S_0^2\right\}.
\end{split}\]
Therefore,
\[\begin{split}
C_n &= \frac{1}{r_1r_0}\left[\frac{S_1^2}{f_{1,n}^2(q_{1,\alpha})}+\frac{S_0^2}{f_{0,n}^2(q_{0,\alpha})}-\frac{2\cdot S_{q_{1}q_{0}}}{f_{1,n}(q_{1,\alpha})f_{0,n}(q_{0,\alpha})}\right]\\
&\leq \frac{1}{r_1r_0}\left[\frac{S_1^2}{f_{1,n}^2(q_{1,\alpha})}+\frac{S_0^2}{f_{0,n}^2(q_{0,\alpha})}+\frac{2\min \left\{\frac{r_1}{r_0}S_1^2,\frac{r_0}{r_1}S_0^2\right\}}{f_{1,n}(q_{1,\alpha})f_{0,n}(q_{0,\alpha})}\right]\\
&= \tilde{C}_n.
\end{split}\]
Moreover, for $\tilde{A}_n = \tilde{C}_n - B_n$, we have 
\[\begin{split}
\tilde{A}_n \geq C_n - B_n = A_n.
\end{split}\]
\end{proof}

\subsection{Proof of Theorem \ref{thm:CI}}
\begin{proof}
For any $\alpha \in (0,1)$, recall that
the covariance matrix of $\bs{u}_{i,n} := (r_0 \one(Y_i(1) \le q_{1,\alpha}), - r_1 \one(Y_i(0) \le q_{0,\alpha}), \bs{X}_i^\top)^\top$ is 
\[\begin{split}
\bs S_{\bm u\bm u} = 
\begin{pmatrix}
S_{q_1q_1} & S_{q_{1}q_{0}} & \bs S_{q_{1}\bx} \\
S_{q_{0}q_{1}} & S_{q_0q_0} & \bs S_{q_{0}\bx} \\
\bs S_{\bx q_{1}} & \bs S_{\bx q_{0}} & \bs S_{\bx\bx}
\end{pmatrix}.
\end{split}\]
For simplicity, for $z\in \{0,1\}$, we redefine 
\[\begin{split}
S_z^2 := S_{q_zq_z},\
\bS_{z \bx} := \bS_{q_z\bx},\
\bS_{\bx z} := \bS_{\bx q_z},\
S_{10} := S_{q_1q_0},\
S_{01} := S_{q_0q_1}.
\end{split}\]
Let the standardized covariates be $\bs W_i = \bs S_{\bx\bx}^{-1/2}(\bs X_i -  \bar{\bs X})$.
Then for $z\in \{0,1\}$, define
\[\begin{split}
\bs S_{zw} :&= \frac{(-1)^{1-z}r_{1-z}}{n-1}\sum_{i=1}^n (\one \{Y_i(z) \leq q_{z,\alpha}\}-{F}_z(q_{z,\alpha}))(\bs W_i-\Bar{\bs W})^\top\\
&= \frac{(-1)^{1-z}r_{1-z}}{n-1}\sum_{i=1}^n (\one \{Y_i(z) \leq q_{z,\alpha}\}-{F}_z(q_{z,\alpha}))(\bs X_i-\Bar{\bs X})^\top\bs S_{\bx\bx}^{-1/2}\\
&= \bs S_{z\bx}\bs S_{\bx\bx}^{-1/2}.
\end{split}\]
Then we have 
\[\begin{split}
S_{q_{z}|\bx}^2 :=
\bS_{q_{z}\bx}\bS_{\bx\bx}^{-1}\bS_{\bx q_{z}}
=\bS_{z\bx}\bS_{\bx\bx}^{-1}\bS_{\bx z}
= 
||\bs S_{zw}||_2^2 \leq S_{z}^2.
\end{split}\]
By definition,
$A_n = C_n - B_n \leq \tilde{A}_n =
\tilde{C}_n - B_n$, where 
\[\begin{split}
B_n 
&= \frac{1}{r_1r_0}\left[\frac{||\bs S_{1w}||_2^2}{f_{1,n}^2(q_{1,\alpha})}+\frac{||\bs S_{0w}||_2^2}{f_{0,n}^2(q_{0,\alpha})}-\frac{2\cdot \bs S_{1w}\bs S_{0w}^\top}{f_{1,n}(q_{1,\alpha})f_{0,n}(q_{0,\alpha})}\right],\\
C_n &= \frac{1}{r_1r_0}\left[\frac{S_{1}^2}{f_{1,n}^2(q_{1,\alpha})}+\frac{S_{0}^2}{f_{0,n}^2(q_{0,\alpha})}-\frac{2\cdot S_{10}}{f_{1,n}(q_{1,\alpha})f_{0,n}(q_{0,\alpha})}\right],\\
\tilde{C}_n &= \frac{1}{r_1r_0}\left[\frac{S_{1}^2}{f_{1,n}^2(q_{1,\alpha})}+\frac{S_{0}^2}{f_{0,n}^2(q_{0,\alpha})}+\frac{2\min \left\{\frac{r_1}{r_0}S_1^2,\frac{r_0}{r_1}S_0^2\right\}}{f_{1,n}(q_{1,\alpha})f_{0,n}(q_{0,\alpha})}\right].
\end{split}\]

\hspace{\fill}\\
\\
Let $\hat{\Bar{\bs X}}_1= n_1^{-1}\sum_{i=1}^n Z_i \bs X_i$, $\hat{\Bar{\bs X}}_0 = n_0^{-1}\sum_{i=1}^n (1-Z_i) \bs X_i$.
Then 
\[\begin{split}
\hat{\Bar{\bs W}}_1 & = n_1^{-1}\sum_{i=1}^n Z_i \bs W_i=\bs S_{\bx\bx}^{-1/2} (\hat{\Bar{\bs X}}_1-\bar{\bX}),\ 
\hat{\Bar{\bs W}}_0  = n_0^{-1}\sum_{i=1}^n (1-Z_i) \bs W_i= \bs S_{\bx\bx}^{-1/2} (\hat{\Bar{\bs X}}_0-\bar{\bX}).
\end{split}\]
Define the finite-sample counterpart of covariances as 
\[\begin{split}
s_1^2 &= \frac{r_0^2}{n_1-1}\sum_{i=1}^n Z_i (\one \{Y_i(1) \leq q_{1,\alpha}\}-\hat{F}_1(q_{1,\alpha}))^2 = \frac{n_1 r_0^2}{n_1-1}\big[\hat{F}_1(q_{1,\alpha})-\hat{F}_1^2(q_{1,\alpha})\big],\\
s_0^2 &= \frac{r_1^2}{n_0-1}\sum_{i=1}^n (1-Z_i) (\one \{Y_i(0) \leq q_{0,\alpha}\}-\hat{F}_0(q_{0,\alpha}))^2 = \frac{n_0 r_1^2}{n_0-1}\big[\hat{F}_0(q_{0,\alpha})-\hat{F}_0^2(q_{0,\alpha})\big],\\
\bs s_{1\bx} &
= \bs s_{\bx 1}^\top = \frac{r_0}{n_1-1}\sum_{i=1}^n Z_i (\one \{Y_i(1) \leq q_{1,\alpha}\}-\hat{F}_1(q_{1,\alpha}))(\bs X_i-\hat{\Bar{\bs X}}_1)^\top,\\
\bs s_{0\bx} & = \bs s_{\bx 0}^\top= \frac{-r_1}{n_0-1}\sum_{i=1}^n (1-Z_i) (\one \{Y_i(0) \leq q_{0,\alpha}\}-\hat{F}_0(q_{0,\alpha}))(\bs X_i-\hat{\Bar{\bs X}}_0)^\top.
\end{split}\]
Then 
\[\begin{split}
\bs s_{1w} &= \bs s_{w1}^\top = \frac{r_0}{n_1-1}\sum_{i=1}^n Z_i (\one \{Y_i(1) \leq q_{1,\alpha}\}-\hat{F}_1(q_{1,\alpha}))(\bs W_i-\hat{\Bar{\bs W}}_1)^\top\\
&= \frac{r_0}{n_1-1}\sum_{i=1}^n Z_i (\one \{Y_i(1) \leq q_{1,\alpha}\}-\hat{F}_1(q_{1,\alpha}))(\bs X_i-\hat{\Bar{\bs X}}_1)^\top\bs S_{\bx\bx}^{-1/2} = \bs s_{1\bx}\bS_{\bx\bx}^{-1/2},\\
\bs s_{0w} &= \bs s_{w0}^\top = \frac{-r_1}{n_0-1}\sum_{i=1}^n (1-Z_i) (\one \{Y_i(0) \leq q_{0,\alpha}\}-\hat{F}_0(q_{0,\alpha}))(\bs W_i-\hat{\Bar{\bs W}}_1)^\top\\
&= \frac{-r_1}{n_0-1}\sum_{i=1}^n (1-Z_i) (\one \{Y_i(0) \leq q_{0,\alpha}\}-\hat{F}_0(q_{0,\alpha}))(\bs X_i-\hat{\Bar{\bs X}}_0)^\top\bs S_{\bx\bx}^{-1/2} = \bs s_{0\bx}\bS_{\bx\bx}^{-1/2}.
\end{split}\]
Then for $z\in \{0,1\}$, 
\[\begin{split}
||\bs s_{1w}||^2_2
= \bs s_{1\bx}\bS_{\bx\bx}^{-1}\bs s_{\bx 1},\ 
||\bs s_{0w}||^2_2
= \bs s_{0 \bx}\bS_{\bx\bx}^{-1}\bs s_{\bx 0}.
\end{split}\]
The corresponding estimators are given by:
\[\begin{split}
\hat{s}_{1}^2 &= \frac{n_1 r_0^2}{n_1-1}\big[\hat{F}_1(\hat{q}_{1,\alpha})-\hat{F}_1^2(\hat{q}_{1,\alpha})\big],\ 
\hat{s}_{0}^2 = \frac{n_0 r_1^2}{n_0-1}\big[\hat{F}_0(\hat{q}_{0,\alpha})-\hat{F}_0^2(\hat{q}_{0,\alpha})\big],\\
\hat{\bs s}_{1\bx} & = \hat{\bs s}_{\bx 1}^\top = \frac{r_0}{n_1-1}\sum_{i=1}^n Z_i (\one \{Y_i(1) \leq \hat{q}_{1,\alpha}\}-\hat{F}_1(\hat{q}_{1,\alpha}))(\bs X_i-\hat{\Bar{\bs X}}_1)^\top,\\
\hat{\bs s}_{0\bx} &= \hat{\bs s}_{\bx 0}^\top = \frac{-r_1}{n_0-1}\sum_{i=1}^n (1-Z_i) (\one \{Y_i(0) \leq \hat{q}_{0,\alpha}\}-\hat{F}_0(\hat{q}_{0,\alpha}))(\bs X_i-\hat{\Bar{\bs X}}_0)^\top,\\
\hat{\bs s}_{zw} &= \hat{\bs s}_{zw}^\top = \hat{\bs s}_{z\bx}\bS_{\bx\bx}^{-1/2}\text{ for }z\in \{0,1\}.
\end{split}\]
Moreover,  
\[\begin{split}
||\hat{\bs s}_{1w}||_2^2 =\hat{\bs s}_{1\bx}\bS_{\bx\bx}^{-1}\hat{\bs s}_{\bx 1},\ 
||\hat{\bs s}_{0w}||_2^2 =\hat{\bs s}_{0\bx}\bS_{\bx\bx}^{-1}\hat{\bs s}_{\bx 0}.
\end{split}\]
By definition, the estimators are identical to those defined in Section \ref{sec:CI}, i.e., for $z\in \{0,1\}$,
\[\begin{split}
\hat{s}_{z}^2 = \hat{s}_{q_z q_z},\ 
\hat{\bs s}_{z\bx} =\hat{\bs s}_{q_z\bx} ,\
||\hat{\bs s}_{zw}||_2^2 = \hat{\bs s}_{q_{z}\mid \bx}^2,\ 
\hat{\bs s}_{1w}\hat{\bs s}_{0w}^\top = \hat{\bs s}_{q_1\bx}\bS_{\bx\bx}^{-1}\hat{\bs s}_{\bx q_0}.
\end{split}\]
Then $\Tilde{C}_n$ is estimated with 
\[\begin{split}
\hat{C}_n &= \frac{1}{r_1r_0}\left[\frac{\hat{s}_{1}^2}{\hat{f}_{1,n}^2(\hat{q}_{1,\alpha})}+\frac{\hat{s}_{0}^2}{\hat{f}_{0,n}^2(\hat{q}_{0,\alpha})}+\frac{2\min \left\{\frac{r_1}{r_0}\hat{s}_1^2,\frac{r_0}{r_1}\hat{s}_0^2\right\}}{\hat{f}_{1,n}(\hat{q}_{1,\alpha})\hat{f}_{0,n}(\hat{q}_{0,\alpha})}\right],
\end{split}\]
$B_n$ is estimated with 
\[\begin{split}
\hat{B}_n &= \frac{1}{r_1r_0}\left[\frac{||\hat{\bs s}_{1w}||_2^2}{\hat{f}_{1,n}^2(\hat{q}_{1,\alpha})}+\frac{||\hat{\bs s}_{0w}||_2^2}{\hat{f}_{0,n}^2(\hat{q}_{0,\alpha})}-\frac{2\cdot \hat{\bs s}_{1w}\hat{\bs s}_{0w}^\top}{\hat{f}_{1,n}(\hat{q}_{1,\alpha})\hat{f}_{0,n}(\hat{q}_{0,\alpha})}\right],
\end{split}\]
and $\tilde{A}_n:=\tilde{C}_n-B_n$ is estimated with $\hat{A}_n = \hat{C}_n-\hat{B}_n$.

\hspace{\fill}\\
 \textbf{Proof of Theorem \ref{thm:CI} (i):}
The proof can be split into the following two steps.
\hspace{\fill}\\
\textbf{Step 1.}
Let 
\[\begin{split}
\hat{C}_n' &= \frac{1}{r_1r_0}\left[\frac{{s}_{1}^2}{f_{1,n}^2(q_{1,\alpha})}+\frac{{s}_{0}^2}{f_{0,n}^2(q_{0,\alpha})}+\frac{2\min \left\{\frac{r_1}{r_0}{s}_1^2,\frac{r_0}{r_1}{s}_0^2\right\}}{f_{1,n}(q_{1,\alpha})f_{0,n}(q_{0,\alpha})}\right].
\end{split}\]
and 
\[\begin{split}
\hat{B}_n' = \frac{1}{r_1r_0}\left[\frac{||{\bs s}_{1w}||_2^2}{f_{1,n}^2(q_{1,\alpha})}+\frac{||{\bs s}_{0w}||_2^2}{f_{0,n}^2(q_{0,\alpha})}-\frac{2\cdot \bs s_{1w}\bs s_{0w}^\top}{f_{1,n}(q_{1,\alpha})f_{0,n}(q_{0,\alpha})}\right],
\end{split}\]
we show that for $\hat{A}_n' = \hat{C}_n'-\hat{B}_n'$,
\[\begin{split}
\max\{|\hat{A}_n'-\tilde{A}_n|,|\hat{B}_n'-B_n|\} = o_p(\tilde{A}_n).
\end{split}\]
\textbf{Step 2.} Show that
\[\begin{split}
\max\{|\hat{A}_n'-\hat{A}_n|,|\hat{B}_n'-\hat{B}_n|\} = o_p(\tilde{A}_n).
\end{split}\]
\begin{proof}[Proof of step 1.]
For $z\in \{0,1\}$,
let 
\[\begin{split}
v_{i,n}(z) &= (-1)^{1-z}r_{1-z} \one\{Y_i(z) \leq q_{z,\alpha}\},\ 
\Bar{v}_n(z) = n^{-1}\sum_{i=1}^n v_{i,n}(z),\
\psi = \max_{z=0,1} \max_{1\leq i\leq n} \{v_{i,n}(z) - \Bar{v}_n(z)\}^2.
\end{split}\]
Since 
\[\begin{split}
\sum_{i=1}^n \{v_{i,n}(1) - \Bar{v}_n(1)\}^2 &= n r_0^2 \cdot F_1(q_{1,\alpha})(1-F_1(q_{1,\alpha})),\\  
\sum_{i=1}^n \{v_{i,n}(0) - \Bar{v}_n(0)\}^2 &= n r_1^2 \cdot F_0(q_{0,\alpha})(1-F_0(q_{0,\alpha})),
\end{split}\]
we have 
\[\begin{split}
\max\big\{r_0^2 \cdot F_1(q_{1,\alpha})(1-F_1(q_{1,\alpha})),r_1^2 \cdot F_0(q_{0,\alpha})(1-F_0(q_{0,\alpha})\big\} \leq \psi \leq \max\{r_0^2,r_1^2\}.
\end{split}\]
Assumption \ref{assumption:replicate} implies that $F_z(q_{z,\alpha}) = \alpha + o(1)$, and thus, for any fixed $\alpha \in (0,1)$,  
\begin{equation}\label{eq:psi}
\max\{\psi,1/\psi\} = O(1).
\end{equation}
By definition,
\[\begin{split}
S_1^2 = \frac{nr_0^2}{n-1}[F_1(q_{1,\alpha})-F_1^2(q_{1,\alpha})] \leq 2\psi,\
S_0^2 = \frac{nr_1^2}{n-1}[F_0(q_{0,\alpha})-F_0^2(q_{0,\alpha})] \leq 2\psi.
\end{split}\]
Hence, $ ||\bs S_{1w}||_2^2 \leq S_1^2 \leq 2 \psi$, $||\bs S_{0w}||_2^2 \leq S_0^2 \leq 2 \psi$
and $|\bs S_{1w}\bs S_{0w}^\top|\leq ||\bs S_{1w}||_2||\bs S_{0w}||_2 \leq 2\psi$.
Moreover, as Assumption \ref{assumption:derivativeF} implies that $f_{z,n}(q_{z,\alpha})= O(1)$ and $1/f_{z,n}(q_{z,\alpha}) = O(1)$, we have
\[\begin{split}
\Tilde{A}_n &= \Tilde{C}_n-B_n
= \frac{1}{r_1r_0}\left[\frac{S_{1}^2-||\bS_{1w}||_2^2}{f_{1}^2(q_{1,\alpha})}+\frac{S_{0}^2-||\bS_{0w}||_2^2}{f_{0}^2(q_{0,\alpha})}+\frac{2\left(\min \left\{\frac{r_1}{r_0}S_{1}^2,\frac{r_0}{r_1}S_{0}^2\right\} + \bs S_{1w}\bs S_{0w}^\top\right)}{f_{1}(q_{1,\alpha})f_{0}(q_{0,\alpha})}\right]\\
&= O(\psi) = O(1).
\end{split}\]
By definition, 
$A_n = n\cdot \tilde{V}_{\bq\bq}(1-\tilde{R}_{\bq}^2)$.
According to Assumption \ref{assumption:R2_2}, we have $\liminf_n A_n > 0$.
Since $\tilde{A}_n \geq A_n$, we have 
$\liminf_n \tilde{A}_n \geq \liminf_n A_n > 0$. Since $\tilde{A}_n = O(1)$, then there exists some $c>0$ and $N_c > 0$ s.t. for $n > N_c$, we have $0 < \tilde{A}_n \leq 1/c$,
i.e., 
$1/\Tilde{A}_n \geq c$.
Since $K_n \geq 1$, Assumption \ref{assumption:delta_p} implies that for $|\Tilde{p}_n - p_n|\leq \Delta_n$, we have $-\log \Tilde{p}_n = -\log p_n + o(1)$, and thus, Assumption \ref{assumption:step1} implies that 
\[\begin{split}
\frac{K_n}{\Tilde{A}_n} \cdot\sqrt{\frac{\max\{1,\log K_n, -\log \Tilde{p}_n\}}{n}} \rightarrow 0.
\end{split}\]
Let $b_n = \max\{1,-\log \tilde{p}_n\}$ and $c_n = \max\{1,\log K_n, -\log \tilde{p}_n\}$, then $b_n/n \leq c_n/n \rightarrow 0$.
According to the proof of Lemmas~A32 and~A33 in \cite{wang2022rerandomization}, under ReM, for $z\in \{0,1\}$, 
\[\begin{split}
&\big|s_z^2-S_z^2\big| = O_p(\psi \sqrt{b_n/n}),\ ||\bs s_{zw}-\bs S_{zw}||_2 = O_p(\sqrt{\psi K_n \cdot c_n/n}),\\
&||\bs S_{zw}||_2^2 = O(\psi K_n),\ 
\big|||\bs s_{zw}||_2^2-||\bs S_{zw}||_2^2\big| = O_p(\psi K_n \sqrt{c_n/n}).
\end{split}\]
Then  
\[\begin{split}
\left|\bs s_{1w}\bs s_{0w}^\top - \bs S_{1w}\bs S_{0w}^\top\right|&=
\left|\bs s_{1w}\bs s_{0w}^\top-\bs s_{1w}\bs S_{0w}^\top+\bs s_{1w}\bs S_{0w}^\top - \bs S_{1w}\bs S_{0w}^\top\right|\\
&\leq ||\bs s_{1w}||_2||\bs s_{0w}-\bs S_{0w}||_2 + ||\bs S_{0w}||_2||\bs s_{1w}-\bs S_{1w}||_2\\
&\leq ||\bs S_{1w}||_2||\bs s_{0w}-\bs S_{0w}||_2+||\bs s_{1w}-\bs S_{1w}||_2||\bs s_{0w}-\bs S_{0w}||_2 + ||\bs S_{0w}||_2||\bs s_{1w}-\bs S_{1w}||_2\\
&= O_p(\psi K_n \sqrt{c_n/n}).
\end{split}\]
Hence, under Assumption \ref{assumption:derivativeF}, 
\[\begin{split}
|\hat{C}_n' - \Tilde{C}_n | &\leq \frac{1}{r_1r_0}\left\{ \left|\frac{s_1^2-S_1^2}{f_{1,n}^2(q_{1,\alpha})} \right|
+\left|\frac{s_0^2-S_0^2}{f_{0,n}^2(q_{0,\alpha})} \right|+2\left|
\frac{\frac{r_1}{r_0}|s_1^2-S_1^2|+\frac{r_0}{r_1}|s_0^2-S_0^2|}{f_{1,n}(q_{1,\alpha})f_{0,n}(q_{0,\alpha})}\right|
\right\} = O_p(\psi \sqrt{b_n/n}).
\end{split}\]
and 
\[\begin{split}
|\hat{B}_n' - B_n | &\leq \frac{1}{r_1r_0}\cdot \frac{1}{f_{1,n}^2(q_{1,\alpha})}\cdot\left|||{\bs s}_{1w}||_2^2-||{\bs S}_{1w}||_2^2\right|+\frac{1}{r_1r_0}\cdot \frac{1}{f_{0,n}^2(q_{0,\alpha})}\cdot\left|||{\bs s}_{0w}||_2^2-||{\bs S}_{0w}||_2^2\right|\\& + \frac{1}{r_1r_0}\cdot\frac{2}{f_{1,n}(q_{1,\alpha})f_{0,n}(q_{0,\alpha})} \cdot \left|\bs s_{1w}\bs s_{0w}^\top - \bs S_{1w}\bs S_{0w}^\top\right|\\
&= O_p(\psi K_n \sqrt{c_n/n}).
\end{split}\]
Under Assumption \ref{assumption:step1},
since $b_n \leq c_n$ and $K_n \geq 1$,
we have 
\[\begin{split}
\max\{|\hat{C}_n' - \Tilde{C}_n | ,|\hat{B}_n' - B_n|\} &= O_p(\psi K_n \sqrt{c_n/n}) = \Tilde{A}_n \cdot O_p\left(\frac{\psi}{\Tilde{A}_n} K_n \sqrt{\frac{c_n}{n}}\right)\\
&= \Tilde{A}_n \cdot O_p\left(\frac{K_n}{\Tilde{A}_n} \sqrt{\frac{c_n}{n}}\right)
= \Tilde{A}_n \cdot o_p(1).
\end{split}\]
Hence, 
\[\begin{split}
\max\{|\hat{A}_n' - \Tilde{A}_n | ,|\hat{B}_n' - B_n|\} = \Tilde{A}_n \cdot o_p(1).
\end{split}\]

\end{proof}
\begin{proof}[Proof of step 2.]

Let 
\[\begin{split}
\hat{v}_{i,n}(z) = (-1)^{1-z}r_{1-z} \one\{Y_i(z) \leq \hat{q}_{z,\alpha}\},\ \hat{\psi} = \max_{z=0,1} \max_{1\leq i\leq n} \{\hat{v}_{i,n}(z) - \Bar{\hat{v}}_n(z)\}^2.
\end{split}\]
Then 
\[\begin{split}
\max\big\{r_0^2 \cdot F_1(\hat{q}_{1,\alpha})(1-F_1(\hat{q}_{1,\alpha})),r_1^2 \cdot F_0(\hat{q}_{0,\alpha})(1-F_0(\hat{q}_{0,\alpha})\big\} &\leq \hat{\psi}  \leq \max\{r_0^2,r_1^2\}.
\end{split}\]
By definition, for $z \in \{0,1\}$,
\[\begin{split}
\left|F_z(\hat{q}_{z,\alpha}) - F_z(q_{z,\alpha})\right| &= \left|\frac{1}{n}\sum_{i=1}^n \left(\one\{Y_i(z) \leq \hat{q}_{z,\alpha}\}-\one\{Y_i(z) \leq q_{z,\alpha}\}\right)\right|\\
&\leq \frac{1}{n}\sum_{i=1}^n \one\{q_{z,\alpha} < Y_i(z) \leq \hat{q}_{z,\alpha}\}+\frac{1}{n}\sum_{i=1}^n \one\{\hat{q}_{z,\alpha} < Y_i(z) \leq q_{z,\alpha}\}.
\end{split}\]
Theorem \ref{thm:qz} shows that under ReM,
\[\begin{split}
\sqrt{n}|\hat{q}_{z,\alpha}-q_{z,\alpha}| = O_p(1),
\end{split}\]
then for any $\epsilon > 0$, there exists constants $M_\epsilon > 0$, $N_\epsilon > 0$ and $c_\epsilon > 0$ s.t. for any $n > N_\epsilon$, we have $M_\epsilon \in [-2n^\gamma,2n^\gamma]$, $M_\epsilon \cdot f_{z,n}(q_{z,\alpha}) + r_{z,n}(q_{z,\alpha}) \leq c_\epsilon < \infty$ and 
\[\begin{split}
\pr \left(\sqrt{n}|\hat{q}_{z,\alpha}-q_{z,\alpha}| \leq M_\epsilon \mid M \leq a_n\right) \geq 1-\epsilon.
\end{split}\]
Hence,
\[\begin{split}
& \pr\left(\sum_{i=1}^n \one\{q_{z,\alpha} < Y_i(z) \leq \hat{q}_{z,\alpha}\}  \leq \sum_{i=1}^n \one\{ q_{z,\alpha} < Y_i(z) \leq q_{z,\alpha} + M_\epsilon/\sqrt{n}\}\mid M \leq a_n\right)\\
&\geq \pr \left(\sqrt{n}|\hat{q}_{z,\alpha}-q_{z,\alpha}| \leq M_\epsilon\mid M \leq a_n\right)  \geq 1-\epsilon,
\end{split}\]
and 
\[\begin{split}
& \pr\left(\sum_{i=1}^n \one\{\hat{q}_{z,\alpha} < Y_i(z) \leq q_{z,\alpha}\}  \leq \sum_{i=1}^n \one\{ q_{z,\alpha} - M_\epsilon/\sqrt{n} < Y_i(z) \leq q_{z,\alpha}\}\mid M \leq a_n\right)\\
&\geq \pr \left(\sqrt{n}|\hat{q}_{z,\alpha}-q_{z,\alpha}| \leq M_\epsilon\mid M \leq a_n\right)  \geq 1-\epsilon.
\end{split}\]
By Assumption \ref{assumption:derivativeF}, for $n > N_\epsilon$,
\[\begin{split}
n^{-1/2}\sum_{i=1}^n \one\{ q_{z,\alpha} < Y_i(z) \leq q_{z,\alpha} + M_\epsilon/\sqrt{n}\} \leq M_\epsilon \cdot f_{z,n}(q_{z,\alpha}) + r_{z,n}(q_{z,\alpha}) \leq c_\epsilon,
\end{split}\]
and 
\[\begin{split}
n^{-1/2}\sum_{i=1}^n \one\{ q_{z,\alpha} - M_\epsilon/\sqrt{n} < Y_i(z) \leq q_{z,\alpha}\} \leq M_\epsilon \cdot f_{z,n}(q_{z,\alpha}) + r_{z,n}(q_{z,\alpha}) \leq c_\epsilon.
\end{split}\]
Hence, for $n > N_\epsilon$,
\[\begin{split}
&\pr\left(n^{-1/2}\sum_{i=1}^n \one\{ q_{z,\alpha} < Y_i(z) \leq \hat{q}_{z,\alpha}\} \leq c_\epsilon\mid M \leq a_n\right) \geq 1-\epsilon,\\
&\pr\left(n^{-1/2}\sum_{i=1}^n \one\{ \hat{q}_{z,\alpha} < Y_i(z) \leq q_{z,\alpha}\} \leq c_\epsilon\mid M \leq a_n\right) \geq 1-\epsilon,
\end{split}\]
i.e., under ReM, 
\begin{equation}
\label{eq:q-qhat} 
\frac{1}{n}\sum_{i=1}^n \one\{q_{z,\alpha} < Y_i(z) \leq \hat{q}_{z,\alpha}\} = O_p(n^{-1/2}),\ 
\frac{1}{n}\sum_{i=1}^n \one\{\hat{q}_{z,\alpha} < Y_i(z) \leq q_{z,\alpha}\} = O_p(n^{-1/2}),
\end{equation}
and thus, 
\[\begin{split}
F_z(\hat{q}_{z,\alpha}) - F_z(q_{z,\alpha}) = O_p(n^{-1/2}).
\end{split}\]
Under Assumption \ref{assumption:replicate}, we have $F_z(q_{z,\alpha}) = \alpha + o(1)$, and thus, $F_z(\hat{q}_{z,\alpha}) = \alpha + o_p(1)$, and thus,
\begin{equation}\label{eq:psihat}
\max\big\{\hat{\psi},1/\hat{\psi}\big\} = O_p(1).
\end{equation}
Let 
\[\begin{split}
\hat{\bs S}_{zw} = \frac{(-1)^{1-z}r_{1-z}}{n-1}\sum_{i=1}^n (\one \{Y_i(z) \leq \hat{q}_{z,\alpha}\}-F_z(\hat{q}_{z,\alpha}))(\bs W_i-\Bar{\bs W})^\top = (e_1,\ldots,e_{K_n}),
\end{split}\]
where for $k=1,\ldots,K_n$,
$$e_k = \frac{(-1)^{1-z}r_{1-z}}{n-1}\sum_{i=1}^n (\one \{Y_i(z) \leq \hat{q}_{z,\alpha}\}-F_z(\hat{q}_{z,\alpha}))(W_{ik}-\Bar{W}_k),$$
and $||\hat{\bs S}_{zw}||_2^2 = \sum_{k=1}^{K_n} e_k^2$.
Since 
\[\begin{split}
e_k^2 &\leq \frac{r_{1-z}^2}{(n-1)^2} \cdot \sum_{i=1}^n \left(\one\{Y_i(z) \leq \hat{q}_{z,\alpha}\} - F_z(\hat{q}_{z,\alpha})\right)^2 \cdot \sum_{i=1}^n (W_{ik}-\Bar{W}_k)^2\\
&\leq \frac{n}{n-1}\cdot \hat{\psi} = O_p(1),
\end{split}\]
then 
\[\begin{split}
||\hat{\bs S}_{zw}||_2^2 = O_p(K_n).
\end{split}\]
By definition,
\[\begin{split}
\bs S_{zw} &= \frac{(-1)^{1-z}r_{1-z}}{n-1}\sum_{i=1}^n \left(\one\{Y_i(z) \leq q_{z,\alpha}\} - F_z(q_{z,\alpha})\right)(\bs W_i-\Bar{\bs W})^\top,
\end{split}\]
then 
\[\begin{split}
\left|||\hat{\bs S}_{zw}||_2^2 - ||\bs S_{zw}||_2^2 \right| &= \left|(\hat{\bs S}_{zw}-\bs S_{zw})(\hat{\bs S}_{zw}+\bs S_{zw})^\top \right| \leq ||\hat{\bs S}_{zw}-\bs S_{zw}||_2 \left(||\hat{\bs S}_{zw}||_2 + ||\bs S_{zw}||_2\right)\\
&= ||\hat{\bs S}_{zw}-\bs S_{zw}||_2 \cdot O_p(\sqrt{K_n}).
\end{split}\]
Since
\[\begin{split}
\hat{\bs S}_{zw}-\bs S_{zw} =& \frac{(-1)^{1-z}r_{1-z}}{n-1}\sum_{i=1}^n \left\{\one \{q_{z,\alpha} < Y_i(z) \leq \hat{q}_{z,\alpha}\} - \frac{1}{n}\sum_{i=1}^n \one \{q_{z,\alpha} < Y_i(z) \leq \hat{q}_{z,\alpha}\}\right\}(\bs W_i-\Bar{\bs W})^\top\\
&- \frac{(-1)^{1-z}r_{1-z}}{n-1}\sum_{i=1}^n \left\{\one \{\hat{q}_{z,\alpha} < Y_i(z) \leq q_{z,\alpha}\} - \frac{1}{n}\sum_{i=1}^n \one \{\hat{q}_{z,\alpha} < Y_i(z) \leq q_{z,\alpha}\}\right\}(\bs W_i-\Bar{\bs W})^\top\\
=& (a_1-b_1,\ldots,a_{K_n}-b_{K_n}),
\end{split}\]
where for $k=1,\ldots,K_n$,
\[\begin{split}
a_k &= \frac{(-1)^{1-z}r_{1-z}}{n-1}\sum_{i=1}^n \left\{\one \{q_{z,\alpha} < Y_i(z) \leq \hat{q}_{z,\alpha}\} - \frac{1}{n}\sum_{i=1}^n \one \{q_{z,\alpha} < Y_i(z) \leq \hat{q}_{z,\alpha}\}\right\}(W_{ik}-\Bar{W}_k),\\
b_k &= \frac{(-1)^{1-z}r_{1-z}}{n-1}\sum_{i=1}^n \left\{\one \{\hat{q}_{z,\alpha} < Y_i(z) \leq q_{z,\alpha}\} - \frac{1}{n}\sum_{i=1}^n \one \{\hat{q}_{z,\alpha} < Y_i(z) \leq q_{z,\alpha}\}\right\}(W_{ik}-\Bar{W}_k),
\end{split}\]
and $||\hat{\bs S}_{zw}-\bs S_{zw}||_2^2 = \sum_{k=1}^{K_n} (a_k - b_k)^2\leq 2\sum_{k=1}^{K_n}a_k^2 + 2\sum_{k=1}^{K_n} b_k^2$. Cauchy-Schwarz inequality implies that 
\[\begin{split}
a_k^2 &\leq \frac{r_{1-z}^2}{(n-1)^2}\cdot \sum_{i=1}^n \left\{\one \{q_{z,\alpha} < Y_i(z) \leq \hat{q}_{z,\alpha}\} - \frac{1}{n}\sum_{i=1}^n \one \{q_{z,\alpha} < Y_i(z) \leq \hat{q}_{z,\alpha}\}\right\}^2 \cdot \sum_{i=1}^n (W_{ik}-\Bar{W}_k)^2\\
&= \frac{r_{1-z}^2}{n-1}\cdot \sum_{i=1}^n \left\{\one \{q_{z,\alpha} < Y_i(z) \leq \hat{q}_{z,\alpha}\} - \frac{1}{n}\sum_{i=1}^n \one \{q_{z,\alpha} < Y_i(z) \leq \hat{q}_{z,\alpha}\}\right\}^2\\
&=\frac{r_{1-z}^2\cdot n}{n-1}\cdot 
\frac{1}{n}\sum_{i=1}^n \one \{q_{z,\alpha} < Y_i(z) \leq \hat{q}_{z,\alpha}\} \cdot \left(1-\frac{1}{n}\sum_{i=1}^n \one \{q_{z,\alpha} < Y_i(z) \leq \hat{q}_{z,\alpha}\}\right),
\end{split}\]
and 
\[\begin{split}
b_k^2 &\leq \frac{r_{1-z}^2\cdot n}{n-1}\cdot 
\frac{1}{n}\sum_{i=1}^n \one \{\hat{q}_{z,\alpha} < Y_i(z) \leq q_{z,\alpha}\} \cdot \left(1-\frac{1}{n}\sum_{i=1}^n \one \{\hat{q}_{z,\alpha} < Y_i(z) \leq q_{z,\alpha}\}\right).
\end{split}\]
By Eq. \eqref{eq:q-qhat},
$a_k^2 = O_p(n^{- 1/2})$, and $b_k^2 = O_p(n^{- 1/2})$ under ReM.
Hence, 
\[\begin{split}
||\hat{\bs S}_{zw}-\bs S_{zw}||_2^2 \leq 2\sum_{k=1}^{K_n}a_k^2 + 2\sum_{k=1}^{K_n} b_k^2 = O_p(K_n\cdot n^{- 1/2}),\ 
\left|||\hat{\bs S}_{zw}||_2^2 - ||\bs S_{zw}||_2^2 \right| = O_p(K_n\cdot n^{-1/4}).
\end{split}\]
By definition,
\[\begin{split}
||\hat{\bs s}_{zw} - \hat{\bs S}_{zw}||_2^2 &= ||\hat{\bs s}_{zw}-{\bs s}_{zw}+{\bs s}_{zw}-{\bs S}_{zw}+{\bs S}_{zw} - \hat{\bs S}_{zw}||_2^2\\
&\leq 3\left[||\hat{\bs s}_{zw}-{\bs s}_{zw}||_2^2 +||{\bs s}_{zw}-{\bs S}_{zw}||_2^2+||{\bs S}_{zw} - \hat{\bs S}_{zw}||_2^2\right]\\
&= 3||\hat{\bs s}_{zw}-{\bs s}_{zw}||_2^2 + O_p(\psi K_n \cdot c_n/n) + O_p(K_n\cdot n^{-1/2}),
\end{split}\]
where 
\[\begin{split}
\hat{\bs s}_{zw}-{\bs s}_{zw} =& \frac{(-1)^{1-z}r_{1-z}}{n_z-1}\sum_{i:Z_i=z} \left\{\one \{q_{z,\alpha} < Y_i(z) \leq \hat{q}_{z,\alpha}\} - \frac{1}{n_z}\sum_{i:Z_i=z} \one \{q_{z,\alpha} < Y_i(z) \leq \hat{q}_{z,\alpha}\}\right\}(\bs W_i-\hat{\Bar{\bs W}}_z)^\top\\
&-\frac{(-1)^{1-z}r_{1-z}}{n_z-1}\sum_{i:Z_i=z} \left\{\one \{\hat{q}_{z,\alpha} < Y_i(z) \leq q_{z,\alpha}\} - \frac{1}{n_z}\sum_{i:Z_i=z} \one \{\hat{q}_{z,\alpha} < Y_i(z) \leq q_{z,\alpha}\}\right\}(\bs W_i-\hat{\Bar{\bs W}}_z)^\top\\
=& (a_1'-b_1',\ldots,a_{K_n}'-b_{K_n}'),
\end{split}\]
where for $k=1,\ldots,K_n$,
\[\begin{split}
a_k' &= \frac{(-1)^{1-z}r_{1-z}}{n_z-1}\sum_{i:Z_i=z} \left\{\one \{q_{z,\alpha} < Y_i(z) \leq \hat{q}_{z,\alpha}\} - \frac{1}{n_z}\sum_{i:Z_i=z} \one \{q_{z,\alpha} < Y_i(z) \leq \hat{q}_{z,\alpha}\}\right\}(W_{ik}-\hat{\Bar{W}}_{k,z})\\
&= \frac{(-1)^{1-z}r_{1-z}}{n_z-1}\sum_{i:Z_i=z} \left\{\one \{q_{z,\alpha} < Y_i(z) \leq \hat{q}_{z,\alpha}\} - \frac{1}{n_z}\sum_{i:Z_i=z} \one \{q_{z,\alpha} < Y_i(z) \leq \hat{q}_{z,\alpha}\}\right\}(W_{ik}-\Bar{W}_{k}),
\end{split}\]
and 
\[\begin{split}
b_k' &= \frac{(-1)^{1-z}r_{1-z}}{n_z-1}\sum_{i:Z_i=z} \left\{\one \{\hat{q}_{z,\alpha} < Y_i(z) \leq q_{z,\alpha}\} - \frac{1}{n_z}\sum_{i:Z_i=z} \one \{\hat{q}_{z,\alpha} < Y_i(z) \leq q_{z,\alpha}\}\right\}(W_{ik}-\hat{\Bar{W}}_{k,z})\\
&= \frac{(-1)^{1-z}r_{1-z}}{n_z-1}\sum_{i:Z_i=z} \left\{\one \{\hat{q}_{z,\alpha} < Y_i(z) \leq q_{z,\alpha}\} - \frac{1}{n_z}\sum_{i:Z_i=z} \one \{\hat{q}_{z,\alpha} < Y_i(z) \leq q_{z,\alpha}\}\right\}(W_{ik}-\Bar{W}_{k}),
\end{split}\]
and $||\hat{\bs s}_{zw}-\bs s_{zw}||_2^2 = \sum_{k=1}^{K_n} (a_k'-b_k')^2$. Cauchy-Schwarz inequality implies that 
\[\begin{split}
a_k'^2 &\leq \frac{r_{1-z}^2}{(n_z-1)^2}\cdot \sum_{i:Z_i=z} \left\{\one \{q_{z,\alpha} < Y_i(z) \leq \hat{q}_{z,\alpha}\} - \frac{1}{n_z}\sum_{i:Z_i=z} \one \{q_{z,\alpha} < Y_i(z) \leq \hat{q}_{z,\alpha}\}\right\}^2 \cdot \sum_{i:Z_i=z} (W_{ik}-\Bar{W}_{k})^2\\
&\leq \frac{r_{1-z}^2}{(n_z-1)^2}\cdot \sum_{i:Z_i=z} \left\{\one \{q_{z,\alpha} < Y_i(z) \leq \hat{q}_{z,\alpha}\} - \frac{1}{n_z}\sum_{i:Z_i=z} \one \{q_{z,\alpha} < Y_i(z) \leq \hat{q}_{z,\alpha}\}\right\}^2 \cdot \sum_{i=1}^n (W_{ik}-\Bar{W}_{k})^2
\\
&= \frac{r_{1-z}^2\cdot (n-1)}{(n_z-1)^2}\cdot \sum_{i;Z_i=z} \left\{\one \{q_{z,\alpha} < Y_i(z) \leq \hat{q}_{z,\alpha}\} - \frac{1}{n_z}\sum_{i:Z_i=z} \one \{q_{z,\alpha} < Y_i(z) \leq \hat{q}_{z,\alpha}\}\right\}^2\\
&=\frac{r_{1-z}^2\cdot (n-1)n_z}{(n_z-1)^2}\cdot \frac{1}{n_z}\sum_{i:Z_i=z} \one \{q_{z,\alpha} < Y_i(z) \leq \hat{q}_{z,\alpha}\}\left(1-\frac{1}{n_z}\sum_{i:Z_i=z} \one \{q_{z,\alpha} < Y_i(z) \leq \hat{q}_{z,\alpha}\}\right),
\end{split}\]
and 
\[\begin{split}
b_k'^2 &\leq \frac{r_{1-z}^2\cdot (n-1)n_z}{(n_z-1)^2}\cdot \frac{1}{n_z}\sum_{i:Z_i=z} \one \{\hat{q}_{z,\alpha} < Y_i(z) \leq q_{z,\alpha}\}\left(1-\frac{1}{n_z}\sum_{i:Z_i=z} \one \{\hat{q}_{z,\alpha} < Y_i(z) \leq q_{z,\alpha}\}\right).
\end{split}\]
By Eq. \eqref{eq:q-qhat}, under ReM,
\begin{align}
\label{eq:Fhat1}
&\frac{1}{n_z}\sum_{i:Z_i=z} \one \{q_{z,\alpha} < Y_i(z) \leq \hat{q}_{z,\alpha}\} \leq \frac{n}{n_z} \cdot \frac{1}{n}\sum_{i=1}^n \one \{q_{z,\alpha} < Y_i(z) \leq \hat{q}_{z,\alpha}\} = O_p(n^{- 1/2}),\\
\label{eq:Fhat2}
&\frac{1}{n_z}\sum_{i:Z_i=z} \one \{\hat{q}_{z,\alpha} < Y_i(z) \leq q_{z,\alpha}\} \leq \frac{n}{n_z} \cdot \frac{1}{n}\sum_{i=1}^n \one \{\hat{q}_{z,\alpha} < Y_i(z) \leq q_{z,\alpha}\} = O_p(n^{- 1/2}).
\end{align}
Hence, $a_k'^2 = O_p(n^{- 1/2})$, $b_k'^2 = O_p(n^{- 1/2})$, and 
\[\begin{split}
||\hat{\bs s}_{zw}-{\bs s}_{zw}||_2^2 \leq 2\sum_{k=1}^{K_n} a_k'^2 + 2\sum_{k=1}^{K_n} b_k'^2 = O_p(K_n\cdot n^{-1/2}),
\end{split}\]
and thus,
\[\begin{split}
||\hat{\bs s}_{zw} - \hat{\bs S}_{zw}||_2^2 
&= O_p(K_n \cdot c_n/n) + O_p(K_n\cdot n^{-1/2}) = O_p\left(K_n \cdot\max\{c_n/n,n^{-1/2}\}\right).
\end{split}\]
Then we have
\[\begin{split}
\left|||\hat{\bs s}_{zw}||_2^2 - ||\hat{\bs S}_{zw}||_2^2 \right| &= \left|(\hat{\bs s}_{zw} - \hat{\bs S}_{zw})(\hat{\bs s}_{zw} - \hat{\bs S}_{zw} + 2\hat{\bs S}_{zw})^\top\right|\\
&\leq ||\hat{\bs s}_{zw} - \hat{\bs S}_{zw}||_2^2 + 2||\hat{\bs s}_{zw} - \hat{\bs S}_{zw}||_2||\hat{\bs S}_{zw}||_2\\
&= O_p\left(K_n \cdot \max\{\sqrt{c_n/n},n^{-1/4}\}\right).
\end{split}\]
Moreover,
\[\begin{split}
\left|||\hat{\bs s}_{zw}||_2^2 - ||\bs s_{zw}||_2^2 \right| &\leq 
\left|||\hat{\bs s}_{zw}||_2^2 - ||\hat{\bs S}_{zw}||_2^2 \right|+
\left|||\hat{\bs S}_{zw}||_2^2 - ||\bs S_{zw}||_2^2 \right|+
\left|||\bs s_{zw}||_2^2 - ||\bs S_{zw}||_2^2 \right|\\
&= O_p\left(K_n \cdot \max\{\sqrt{c_n/n},n^{-1/4}\}\right).
\end{split}\]
Under Assumption \ref{assumption:step1},  
\begin{equation}\label{eq:shat}
\left|||\hat{\bs s}_{zw}||_2^2 - ||\bs s_{zw}||_2^2 \right| = \Tilde{A}_n\cdot o_p(1).
\end{equation}
Furthermore, by triangle inequality,
\[\begin{split}
\left|\hat{\bs s}_{1w}\hat{\bs s}_{0w}^\top - \bs s_{1w}\bs s_{0w}^\top \right| &\leq \left|\hat{\bs s}_{1w}\hat{\bs s}_{0w}^\top-\hat{\bs s}_{1w}\hat{\bs S}_{0w}^\top+\hat{\bs s}_{1w}\hat{\bs S}_{0w}^\top-\hat{\bs S}_{1w}\hat{\bs S}_{0w}^\top\right|\\
&+ \left|\hat{\bs S}_{1w}\hat{\bs S}_{0w}^\top-\hat{\bs S}_{1w}\bs S_{0w}^\top+\hat{\bs S}_{1w}\bs S_{0w}^\top - \bs S_{1w}\bs S_{0w}^\top \right|\\ 
&+\left|\bs S_{1w}\bs S_{0w}^\top-{\bs S}_{1w}\bs s_{0w}^\top+{\bs S}_{1w}\bs s_{0w}^\top - \bs s_{1w}\bs s_{0w}^\top \right|\\ &\leq ||\hat{\bs s}_{1w}||_2 ||\hat{\bs s}_{0w}-\hat{\bs S}_{0w}||_2 + ||\hat{\bs S}_{0w}||_2||\hat{\bs s}_{1w}-\hat{\bs S}_{1w}||_2\\
&+ ||\hat{\bs S}_{1w}||_2||\hat{\bs S}_{0w}-\bs S_{0w}||_2 + ||{\bs S}_{0w}||_2||\hat{\bs S}_{1w}-\bs S_{1w}||_2\\
&+ ||{\bs S}_{1w}||_2||{\bs s}_{0w}-\bs S_{0w}||_2 + ||{\bs s}_{0w}||_2||{\bs s}_{1w}-\bs S_{1w}||_2.
\end{split}\]
Since 
\[\begin{split}
&||\hat{\bs S}_{zw}-\bs S_{zw}||_2^2 = O_p(K_n\cdot n^{-1/2}),\ 
||{\bs s}_{zw}-\bs S_{zw}||_2^2 = O_p( K_n\cdot c_n/n),\\
&||\hat{\bs s}_{zw}-\hat{\bs S}_{zw}||_2^2 = O_p\left(K_n \cdot\max\{c_n/n,n^{-1/2}\}\right),\ 
||\bs S_{zw}||_2^2 = O(K_n),\ 
||\hat{\bs S}_{zw}||_2^2 = O_p(K_n),
\end{split}\]
we have 
\[\begin{split}
||\bs s_{zw}||_2^2 \leq 2||\bs S_{zw}||_2^2+2||{\bs s}_{zw}-\bs S_{zw}||_2^2 = O_p(K_n),\ 
||\hat{\bs s}_{zw}||_2^2 = O_p(K_n),
\end{split}\]
and thus, 
\[\begin{split}
\left|\hat{\bs s}_{1w}\hat{\bs s}_{0w}^\top - \bs s_{1w}\bs s_{0w}^\top \right| = O_p(K_n \cdot \max\{\sqrt{c_n/n},n^{-1/4}\}).
\end{split}\]
Under Assumption \ref{assumption:step1},  
\begin{equation}\label{eq:shat_shat}
\left|\hat{\bs s}_{1w}\hat{\bs s}_{0w}^\top - \bs s_{1w}\bs s_{0w}^\top \right| = \Tilde{A}_n\cdot o_p(1).
\end{equation}

Since 
\[\begin{split}
s_z^2 = \frac{n_z r_{1-z}^2}{n_z-1}\big[\hat{F}_z(q_{z,\alpha})-\hat{F}_z^2(q_{z,\alpha})\big],\ 
\hat{s}_z^2 = \frac{n_z r_{1-z}^2}{n_z-1}\big[\hat{F}_z(\hat{q}_{z,\alpha})-\hat{F}_z^2(\hat{q}_{z,\alpha})\big],
\end{split}\]
Then 
\[\begin{split}
|\hat{s}_z^2-s_z^2| &= \frac{n_z r_{1-z}^2}{n_z-1}\cdot \left|\hat{F}_z(\hat{q}_{z,\alpha}) - \hat{F}_z(q_{z,\alpha})\right|\cdot \left|1-\hat{F}_z(\hat{q}_{z,\alpha}) - \hat{F}_z(q_{z,\alpha})\right|\\ &= \left|\hat{F}_z(\hat{q}_{z,\alpha}) - \hat{F}_z(q_{z,\alpha})\right|\cdot O(1),
\end{split}\]
where Eq.'s~\eqref{eq:Fhat1} and \eqref{eq:Fhat2} imply that under ReM,
\[\begin{split}
|\hat{F}_z(\hat{q}_{z,\alpha}) - \hat{F}_z(q_{z,\alpha})| &= \left|\frac{1}{n_z}\sum_{i: Z_i=z} \left(\one\{Y_i(z) \leq \hat{q}_{z,\alpha}\}-\one\{Y_i(z) \leq q_{z,\alpha}\}\right)\right|\\
&\leq \frac{1}{n_z}\sum_{i: Z_i=z} \one\{q_{z,\alpha} < Y_i(z) \leq \hat{q}_{z,\alpha}\}
+ \frac{1}{n_z}\sum_{i: Z_i=z} \one\{\hat{q}_{z,\alpha} < Y_i(z) \leq q_{z,\alpha}\}
\\
&= O_p(n^{-1/2}).
\end{split}\]
Since Assumption \ref{assumption:step1} implies $n^{-1/4}/\Tilde{A}_n = o(1)$,
we have 
\begin{equation}\label{eq:s_hatz}
|\hat{s}_z^2-s_z^2| = O_p(n^{-1/2}) = \Tilde{A}_n \cdot o_p(1).
\end{equation}

\hspace{\fill}\\


By definition,
\[\begin{split}
|\hat{f}_{z,n}(\hat{q}_{z,\alpha}) - f_{z,n}(q_{z,\alpha})| 
\leq |\hat{f}_{z,n}(\hat{q}_{z,\alpha}) - \hat{f}_{z,n}(q_{z,\alpha})|
+ |\hat{f}_{z,n}(q_{z,\alpha})- f_{z,n}(q_{z,\alpha})|,
\end{split}\]
where 
\[\begin{split}
&\hat{f}_{z,n}({q}_{z,\alpha})\\
=& \frac{1}{n_z}\sum_{i:Z_i=z} \frac{1}{h_n}K\left(\frac{Y_i(z) - {q}_{z,\alpha}}{h_n}\right)\\
=& \frac{1}{n_z}\sum_{i:Z_i=z} \frac{1}{h_n}K\left(\frac{Y_i(z) - {q}_{z,\alpha}}{h_n}\right)-
\frac{1}{n}\sum_{i=1}^n \frac{1}{h_n}K\left(\frac{Y_i(z) - {q}_{z,\alpha}}{h_n}\right)+
\frac{1}{n}\sum_{i=1}^n \frac{1}{h_n}K\left(\frac{Y_i(z) - {q}_{z,\alpha}}{h_n}\right).
\end{split}\]
Hence, 
\[\begin{split}
&|\hat{f}_{z,n}(\hat{q}_{z,\alpha}) - f_{z,n}(q_{z,\alpha})| \\
\leq & \left|
\frac{1}{n_z}\sum_{i:Z_i=z} \frac{1}{h_n}K\left(\frac{Y_i(z) - \hat{q}_{z,\alpha}}{h_n}\right)-
\frac{1}{n_z}\sum_{i:Z_i=z} \frac{1}{h_n}K\left(\frac{Y_i(z) - {q}_{z,\alpha}}{h_n}\right)\right|\\
& + \left|\frac{1}{n_z}\sum_{i:Z_i=z} \frac{1
}{h_n}K\left(\frac{Y_i(z) - {q}_{z,\alpha}}{h_n}\right)-
\frac{1}{n}\sum_{i=1}^n \frac{1
}{h_n}K\left(\frac{Y_i(z) - {q}_{z,\alpha}}{h_n}\right)\right|\\
& + \left|\frac{1}{n}\sum_{i=1}^n \frac{1
}{h_n}K\left(\frac{Y_i(z) - {q}_{z,\alpha}}{h_n}\right)-
f_{z,n}(q_{z,\alpha})\right|\\
=& \textbf{I} + \textbf{II} + \textbf{III}.
\end{split}\]

\hspace{\fill}\\
\textbf{I:}
By definition, for Gaussian kernel $K(u) = \frac{1}{\sqrt{2\pi}}e^{-u^2/2}$, we have
\[\begin{split}
\frac{1}{n_z}\sum_{i:Z_i=z} \frac{1
}{h_n}K\left(\frac{Y_i(z) - {q}_{z,\alpha}}{h_n}\right)
&= \int \frac{1}{h_n}
K\left(\frac{y - {q}_{z,\alpha}}{h_n}\right)
d\hat{F}_z(y)\\
&= \int \frac{1}{h_n}
K\left(u\right)
d\hat{F}_z({q}_{z,\alpha} + h_n u)\\
&= \frac{1}{h_n}
K\left(u\right)
\hat{F}_z({q}_{z,\alpha} + h_n u)|_{-\infty}^{\infty} - 
\int \frac{1}{h_n}
\hat{F}_z({q}_{z,\alpha} + h_n u)K'(u)
d u\\
&= -\int \frac{1}{h_n}
\hat{F}_z({q}_{z,\alpha} + h_n u)K'(u)du.
\end{split}\]
The normal kernel satisfies that $K'(u) = -uK(u)$, 
\[\begin{split}
\int -uK'(u) du = \int u^2K(u) du = 1,\       
\int K'(u) du = 0.
\end{split}\]
Hence, 
\[\begin{split}
\frac{1}{n_z}\sum_{i:Z_i=z} \frac{1
}{h_n}K\left(\frac{Y_i(z) - {q}_{z,\alpha}}{h_n}\right)
&= \int \frac{\hat{F}_z({q}_{z,\alpha} + h_n u)-
\hat{F}_z({q}_{z,\alpha})}{h_n u}
(-uK'(u))du.
\end{split}\]
Similarly, 
\[\begin{split}
\frac{1}{n_z}\sum_{i:Z_i=z} \frac{1
}{h_n}K\left(\frac{Y_i(z) - \hat{q}_{z,\alpha}}{h_n}\right)
&= \int \frac{\hat{F}_z(\hat{q}_{z,\alpha} + h_n u)-
\hat{F}_z(\hat{q}_{z,\alpha})}{h_n u}
(-uK'(u))du.
\end{split}\]
Therefore, 
\[\begin{split}
&\left|\frac{1}{n_z}\sum_{i:Z_i=z} \frac{1
}{h_n}K\left(\frac{Y_i(z) - {q}_{z,\alpha}}{h_n}\right)
- \frac{1}{n_z}\sum_{i:Z_i=z} \frac{1
}{h_n}K\left(\frac{Y_i(z) - \hat{q}_{z,\alpha}}{h_n}\right)\right|\\
\leq & \int \left|\frac{\hat{F}_z({q}_{z,\alpha} + h_n u)-
\hat{F}_z({q}_{z,\alpha})}{h_n u}-
\frac{\hat{F}_z(\hat{q}_{z,\alpha} + h_n u)-
\hat{F}_z(\hat{q}_{z,\alpha})}{h_n u}\right|
|uK'(u)|du\\
\leq & \int \left|\frac{\hat{F}_z({q}_{z,\alpha} + h_n u)-
\hat{F}_z(\hat{q}_{z,\alpha} + h_n u)}{h_n}\right|
|K'(u)|du
+ \int \left|\frac{\hat{F}_z({q}_{z,\alpha})-
\hat{F}_z(\hat{q}_{z,\alpha})}{h_n}\right|
|K'(u)|du.
\end{split}\]
By Eq.'s~\eqref{eq:Fhat1} and \eqref{eq:Fhat2}, under ReM,
\[\begin{split}
\left|\hat{F}_z({q}_{z,\alpha})-
\hat{F}_z(\hat{q}_{z,\alpha})\right| = O_p(n^{-1/2}).
\end{split}\]
Since $\sqrt{n}|\hat{q}_{z,\alpha}-q_{z,\alpha}| = O_p(1)$ under ReM, 
then for any $\epsilon > 0$, there exists some $M_\epsilon >0$ and $N_\epsilon > 0$
s.t. for any $n > N_\epsilon$,
\[\begin{split}
\pr\left(\sqrt{n}|\hat{q}_{z,\alpha}-q_{z,\alpha}| \leq M_\epsilon \mid M\leq a_n\right) \geq 1-\epsilon.
\end{split}\]
According to Assumption \ref{assumption:derivativeF}, 
there exists some constant $c_\epsilon > 0$ such that for any $x$, 
\[\begin{split}
\sqrt{n}(F_z(q_{z,\alpha} + x + M_\epsilon/\sqrt{n}) - F_z({q}_{z,\alpha} + x))
& \leq M_\epsilon\cdot f_{z,n}(q_{z,\alpha} + x) + r_{z,n}(q_{z,\alpha} + x) \leq c_\epsilon,\\
\sqrt{n}(F_z(q_{z,\alpha} + x) - F_z({q}_{z,\alpha} + x - M_\epsilon/\sqrt{n}))
& \leq M_\epsilon\cdot f_{z,n}(q_{z,\alpha} + x - M_\epsilon/\sqrt{n}) \\&\ \ \ \  + r_{z,n}(q_{z,\alpha} + x - M_\epsilon/\sqrt{n}) \leq c_\epsilon.
\end{split}\]
Hence, for any $n > N_\epsilon$,
\[\begin{split}
&\pr\left(\sup_x n^{-1/2}\sum_{i=1}^n 1\{\hat{q}_{z,\alpha} + x < Y_i(z) \leq {q}_{z,\alpha}+x\}
\leq c_\epsilon \mid M\leq a_n\right)\\
= & \pr\left(n^{-1/2}\sum_{i=1}^n 1\{\hat{q}_{z,\alpha} + x < Y_i(z) \leq {q}_{z,\alpha}+x\}
\leq c_\epsilon,\forall\ x \mid M\leq a_n\right)\\
\geq & \pr\left(\sum_{i=1}^n 1\{\hat{q}_{z,\alpha} + x < Y_i(z) \leq {q}_{z,\alpha}+x\}
\leq \sum_{i=1}^n 1\{q_{z,\alpha} + x - M_\epsilon/\sqrt{n} < Y_i(z) \leq {q}_{z,\alpha}+x\},\forall\ x \mid M\leq a_n\right)\\
\geq & \pr\left(\sqrt{n}|\hat{q}_{z,\alpha}-q_{z,\alpha}| \leq M_\epsilon \mid M\leq a_n\right) \geq 1-\epsilon.
\end{split}\]
Similarly, we have 
\[\begin{split}
&\pr\left(\sup_x n^{-1/2}\sum_{i=1}^n 1\{{q}_{z,\alpha} + x < Y_i(z) \leq \hat{q}_{z,\alpha}+x\}
\leq c_\epsilon \mid M\leq a_n\right) \geq 1-\epsilon.
\end{split}\]
Hence, under ReM,
\[\begin{split}
\sup_u  \left|F_z(\hat{q}_{z,\alpha} + h_n u) - F_z({q}_{z,\alpha}+ h_n u)\right|
= O_p(n^{-1/2}).
\end{split}\]
By definition, 
\[\begin{split}
\frac{1}{n_z}\sum_{i:Z_i=z} 1\{\hat{q}_{z,\alpha}+ h_n u < Y_i(z) \leq {q}_{z,\alpha}+ h_n u\}
&\leq \frac{n}{n_z}\frac{1}{n}\sum_{i=1}^n 
1\{\hat{q}_{z,\alpha}+ h_n u < Y_i(z) \leq {q}_{z,\alpha}+ h_n u\},\\
\frac{1}{n_z}\sum_{i:Z_i=z} 1\{{q}_{z,\alpha}+ h_n u < Y_i(z) \leq \hat{q}_{z,\alpha}+ h_n u\}
&\leq \frac{n}{n_z}\frac{1}{n}\sum_{i=1}^n 
1\{{q}_{z,\alpha}+ h_n u < Y_i(z) \leq \hat{q}_{z,\alpha}+ h_n u\}.
\end{split}\]
Hence, under ReM,
\[\begin{split}
\sup_u \left|\hat{F}_z({q}_{z,\alpha} + h_n u) - \hat{F}_z(\hat{q}_{z,\alpha} + h_n u)\right| = O_p(n^{-1/2}).
\end{split}\]
Then we have 
\[\begin{split}
\int \left|\frac{\hat{F}_z({q}_{z,\alpha} + h_n u)-
\hat{F}_z(\hat{q}_{z,\alpha} + h_n u)}{h_n}\right|
|K'(u)|du = O_p\left(\frac{1}{\sqrt{n}h_n}\right),
\end{split}\]
and 
\[\begin{split}
\int \left|\frac{\hat{F}_z({q}_{z,\alpha})-
\hat{F}_z(\hat{q}_{z,\alpha})}{h_n}\right|
|K'(u)|du = O_p\left(\frac{1}{\sqrt{n}h_n}\right).
\end{split}\]
Therefore, under ReM, {for $\sqrt{n}h_n \rightarrow \infty$},
\[\begin{split}
&\textbf{I} = \left|
\frac{1}{n_z}\sum_{i:Z_i=z} \frac{1}{h_n}K\left(\frac{Y_i(z) - \hat{q}_{z,\alpha}}{h_n}\right)-
\frac{1}{n_z}\sum_{i:Z_i=z} \frac{1}{h_n}K\left(\frac{Y_i(z) - {q}_{z,\alpha}}{h_n}\right)\right|\\
\leq & O_p\left(\frac{1}{\sqrt{n}h_n}\right)
= o_p(1).
\end{split}\]

\hspace{\fill}\\
\textbf{II:}
Let $a_i = \frac{1
}{h_n}K\left(\frac{Y_i(z) - {q}_{z,\alpha}}{h_n}\right)$,
and define 
\[\begin{split}
s^2 = \frac{1}{n-1}\sum_{i=1}^n (a_i - \bar{a})^2 = O(h_n^{-2}).
\end{split}\]
Then by Lemma \ref{lem:bloniarz2016lasso}, under CRE, for any $t > 0$, 
there exists a constant $M > 0$ such that
\[\begin{split}
&\pr\left(
\frac{1}{n_z}\sum_{i:Z_i=z} \frac{1
}{h_n}K\left(\frac{Y_i(z) - {q}_{z,\alpha}}{h_n}\right)-
\frac{1}{n}\sum_{i=1}^n \frac{1
}{h_n}K\left(\frac{Y_i(z) - {q}_{z,\alpha}}{h_n}\right)
\geq t\right)\\
\leq & \exp\left\{-\frac{n r_z^2 t^2}{(1+\tau)^2s^2}\right\}
\leq \exp\left\{-M\cdot nh_n^2 t^2\right\}.
\end{split}\]
Moreover, 
\[\begin{split}
&\pr\left(
\frac{1}{n_z}\sum_{i:Z_i=z} \frac{1
}{h_n}K\left(\frac{Y_i(z) - {q}_{z,\alpha}}{h_n}\right)-
\frac{1}{n}\sum_{i=1}^n \frac{1
}{h_n}K\left(\frac{Y_i(z) - {q}_{z,\alpha}}{h_n}\right)
\leq -t\right)\\
=& \pr\left(
\frac{1}{n_{1-z}}\sum_{i:Z_i=1-z} \frac{1
}{h_n}K\left(\frac{Y_i(z) - {q}_{z,\alpha}}{h_n}\right)-
\frac{1}{n}\sum_{i=1}^n \frac{1
}{h_n}K\left(\frac{Y_i(z) - {q}_{z,\alpha}}{h_n}\right)
\geq \frac{n_z t}{n_{1-z}}
\right)
\\
\leq & \exp\left\{-\frac{n r_z^2 t^2}{(1+\tau)^2s^2}\right\}
\leq \exp\left\{-M\cdot nh_n^2 t^2\right\}.
\end{split}\]
Hence, 
\[\begin{split}
&\pr\left(\left|
\frac{1}{n_z}\sum_{i:Z_i=z} \frac{1
}{h_n}K\left(\frac{Y_i(z) - {q}_{z,\alpha}}{h_n}\right)-
\frac{1}{n}\sum_{i=1}^n \frac{1
}{h_n}K\left(\frac{Y_i(z) - {q}_{z,\alpha}}{h_n}\right)\right|
\geq t\right)\\
\leq & \exp\left\{-\frac{n r_z^2 t^2}{(1+\tau)^2s^2}\right\}
\leq \exp\left\{-M\cdot nh_n^2 t^2\right\}.
\end{split}\]
According to Assumption \ref{assumption:pn}, under ReM, we have 
\[\begin{split}
&\pr\left(\left|
\frac{1}{n_z}\sum_{i:Z_i=z} \frac{1
}{h_n}K\left(\frac{Y_i(z) - {q}_{z,\alpha}}{h_n}\right)-
\frac{1}{n}\sum_{i=1}^n \frac{1
}{h_n}K\left(\frac{Y_i(z) - {q}_{z,\alpha}}{h_n}\right)\right|
\geq t \mid M \leq a_n\right)\\
\leq & \frac{\exp\left\{-M\cdot nh_n^2 t^2\right\}}{\pr(M \leq a_n)}
= \exp\left\{-M\cdot nh_n^2 t^2 - \log p_n + o(1)\right\}\\
=& \exp\left\{-M\cdot nh_n^2 t^2 - O(n^m) + o(1)\right\}.
\end{split}\]
For {$\sqrt{n}h_n \rightarrow \infty$} and under ReM, we have 
\[\begin{split}
\textbf{II} = \left|
\frac{1}{n_z}\sum_{i:Z_i=z} \frac{1
}{h_n}K\left(\frac{Y_i(z) - {q}_{z,\alpha}}{h_n}\right)-
\frac{1}{n}\sum_{i=1}^n \frac{1
}{h_n}K\left(\frac{Y_i(z) - {q}_{z,\alpha}}{h_n}\right)\right| = o_p(1).
\end{split}\]

\hspace{\fill}\\
\textbf{III:}
By definition,
\[\begin{split}
\frac{1}{n}\sum_{i=1}^n \frac{1
}{h_n}K\left(\frac{Y_i(z) - {q}_{z,\alpha}}{h_n}\right)
&= \int \frac{1}{h_n}
K\left(\frac{y - {q}_{z,\alpha}}{h_n}\right)
dF_z(y)\\
&= \int \frac{1}{h_n}
K\left(u\right)
dF_z({q}_{z,\alpha} + h_n u)\\
&= \frac{1}{h_n}
K\left(u\right)
F_z({q}_{z,\alpha} + h_n u)|_{-\infty}^{\infty} - 
\int \frac{1}{h_n}
F_z({q}_{z,\alpha} + h_n u)K'(u)
d u\\
&= -\int \frac{1}{h_n}
F_z({q}_{z,\alpha} + h_n u)K'(u)du\\
&= \int \frac{F_z({q}_{z,\alpha} + h_n u)-
F_z({q}_{z,\alpha})}{h_n u}
(-uK'(u))du.
\end{split}\]
Since
\[\begin{split}
f_{z,n}(q_{z,\alpha})
&= \int f_{z,n}(q_{z,\alpha}) (-uK'(u))du,
\end{split}\]
and $K'(u) = -uK(u)$,
we have
\[\begin{split}
&\frac{1}{n}\sum_{i=1}^n \frac{1
}{h_n}K\left(\frac{Y_i(z) - {q}_{z,\alpha}}{h_n}\right)-
f_{z,n}(q_{z,\alpha})\\
=& \int \left[\frac{F_z({q}_{z,\alpha} + h_n u)-
F_z({q}_{z,\alpha})}{h_n u}
- f_{z,n}(q_{z,\alpha})
\right]
u^2K(u)du.
\end{split}\]
Then we have 
\[\begin{split}
&\left|\frac{1}{n}\sum_{i=1}^n \frac{1
}{h_n}K\left(\frac{Y_i(z) - {q}_{z,\alpha}}{h_n}\right)-
f_{z,n}(q_{z,\alpha})\right|\\
\leq & \int \left|\frac{F_z({q}_{z,\alpha} + h_n u)-
F_z({q}_{z,\alpha})}{h_n u}  
- f_{z,n}(q_{z,\alpha})\right|
u^2K(u)du\\
=& \int_{-2n^{\gamma-1/2}/h_n}^{2n^{\gamma-1/2}/h_n}
\left|\frac{F_z({q}_{z,\alpha} + h_n u)-
F_z({q}_{z,\alpha})}{h_n u}  
- f_{z,n}(q_{z,\alpha})\right|
u^2K(u)du\\
&+ \int_{|u|>2n^{\gamma-1/2}/h_n}
\left|\frac{F_z({q}_{z,\alpha} + h_n u
)-F_z({q}_{z,\alpha})}{h_n u}
- f_{z,n}(q_{z,\alpha})\right|
u^2K(u)du.
\end{split}\]
For the first term, by Assumption \ref{assumption:derivativeF},
\[\begin{split}
&\int_{-2n^{\gamma-1/2}/h_n}^{2n^{\gamma-1/2}/h_n}
\left|\frac{F_z({q}_{z,\alpha} + h_n u)-
F_z({q}_{z,\alpha})}{h_n u}   
- f_{z,n}(q_{z,\alpha})\right|
u^2K(u)du\\
\leq & \int_{-2n^{\gamma-1/2}/h_n}^{2n^{\gamma-1/2}/h_n}
\frac{r_{z,n}(q_{z,\alpha})}{\sqrt{n} h_n |u|}
u^2K(u)du\\
=& \frac{r_{z,n}(q_{z,\alpha})}{\sqrt{n} h_n}
\int_{-2n^{\gamma-1/2}/h_n}^{2n^{\gamma-1/2}/h_n}
|u| K(u)du
= O\left(\frac{1}{\sqrt{n}h_n}\right).
\end{split}\]
For $\sqrt{n}h_n \rightarrow \infty$, we have 
\[\begin{split}
&\int_{-2n^{\gamma-1/2}/h_n}^{2n^{\gamma-1/2}/h_n}
\left|\frac{F_z({q}_{z,\alpha} + h_n u)-
F_z({q}_{z,\alpha})}{h_n u}   
- f_{z,n}(q_{z,\alpha})\right|
u^2K(u)du = o(1).
\end{split}\]
For the second term, 
since $n^\gamma/(\sqrt{n}h_n) = \Theta(n^v)$ for some $v > 0$, we have 
\[\begin{split}
&\int_{|u|>2n^{\gamma-1/2}/h_n}
\left|\frac{F_z({q}_{z,\alpha} + h_n u
)-F_z({q}_{z,\alpha})}{h_n u}
- f_{z,n}(q_{z,\alpha})\right|
u^2K(u)du\\
\leq& \frac{2}{h_n}\int_{|u|>2n^{\gamma-1/2}/h_n}
|u|K(u)du + 
f_{z,n}(q_{z,\alpha})
\int_{|u|>2n^{\gamma-1/2}/h_n}
u^2K(u)du\\
=& \frac{4}{h_n}\int_{u>2n^{\gamma-1/2}/h_n}
-K'(u)du + f_{z,n}(q_{z,\alpha})
\int_{|u|>2n^{\gamma-1/2}/h_n}
u^2K(u)du\\
=& \frac{4}{h_n}K\left(\frac{2n^{\gamma-1/2}}{h_n}\right)
+ f_{z,n}(q_{z,\alpha})
\int_{|u|>2n^{\gamma-1/2}/h_n}
u^2K(u)du\\
=& o(1).
\end{split}\]
Hence, for {$\sqrt{n}h_n \rightarrow \infty$ and 
$n^\gamma/(\sqrt{n}h_n)  = \Theta(n^v)$ for some $v > 0$}, we have 
\[\begin{split}
&\textbf{III} = \left|\frac{1}{n}\sum_{i=1}^n \frac{1
}{h_n}K\left(\frac{Y_i(z) - {q}_{z,\alpha}}{h_n}\right)-
f_{z,n}(q_{z,\alpha})\right|
= o(1).
\end{split}\]
Above all, under ReM, we have
\[\begin{split}
|\hat{f}_{z,n}(\hat{q}_{z,\alpha}) - f_{z,n}(q_{z,\alpha})|
\leq \textbf{I}+\textbf{II}+\textbf{III}
= o_p(1) + o_p(1) + o(1) = o_p(1).
\end{split}\]
Under Assumption \ref{assumption:derivativeF}, 
\begin{equation}\label{eq:fhat}
|\hat{f}_{z,n}(\hat{q}_{z,\alpha})-f_{z,n}(q_{z,\alpha})| = f_{z,n}(q_{z,\alpha}) \cdot o_p(1).
\end{equation}

\hspace{\fill}\\
Combining Eq.'s \eqref{eq:shat}, \eqref{eq:shat_shat}, \eqref{eq:s_hatz} and \eqref{eq:fhat}, we have
\[\begin{split}
&r_1r_0 \cdot \hat{C}_n = \frac{\hat{s}_{1}^2}{\hat{f}_{1,n}^2(\hat{q}_{1,\alpha})}+\frac{\hat{s}_{0}^2}{\hat{f}_{0,n}^2(\hat{q}_{0,\alpha})}+\frac{2\min \left\{\frac{r_1}{r_0}\hat{s}_1^2,\frac{r_0}{r_1}\hat{s}_0^2\right\}}{\hat{f}_{1,n}(\hat{q}_{1,\alpha})\hat{f}_{0,n}(\hat{q}_{0,\alpha})}\\
=& \frac{1+ o_p(1)}{{f}_{1,n}^2({q}_{1,\alpha})} \big[s_1^2 + \Tilde{A}_n o_p(1)\big] + 
\frac{1+ o_p(1)}{{f}_{0,n}^2({q}_{0,\alpha})} \big[s_0^2 + \Tilde{A}_n o_p(1)\big]
+ \frac{2(1+o_p(1))\left(\min \left\{\frac{r_1}{r_0}{s}_1^2,\frac{r_0}{r_1}{s}_0^2\right\} + \tilde{A}_n o_p(1)\right)}{{f}_{1,n}({q}_{1,\alpha}){f}_{0,n}({q}_{0,\alpha})}\\
=& r_1r_0\cdot \hat{C}_n'(1+o_p(1)) + \Tilde{A}_n \cdot o_p(1),
\end{split}\]
and 
\[\begin{split}
r_1r_0 \cdot \hat{B}_n &= \frac{||\hat{\bs s}_{1w}||_2^2}{\hat{f}_{1,n}^2(\hat{q}_{1,\alpha})}+\frac{||\hat{\bs s}_{0w}||_2^2}{\hat{f}_{0,n}^2(\hat{q}_{0,\alpha})}-\frac{2\cdot \hat{\bs s}_{1w}\hat{\bs s}_{0w}^\top}{\hat{f}_{1,n}(\hat{q}_{1,\alpha})\hat{f}_{0,n}(\hat{q}_{0,\alpha})}\\
&= \frac{1+o_p(1)}{f_{1,n}^2(q_{1,\alpha})} \cdot \left[||{\bs s}_{1w}||_2^2 + \Tilde{A}_n \cdot o_p(1)\right] + \frac{1+o_p(1)}{f_{0,n}^2(q_{0,\alpha})} \cdot \left[||{\bs s}_{0w}||_2^2 + \Tilde{A}_n \cdot o_p(1)\right]\\
&- \frac{2(1+o_p(1))}{f_{1,n}(q_{1,\alpha})f_{0,n}(q_{0,\alpha})} \cdot \left[\bs s_{1w}\bs s_{0w}^\top + \Tilde{A}_n \cdot o_p(1)\right]\\
&= r_1r_0 \cdot \hat{B}_n' +\hat{C}_n' o_p(1) + \Tilde{A}_n \cdot o_p(1),
\end{split}\]
where the last equation holds since 
$|\bs s_{1w}\bs s_{0w}^\top| \leq (||\bs s_{1w}||_2^2+||\bs s_{0w}||_2^2)/2$, and for $z=0,1$,
\[\begin{split}
||\bs s_{zw}||_2^2 &\leq ||\bs S_{zw}||_2^2 + O_p( K_n \sqrt{c_n/n}) \leq S_{z}^2 + \Tilde{A}_n \cdot o_p(1)\\
&= s_{z}^2 + O_p(\sqrt{b_n/n}) + \Tilde{A}_n \cdot o_p(1)
= s_{z}^2 + \Tilde{A}_n \cdot o_p(1).
\end{split}\]
The proof of step 1 shows that 
$\hat{C}_n' = \tilde{C}_n + \tilde{A}_n\cdot o_p(1)$.
Hence, $\hat{C}_n' o_p(1) = \tilde{C}_n\cdot o_p(1) + \tilde{A}_n\cdot o_p(1)$.
Assumption \ref{assumption:R2_2} implies that 
\[\begin{split}
\frac{\tilde{C}_n}{\tilde{A}_n}
= 1 + \frac{B_n}{\tilde{A}_n} \leq 1 + \frac{B_n}{A_n}
= 1 + \frac{\tilde{R}^2_{\bq}}{1-\tilde{R}^2_{\bq}} = O(1),
\end{split}\]
and thus, $\hat{C}_n' o_p(1) = \tilde{A}_n\cdot o_p(1)$.
Then we have
\[\begin{split}
\hat{A}_n = \hat{C}_n - \hat{B}_n = \hat{A}_n' + \Tilde{A}_n \cdot o_p(1),
\end{split}\]
i.e.,
\[\begin{split}
|\hat{A}_n-\hat{A}_n'| = \Tilde{A}_n \cdot o_p(1).
\end{split}\]
Similarly, since 
\[\begin{split}
|\hat{B}_n-\hat{B}_n'| = \Tilde{A}_n \cdot o_p(1),
\end{split}\]
we have 
\[\begin{split}
\max\{|\hat{A}_n-\hat{A}_n'|,|\hat{B}_n-\hat{B}_n'|\} = \Tilde{A}_n \cdot o_p(1).
\end{split}\]
\end{proof}
Combining the results in Step 1 and Step 2, i.e.,
\[\begin{split}
\max\{|\hat{A}_n'-\tilde{A}_n|,|\hat{B}_n'-B_n|\} = \Tilde{A}_n \cdot o_p(1),
\end{split}\]
and 
\[\begin{split}
\max\{|\hat{A}_n-\hat{A}_n'|,|\hat{B}_n-\hat{B}_n'|\} = \Tilde{A}_n \cdot o_p(1).
\end{split}\]
Then we have 
\[\begin{split}
\max\{|\hat{A}_n-\tilde{A}_n|,|\hat{B}_n-B_n|\} = \Tilde{A}_n \cdot o_p(1).
\end{split}\]
We complete the proof.

\hspace{\fill}\\
\textbf{Proof of Theorem \ref{thm:CI} (ii):}
Define
\[\begin{split}
\theta_n &= A_n^{1/2}\varepsilon + B_n^{1/2}L_{K_n,a_n},\\
\tilde{\theta}_n &= \tilde{A}_n^{1/2}\varepsilon + B_n^{1/2}L_{K_n,a_n},\\
\hat{\theta}_n &= \hat{A}_n^{1/2}\varepsilon + \hat{B}_n^{1/2}L_{K_n,a_n},
\end{split}\]
and let $q_{n,\alpha} = q_\alpha(A_n,B_n,K_n,a_n)$ denote the $\alpha$-th quantile of $\theta_n$, 
$\tilde{q}_{n,\alpha} = q_\alpha(\tilde{A}_n,{B}_n,K_n,a_n)$ denote the $\alpha$-th $\tilde\theta_n$,
and 
$\hat{q}_{n,\alpha} = q_\alpha(\hat{A}_n,\hat{B}_n,K_n,a_n)$ denote the $\alpha$-th $\hat\theta_n$.
According to Theorem \ref{thm:CI} (i), under ReM, $\max\{|\hat{A}_n-\tilde{A}_n|,|\hat{B}_n-B_n|\} = \Tilde{A}_n \cdot o_p(1).$
Then for any subsequence $\{n_j:j=1,2,\ldots\}$, there exists a further subsequence $\{m_j:j=1,2,\ldots\}\subset \{n_j:j=1,2,\ldots\}$ such that $\hat{A}_{m_j}-\tilde{A}_{m_j}/\tilde{A}_{m_j} \xrightarrow{a.s.} 0$, and $\hat{B}_{m_j}-\tilde{B}_{m_j}/\tilde{A}_{m_j} \xrightarrow{a.s.} 0$ as $j\rightarrow\infty$.
Then Lemma A36 in \cite{wang2022rerandomization} implies that for any $0 < \alpha < \beta <1$,
we have $\one\{\hat{q}_{m_j,\beta} \leq \tilde{q}_{m_j,\alpha}\} \xrightarrow{a.s.} 0$.
Therefore, under ReM, $\one\{\hat{q}_{n,\beta} \leq \tilde{q}_{n,\alpha}\} \xrightarrow{\pr} 0$.
Consequently, 
\[\begin{split}
\E\left\{\one\{\hat{q}_{n,\beta} \leq \tilde{q}_{n,\alpha}\}\mid M\leq a_n\right\} = \pr\left(\hat{q}_{n,\beta} \leq \tilde{q}_{n,\alpha}\mid M\leq a_n\right)
\rightarrow 0.
\end{split}\]
Then for any $\eta \in (0,(1-\alpha)/2)$,
\[\begin{split}
&\pr\left(\tau_\alpha \in \hat{\cC}_\alpha \mid M \leq a_n\right)= \pr\left(\sqrt{n}|\tau_\alpha - \hat{\tau}_\alpha|\leq \hat{q}_{n,1-\alpha/2} \mid M \leq a_n\right)\\
\geq & \pr\left(\sqrt{n}|\tau_\alpha - \hat{\tau}_\alpha|\leq \hat{q}_{n,1-\alpha/2},
\hat{q}_{n,1-\alpha/2}\geq \tilde{q}_{n,1-\alpha/2-\eta}\mid M \leq a_n\right)\\
\geq & \pr\left(\sqrt{n}|\tau_\alpha - \hat{\tau}_\alpha|\leq \tilde{q}_{n,1-\alpha/2-\eta},
\hat{q}_{n,1-\alpha/2}\geq \tilde{q}_{n,1-\alpha/2-\eta}\mid M \leq a_n\right)\\
\geq & \pr\left(\sqrt{n}|\tau_\alpha - \hat{\tau}_\alpha|\leq \tilde{q}_{n,1-\alpha/2-\eta}\mid M \leq a_n\right)-
\pr\left(\hat{q}_{n,1-\alpha/2}< \tilde{q}_{n,1-\alpha/2-\eta} \mid M \leq a_n\right)\\
=& \pr\left(\sqrt{n}|\tau_\alpha - \hat{\tau}_\alpha|\leq \tilde{q}_{n,1-\alpha/2-\eta}\mid M \leq a_n\right) + o(1).
\end{split}\]
Theorem \ref{thm:QTE} shows that 
\[\begin{split}
&\pr\left(\sqrt{n}|\tau_\alpha - \hat{\tau}_\alpha|\leq \tilde{q}_{n,1-\alpha/2-\eta}\right)
= \pr\left(|\theta_n|\leq \tilde{q}_{n,1-\alpha/2-\eta}\right) + o(1)\\
\geq & \pr\left(|\tilde\theta_n|\leq \tilde{q}_{n,1-\alpha/2-\eta}\right) + o(1)
= 1-\alpha-2\eta + o(1).
\end{split}\]
These imply that for any $\eta \in (0,(1-\alpha)/2)$,
\[\begin{split}
\liminf_{n\rightarrow \infty}\pr\left(\tau_\alpha \in \hat{\cC}_\alpha \mid M \leq a_n\right) \geq 1-\alpha-2\eta.
\end{split}\]
Then we have 
\[\begin{split}
\liminf_{n\rightarrow \infty}\pr\left(\tau_\alpha \in \hat{\cC}_\alpha \mid M \leq a_n\right) \geq 1-\alpha.
\end{split}\]
We complete the proof.
\end{proof}

\end{document}